\documentclass[runningheads]{llncs}
\usepackage{graphicx}
\usepackage{float}
\usepackage[
backend=biber,
style=lncs,
sorting=ynt,
]{biblatex}

\AtEveryBibitem{\clearfield{issn}}
\AtEveryBibitem{\clearfield{isbn}}

\usepackage{xpatch}
\xpretobibmacro{bib+doi+url}
  {\iffieldundef{doi}
     {}
     {\clearfield{url}}}
  {}{}

\usepackage[]{enumitem}

\graphicspath{{figures/}}
\usepackage{caption}
\usepackage{subcaption}
\usepackage{array}
\usepackage{multirow}
\newcolumntype{P}[1]{>{\centering\arraybackslash}p{#1}}
\newcolumntype{M}[1]{>{\centering\arraybackslash}m{#1}}
\newcolumntype{Y}{>{\centering\arraybackslash}X}
\usepackage{tabularx}

\begin{document}

\title{sAirflow: Adopting Serverless in a Legacy Workflow Scheduler\thanks{Accepted to Euro-Par 2024. Compared with the camera-ready version, this manuscript contains additional experimental results.}}
\author{Filip Mikina\inst{1} \and
Pawel Zuk\inst{2}\orcidID{0000-0002-4904-7171} \and
Krzysztof Rzadca\inst{3}\orcidID{0000-0002-4176-853X}}
\authorrunning{F.Mikina, P. Zuk, K. Rzadca}
\institute{Institute of Informatics, University of Warsaw
\email{fmikina@gmail.com}\and
University of Southern California 
\email{pawelzuk@isi.edu}\and
Institute of Informatics, University of Warsaw 
\email{krzadca@mimuw.edu.pl}}

\maketitle
\begin{abstract}
Serverless clouds promise efficient scaling, reduced toil and monetary costs.
Yet, serverless-ing a complex, legacy application might require major refactoring and thus is risky.
As a case study, we use Airflow, an industry-standard workflow system.
To reduce migration risk, we propose to limit code modifications by relying on change data capture (CDC) and message queues for internal communication. 
To achieve serverless efficiency, we rely on Function-as-a-Service (FaaS).
Our system, sAirflow, is the first adaptation of the control plane and workers to the serverless cloud --- and it maintains the same interface and most of the code.
Experimentally, we show that sAirflow delivers the key serverless benefits: scaling and cost reduction. We compare sAirflow to MWAA, a managed (SaaS) Airflow.
On Alibaba benchmarks on warm systems, sAirflow performs similarly while halving the monetary cost. On highly parallel workflows on cold systems, sAirflow scales out in seconds to 125 workers, reducing makespan by 2x-7x.
\end{abstract}

\section{Introduction}
Serverless cloud~\cite{10.1145/3368454} products (AWS Lambda, GCP Cloud Run, Azure Functions, OpenWhisk) propose a new kind of contract between the provider and the customer. The customer supplies just the code of the function to execute, while the provider manages resources at the granularity of individual invocations. For customers, this reduces the toil of maintaining infrastructure, and often reduces monetary costs, as typically, the customer pays only for the consumed resources. 
The providers can optimize hardware while providing highly dynamic horizontal scaling: from zero when a function is not invoked to hundreds of concurrent invocations.
This invocation-by-invocation management of resources \emph{by providers} has received considerable attention~\cite{10.1145/3368454}:
e.g., avoiding cold-starts by optimizing the environment pre-warming and evictions~\cite{shahrad2020characterization}; reducing response latency in a warm system by changing how invocations are allocated to workers~\cite{aumala2019beyond}, or by more efficient scheduling at a worker node~\cite{zuk2022call}.
Comparatively, \emph{applications} received less attention; and most of the serverless applications described in the literature~\cite{eismann2021state} seem to be built from scratch. 
Yet, research and industry alike operate on proven, tested and well-understood legacy systems. Throwing away old code and starting from scratch is risky~\cite{fairbanks19ignore-refactor-rewrite}.
Yet, there are few blueprints for refactoring towards serverless~\cite{hamza2024JourneyServerlessMigration}.

The research question we address in this paper is: How can we effectively refactor a complex, legacy application to reap the benefits of serverless computing, including seamless scalability and cost-effectiveness, without introducing significant disruptions to its existing code structure?

As an example of a legacy application we take Airflow, an industry standard for authoring, scheduling, running and monitoring workflows, in particular, data processing pipelines~\cite{harenslak2021data}. Airflow 
is widely used in the industry, both on-premises and as a SaaS offering (AWS' MWAA, Google's Cloud Composer, and Azure's Data Factory Managed Airflow).

The primary challenge in refactoring Airflow lies in its reliance on a metadata database updated with SQL queries from many code locations. To overcome this obstacle, we utilize database-level change data capture (CDC) to transform metadata updates into events then transmitted to the control plane. This pattern allows us to transform Airflow's architecture into event-drive one. Functions from the original Airflow control plane execute as serverless functions (lambdas) that are triggered by events delivered through message queues.
For example, when a user submits a new DAG, a CDC-triggered event invokes the scheduler.
Similarly, we launch user-defined work on serverless offerings: a FaaS executor for shorter tasks (up to 15\,min.); and a universal Container-as-a-Service (CaaS) executor. When a task ends, an event triggers a metadata update, that, in turn, triggers the scheduler. 
No sAirflow code continuously pulls or runs in the background.

\noindent \textbf{The contribution of this paper is as follows:} 
(1) We propose a new software design pattern for adapting to serverless legacy, database-driven applications: to minimize changes in the code, keep the database interactions; and rely on change data capture (CDC) to produce events driving the control plane.
(2) sAirflow is the first serverless adaptation of Airflow's control plane and executors. This enables sAirflow to scale horizontally in seconds to 125 executors and to halve monetary costs. sAirflow thus efficiently surfaces through a legacy system the key advantages of FaaS: elasticity and usage-based pricing.

This paper is organized as follows. We start by reviewing related work in Section~\ref{sec:related-work}. We then describe Function as a Service (FaaS) and Airflow in Section~\ref{sec:system-context}. Section~\ref{sec:design} describes the design and the key implementation details of sAirflow. Section~\ref{sec:eval-method} describes how sAirflow is deployed to the cloud; it also describes the evaluation method. Section~\ref{sec:results} describes results of experiments comparing sAirflow with MWAA.

The source code is available at \url{https://github.com/fiffeek/beeflow} .

\section{Related work}\label{sec:related-work}

\noindent\textbf{Workflow management systems (WMSs):}
\cite{9005494} is a recent survey. 
Popular WMS include Airflow, Pegasus~\cite{deelman2019evolution}, FuncX~\cite{li2022funcx,chard2020funcx} (both mostly used in scientific computing), Pachyderm~\cite{novella2019container} (bioinformatics), 
Argo Workflows 
(big data) and  Kubeflow 
(ML).
We evaluate sAirflow on workflows derived from Alibaba Cloud~\cite{alibaba-clusterdata}, following~\cite{9066946}, advocating real-world-based traces.

\noindent\textbf{Running workflows in the cloud:} 
A survey~\cite{9393895} classifies the available approaches, challenges, and evaluation techniques for scientific workflows in the cloud.
In particular,~\cite{9837056, 7839905, 10.1145/3401025.3401731} focus on systems approaches to achieve a reliable and scalable scientific WMS.
Improvements often concentrate on the scheduling algorithm~\cite{9812609,kijak2018challenges} with, e.g., reinforcement learning~\cite{8976136}, or prediction of task execution times\cite{8013738}.
sAirflow uses a complementary approach: by switching to a different execution model (serverless), we reduce the platform and worker costs.

 \noindent\textbf{Serverless computing:}
In contrast to many papers optimizing serverless back-ends\cite{shahrad2020characterization,aumala2019beyond,zuk2022call}, we take the perspective of a developer of serverless applications.
Our contribution is analogous to adaptations of existing systems to FaaS: Unix shell~\cite{maheo2021serverless}, MapReduce~\cite{gimenez2019framework}, or parallelizing Python~\cite{jonas2017occupy}.
A survey~\cite{eismann2021state} of 89 serverless applications does not analyze whether an application was migrated to serverless. 

A serverless blueprint~\cite{copik22faaskeeper} does not address the challenges of starting from a legacy system. CDC was proposed as one of possible methods to \emph{interoperate} with legacy systems in the software architecture context~\cite{gilbert2021software}, but they do not specifically target \emph{migrating} large, legacy code, nor they quantify performance.

\noindent\textbf{Serverless Workflow Management Frameworks (WMF):}
Some serverless WMFs are built from scratch~\cite{jiang2017serverless,10.1145/3415958.3433082}, thus introducing migration risks and incompatibilities. \cite{10.1145/3415958.3433082} introduces a container-based WMF; similarly to sAirflow, they rely on messages for internal communication.  analyzes serverless-based WMF. 
\cite{burckhardt2022netherite} extends FaaS with stateful, addressable instances to run workflows (compatible with Durable Functions, not Airflow). 
\cite{9582324}~uses serverless containers in HyperFlow 
by extending the executors to run on AWS Lambda and AWS Fargate; HyperFlow orchestrates a one-off workflow, in contrast to Airflow's continously-running control plane coordinating recurring runs of multiple workflows.
\cite{10.1145/3366623.3368137} states that FaaS-based workflow orchestrators (AWS Step Functions or Apache OpenWhisk Composer) are more cost-efficient and easier to scale than Airflow. Our sAirflow addresses this exact shortcoming.

\noindent\textbf{Apache Airflow extensions and serverless:}
Airflow scheduler improvements include~\cite{lin2022global, https://doi.org/10.48550/arxiv.1905.10270}; they propose adding components to predict resource requirements.
\cite{airflow_native_aws_executors} provides an
Airflow executor (a plugin) allowing job scheduling on AWS Batch and managed Kubernetes.
While this plug-in partly addresses the scaling of executors, the control plane is always running --- in contrast to sAirflow that additionally uses serverless architecture for the control plane.

\section{System context: FaaS and Airflow}\label{sec:system-context}

\noindent\textbf{FaaS and related serverless offerings:} A \textit{function} is the core building block in a FaaS platform.
A programmer defines a function by writing code, packaging it, specifying its memory limits, and defining the invoking triggers.
As serverless applications are event-driven, 
the programmer must bind the invocation of a function to an event: e.g., a direct HTTP call; or a periodic, cron-like schedule.
For resiliency, event producers should be decoupled from consumers. A queuing broker (e.g. Kafka) temporary queues events, thus allowing multiple producers and consumers, or consumers to go briefly offline. 

FaaS has, however, certain limitations. First, the maximum time of a single function run is limited (e.g., 15~min in AWS).
Second, a \textit{cold start} occurs when new resources are assigned for an invocation; a cold start increases the response time by between 300~ms and 24~s~\cite{manner2018cold} ---  a significant delay for short functions.
Third, the functions are expected to be stateless, as the environment is ephemeral.
Therefore, state must be stored externally (e.g., in blob storage) 
which may significantly slow down some applications.

\noindent\textbf{Apache Airflow: a workflow scheduler:}
Airflow processes \emph{DAG}s (workflows) of \emph{tasks}.
Tasks are the smallest units of work, with user-defined processing, 
e.g., 
copying files or creating a database table from a query.
Dependencies between tasks are defined
through an API or special operators in the task code. 
A workflow can be launched manually or scheduled to run, e.g., every day at 4 am.
A single execution of a workflow is internally represented by a \emph{DAG run}. 
Similarly, a single execution of a task is a \emph{task instance}.

The scheduler monitors and orchestrates all tasks and DAGs. 
By default, once per minute, the scheduler collects DAGs, 
parses tasks' statuses, 
and adjusts the metadata on the DAGs and the tasks.

Tasks are launched through local or remote \emph{executors}. A \emph{local executor} launches a task in a child OS process.
A \emph{remote executor} sends a task to an external service responsible for queueing and launching. 
For example, the Kubernetes executor contacts a pre-configured Kubernetes cluster and uses this cluster's API to request a pod that will execute the task.

\section{sAirflow: Design and Implementation}\label{sec:design}

We had the following design requirements for sAirflow. 
    (1) \emph{Compatibility} with all Airflow interfaces.
    (2) \emph{Scalability, Performance and Availability} at least on par with the managed Airflow.
    (3) \emph{Reproducibility} by persisting infrastructure in code. 
    (4) \emph{Precise intervention} by limiting the modifications of the Airflow source code.
    (5) \emph{Pay-as-you-use:} Minimize fixed costs. 
sAirflow achieves all but the last goal, as the database (with the accompanying CDC mechanism), an external dependency, are not pay-as-you-use. Currently, even the AWS serverless database, Aurora, does not scale down to 0; and additionally, it does not support CDC, thus requiring extensive changes in Airflow code (Section~\ref{sec:cdc}), contradicting our penultimate goal. However, we emphasize these are just external dependencies that, in the future, could be replaced with pay-as-you-use products, if the cloud provider introduces them (as, e.g., Google's recent Datastream for CDC).

\begin{figure}[tb]
    \centering
    \includegraphics[width=1.0\textwidth]{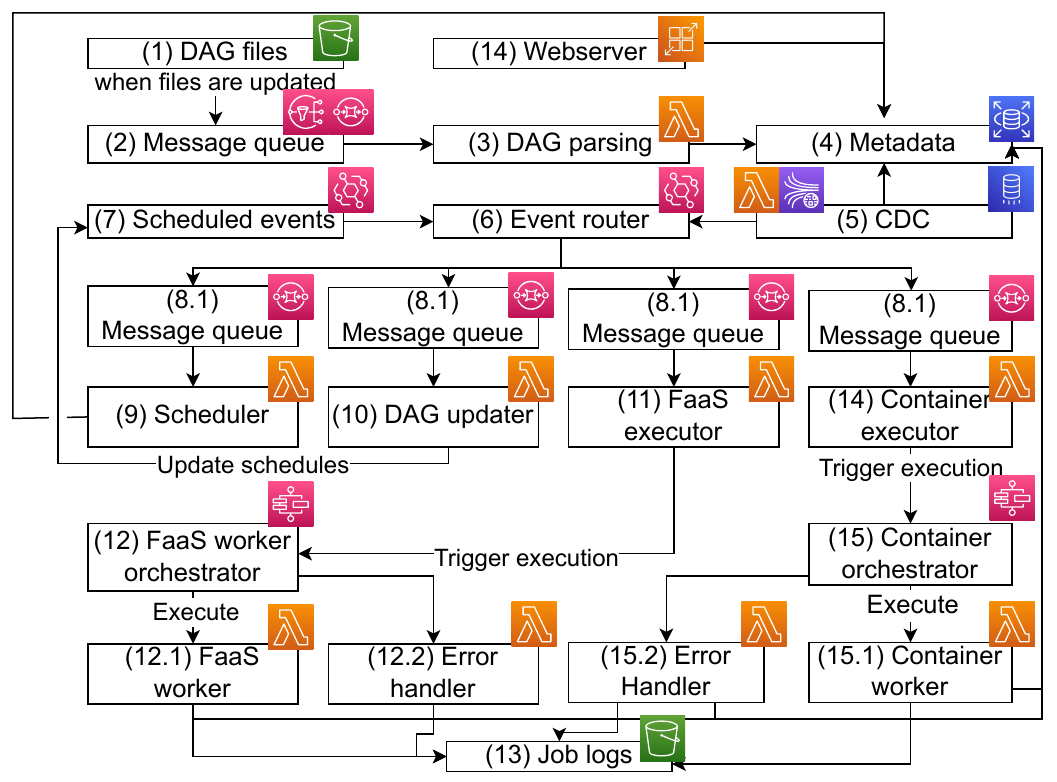}
    \caption{sAirflow on AWS (icons' source: AWS)}
    \label{fig:serverless_airflow_aws_implementation}
\end{figure}

Fig.~\ref{fig:serverless_airflow_aws_implementation} presents the high-level design of sAirflow. The figure maps sAirflow architecture to the standard serverless idioms (e.g., a message queue); and then to concrete AWS products (e.g., SQS). 
As other providers are comparable, porting sAirflow to, e.g., GCP, would require minor code changes.

\subsection{Control flow}
We introduce the event-based architecture of sAirflow by describing the flow of control in the system, with component numbers referring to Figure~\ref{fig:serverless_airflow_aws_implementation}.
An Airflow user submits a new (or updated) workflow which is persisted as a DAG file in the blob storage (1). The storage sends notifications via a message queue (2).
The notification triggers a function (3) that parses the DAG files and updates the Metadata DB (4).
To reduce the load when multiple DAGs are uploaded, we batch these invocations.
A similar flow is triggered from the web UI (14).

Changes in the metadata DB are captured by the Change Data Capture (CDC) (5) that produces an event routed to the Event Router (6). Applying CDC to Airflow was the key architectural difficulty we had to solve (Section~\ref{sec:cdc})

The Event Router (6) routes the following events:

\noindent A \emph{change of a parsed DAG} is routed to a function (10) that parses the schedule and updates the scheduled events in EventBridge, a cron-like module~(7).\\
\noindent A \emph{periodic event} represents a single launch of a workflow that, e.g.,  is scheduled daily at 4am. This event is routed to the Airflow scheduler (9). The scheduler creates a new \emph{DAG run} in the metadata DB (4).\\
\noindent A \emph{DAG run} event is routed to the scheduler (9), Section~\ref{section:scheduler_sairflow}. The scheduler determines if new tasks should be created and which tasks should be executed. Upon deciding to run a task, the scheduler changes the task's status in the metadata DB (4) to queued, triggering a \emph{task queued} event.\\
\noindent A \emph{task queued} event is routed to one of two executors (workers): the Function Executor (11) for invocations of up to 15~min.; or the Container Executor (14). 
Executors start the user-defined task (12.1) and handle failures (12.2), Section~\ref{section:executors_and_workers}. We stress that executors do not actively wait for the completion of the user work. Once the task is completed (or fails), the executor saves logs (13) and updates the metadata DB (4), triggering a \emph{task finished/failed} event.\\
\noindent A \emph{task finished/failed} event is routed to the scheduler (9), that reruns a task, queues its successors, or marks the DAG run as complete.

\subsection{Change Data Capture (CDC)}\label{sec:cdc}
In serverless architecture, propagating database changes in an event-driven manner requires additional effort. 
A Change Data Capture (CDC) pattern allows us to reuse most Airflow code (rather than reimplement all database interactions). In AWS cloud, CDC is provided through Database Migration Service (DMS), an external dependency.
DMS creates an event on a change in the SQL database and forwards it to a pre-configured destination. We use Amazon Kinesis Data Streams, coupled with a short function (executed as AWS Lambda) to pre-parse the event (e.g., remove redundancies).

DMS introduces a significant delay to the control loop. Typically, it takes 1-1.5~s between the change in the database and the event being delivered to the event router. Our experiments will later show this delay considerably affects sAirflow performance.

The alternative to CDC is to manually inject code that generates events near each DB modification. Apart from the volume of the code modifications needed, the problem with that approach is the joint, transactional nature of the event and the database change. 
A naive implementation has a \emph{dual write problem}: if the process fails after the database change but before the event is sent, the event might be lost; and if we reverse the order, the event might be consumed by a reader before the change in the database is committed (and visible to the reader).

\subsection{Scheduler} \label{section:scheduler_sairflow}
Airflow scheduler determines which tasks to launch. Airflow runs the scheduler as a separate, always-running thread, even if workflows are executed only sporadically. Moreover, all previous attempts to serverless Airflow kept this always-on scheduler (Section~\ref{sec:related-work}). The change of the Airflow scheduler's architecture to event-based --- without major refactoring, and retaining Airflow scheduling semantics --- was one of the key difficulties while working on sAirflow.

In sAirflow, an event triggers the execution of a scheduling algorithm --- for example, an event produced when a task has just been completed.
A single pass of the scheduler is executed in a single FaaS invocation.
To increase reliability, Airflow can be configured to run multiple, redundant schedulers. 
In contrast, sAirflow's reliability directly relies on the guarantees provided by FaaS (e.g., multiple availability zones for AWS Lambda).
In Airflow, most of the scheduler code executes in a critical section.
For consistency, sAirflow feeds the scheduler from a single-shard message queue.
The algorithm, however, is largely not modified:

\begin{enumerate}[nosep,leftmargin=*]
    \item For each DAG ready to execute: create a DAG run.
    \item For each task in each DAG run with all predecessors completed: create a scheduled task instance.
    \item For each scheduled task instance, label it queued.
\end{enumerate}
In contrast to Airflow which might launch some (short) tasks and then actively poll their state running the scheduling loop, sAirflow consistently relies on changes in the metadata database, delivered to external executors.

\subsection{Executors and workers}\label{section:executors_and_workers}
An executor starts a task instance and then monitors its execution.
sAirflow moves the task handling logic to AWS Step Functions; this enables sAirflow to avoid always-on workers polling the state of the user task. AWS Step Functions executes the user code (as a lambda or in a container, details follow). If the user code fails, AWS Step Functions calls a short sAirflow lambda that handles this failure. 

sAirflow provides two executors (Function and Container), but the framework algorithm is common:
\begin{enumerate}[nosep,leftmargin=*]
    \item \textbf{Invoke execution}: AWS services invoke sAirflow code in an isolated environment.
    The environment contains a handler that intakes the metadata about the task to execute. The metadata is then passed to the worker. 
    \item \textbf{Pull configuration}: 
    The worker downloads the deployment configuration from the blob storage.
    \item \textbf{Pull DAG files}: The worker downloads the DAG files defining the workload.
    \item \textbf{Start task}: The worker starts the task using LocalTaskJob, a standard Airflow component that executes the task in the process of the worker.
    When a task completes (or fails), this component modifies the metadata DB.
    \item \textbf{Push logs}: When a task completes, logs are collected throughout the runtime and sent to the blob storage. sAirflow needs minor modifications to push the logs and not close the logging sinks (thus, a single Lambda instance can serve for multiple invocations).
\end{enumerate}

The function executor and the container executor differ by what service they use to run the worker. The \emph{Function Executor} uses FaaS, AWS Lambda. While FaaS scales extremely well, the execution duration is limited (in AWS, to 15~min). This executor 
forwards task instances from an AWS SQS to a serverless orchestrator, AWS Step Functions. 

In the \emph{Container Executor}, the worker launches code in a container through an external service, AWS Batch with AWS Fargate. 
Containers typically have unbounded execution duration.
As in a standard managed container service, a container must specify the limits on the memory, CPU and number of copies.
AWS Batch on AWS Fargate supports horizontal scaling (including to 0), thus is consistent with our pay-per-use requirement.
AWS Fargate does not limit the duration of execution but typically scales out slower than AWS Lambda~\cite{9582324} and with a significant start-up overhead. On each invocation, the start-up might involve downloading an image and initializing a container, which results in a minutes-long delay. While~\cite{9582324} reports a 1-minute start time, 
in our experiments, we additionally observed significant variance.

\section{Deployment and Evaluation Method}\label{sec:eval-method}

To benchmark the performance, we deploy sAirflow in the cloud and compare it to a SaaS solution, Amazon Managed Workflows for Apache Airflow (MWAA). Cloud has myriad configuration and deployment options; our goal is to create environments that are as similar to each other as possible to achieve fair comparison at a reasonable cost.
In this section, we describe how we deploy and configure both systems and then how we generate the workflows. 

\noindent\textbf{Managed Workflows for Apache Airflow:}
We run MWAA with a \emph{small} environment (as the large environment is four times more expensive).
By default, MWAA uses the Celery executor, with 5 tasks on each worker node.
Unless explicitly specified, this parameter was not changed.
The environment starts with one worker and might be scaled to 25 workers.
Thus, MWAA can run up to 125 tasks concurrently (a \emph{large} environment might support more tasks, but the scalability issues would simply be deferred).
Each worker has 1vCPU and 2GB of RAM, so each task gets roughly 0.2vCPU and 400MB of RAM. 
MWAA runs two schedulers in parallel (high availability setting), which 
might be an advantage compared with sAirflow's single scheduler: both schedulers run the scheduling loop processing the workload, which could improve the task throughput.

\noindent\textbf{sAirflow:}
For a fair comparison, we match sAirflow's configuration to MWAA.
Both systems use Airflow 2.4.3.
The resources used by sAirflow's services are scaled to ones used by MWAA.
The worker functions have a memory limit of 340MB (which corresponds to vCPU of around 0.2 as AWS allocates 1vCPU per 1769MB of memory.
The scheduler uses 512MB of RAM (around 0.35vCPU).
We use \textit{db.t3.small} (2vCPU, 2GB memory) instance with PostgreSQL (without high availability and \textit{SQL proxy}).

\noindent\textbf{Workloads:} We experiment both on synthetic and on realistic workflows, characterized by:
$n$, the number of tasks in the DAG;
$p_i$, the duration of execution of a task in seconds;
and $T$, the period: the DAG executes every $T$ minutes.
When measuring \emph{warm} starts, we use $T=5$
which allows AWS Lambda to reuse previously allocated resources. 
In contrast, when measuring \emph{cold} starts, we use $T=30$: AWS Lambda should always spawn new resources for each invocation.
When $T=5$, we run the given DAG for an hour (12 invocations);
when $T=10$, we also run the DAG for an hour (but with 6 invocations);
finally, when $T=30$, we run the DAG for 1.5 hours (and get 3 invocations). We do not increase the runtime of experiments (to get more invocations), as MWAA would get too expensive.

As common in evaluating schedulers~\cite{DBLP:conf/sc/PrzybylskiPZLMR22}, tasks in both realistic and synthetic DAGs \texttt{sleep()} for time $p$: this does not influence the results, as both MWAA and sAirflow execute tasks in isolated environments with static CPU and memory limits. Additionally, workload traces do not contain all information needed to run the tasks (e.g., binaries, environments or parameters).

We use synthetic \emph{chain} and {parallel} DAGs; and realistic DAGs generated from Alibaba cloud traces~\cite{alibaba-clusterdata}.
A \emph{chain DAG} has tasks sequentially executing one after another (no parallelism).
The optimal execution time of a chain DAG is $n * p$.
A \emph{parallel DAG} models highly-parallelizable workloads: after a short startup task, $n$ tasks can be executed in parallel. The optimal execution time is $p$.
Finally, for \emph{Alibaba} DAGs, we extract the DAG shapes and task durations from the batch jobs of Alibaba cloud traces~\cite{alibaba-clusterdata}. 
After filtering out chain and parallel DAGs,
we select 30 different DAGs at random. 
To reduce the costs of experiments, we limit the runtime of any task to at most 60~s.

\begin{figure}[tbp]
  \centering
  \subfloat[job 3441830\label{fig:alibaba_example}\label{fig:alibaba_j_3441830}]{\includegraphics[width=0.3\textwidth]{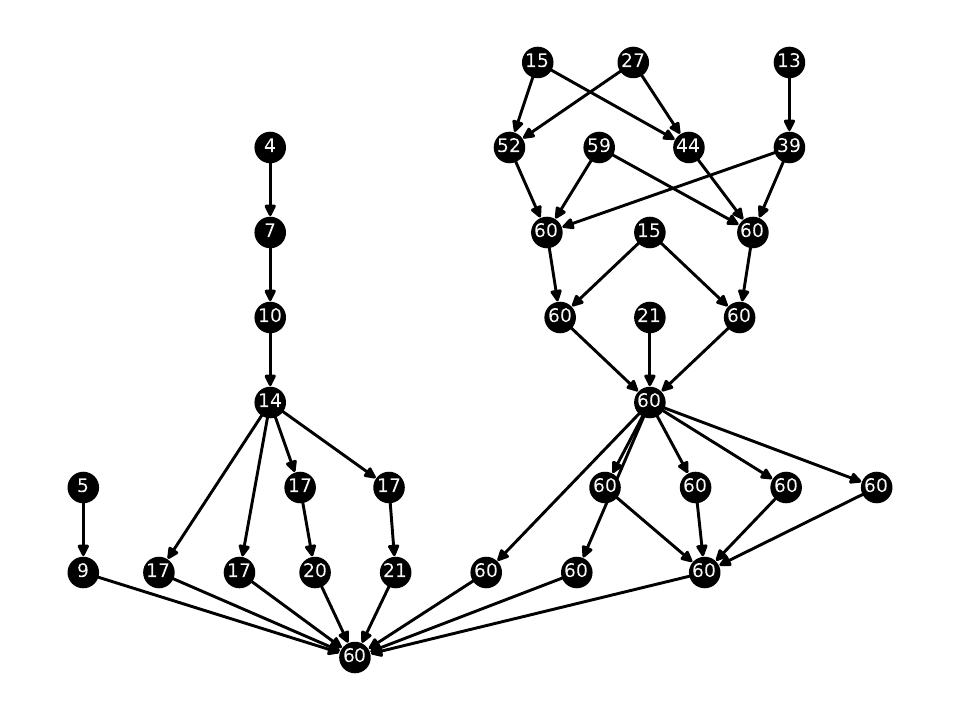}}
  \subfloat[job 581851\label{fig:alibaba_j_581851}]{\includegraphics[width=0.3\textwidth]{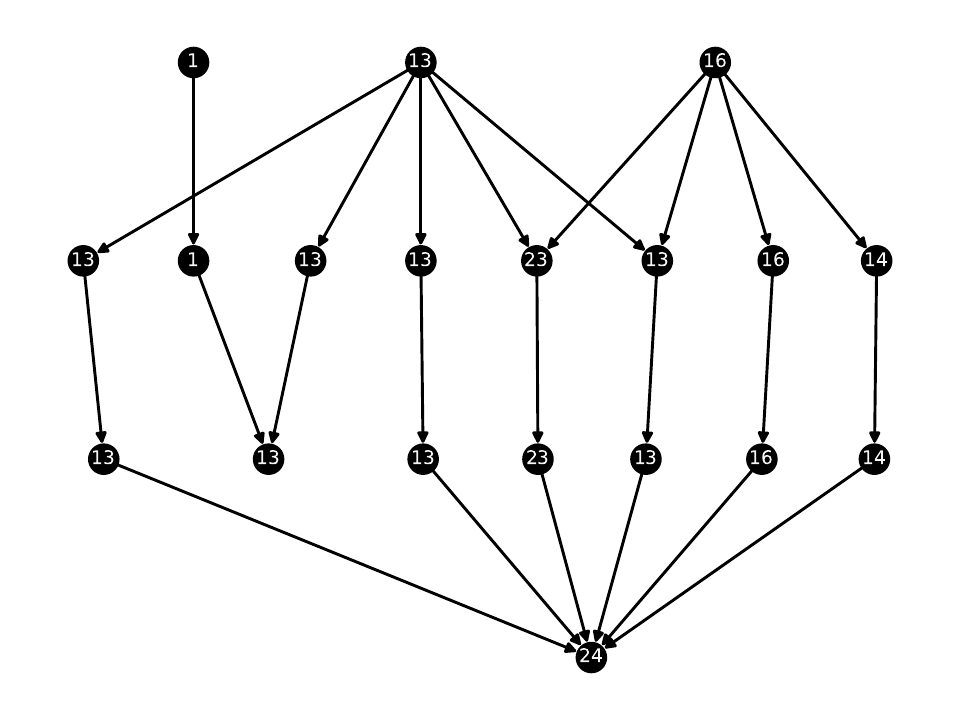}}
  \subfloat[job 3302772\label{fig:alibaba_j_3302772}]{\includegraphics[width=0.3\textwidth]{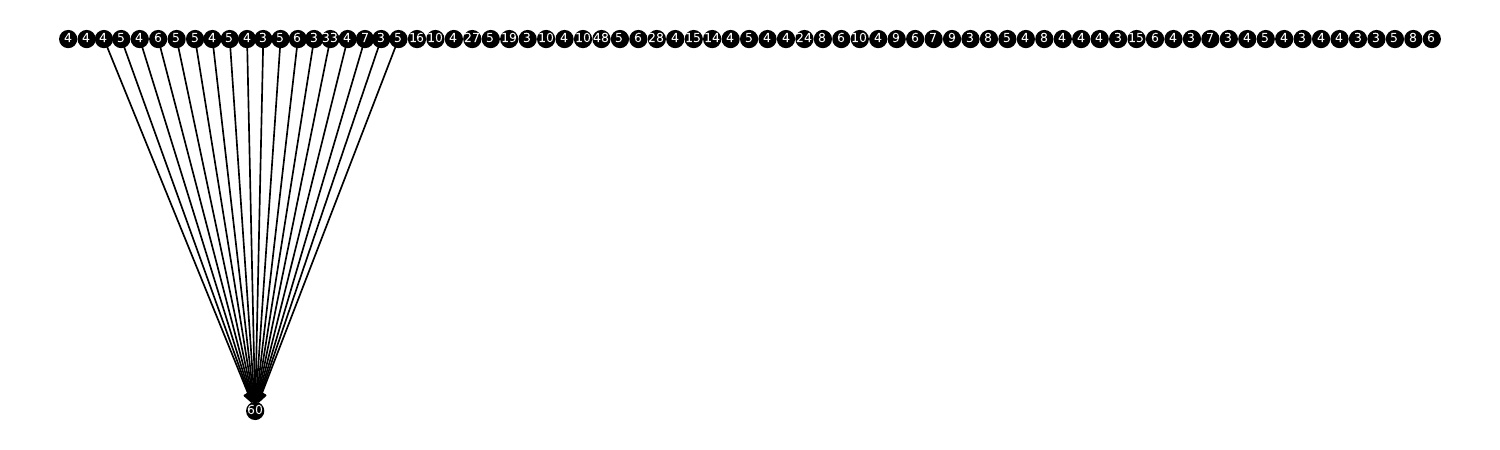}}
  \vspace{-1em}\caption{
  Sample DAGs obtained from jobs in the Alibaba trace.
    Note that the last DAG is highly parallel (77 tasks in total, 76 of which run in parallel on start-up). Some of the tasks do not have a downstream dependency.   
  }\label{fig:alibaba-4examples}
  \label{fig:example_dags}
\end{figure}

Fig.~\ref{fig:alibaba-4examples} shows three examples of derived DAGs. For example, the DAG in (\ref{fig:alibaba_j_3441830}) has $n=34$ tasks: 13 tasks were shortened to 60~s. The \emph{critical path}, i.e., the path with the longest duration, is 439~s, while the \emph{longest path}, i.e., the path with the maximum number of nodes, is 8 nodes.

\noindent\textbf{Metrics:} The key metric, measuring the overall efficiency of the system, is the \emph{DAG makespan}, the difference between DAG's start and end times reported by Airflow. More formally, denoting by $v_i$ the task's ready time, $s_i$ the start time, and by $c_i$ the completion time, the makespan is $C_{\max}(D) =  \max_{i \in D} c_i - \min_{i \in D} v_i$. Additionally, to better understand performance, we also report the \emph{task duration}, $(c_i - s_i)$: the difference between the duration and the workload $p_i$ shows the per-task overhead of the system; and the \emph{task wait time}, $(s_i - v_i)$, showing the start-up overhead.
    
\section{Experimental Evaluation Results}\label{sec:results}
For a comprehensive performance comparison, we consider three setups. We start with the \textit{function (FaaS) executor} in two variants, cold (Section~\ref{section:experiments:faas:cold}) and warm (Section~\ref{section:experiments:faas:warm}). Then, Section~\ref{section:experiments:containers}, we run sAirflow with the \textit{container executor}: as it cannot reuse environments, all the executions are cold. We conclude with cost estimation (Section~\ref{sec:cost-comparison}). Due to limited space, we show only the key trends and results; for transparency, all metrics and all results are in the Appendix.

\begin{figure}[tbp]
  \centering
    \includegraphics[width=0.3\textwidth]{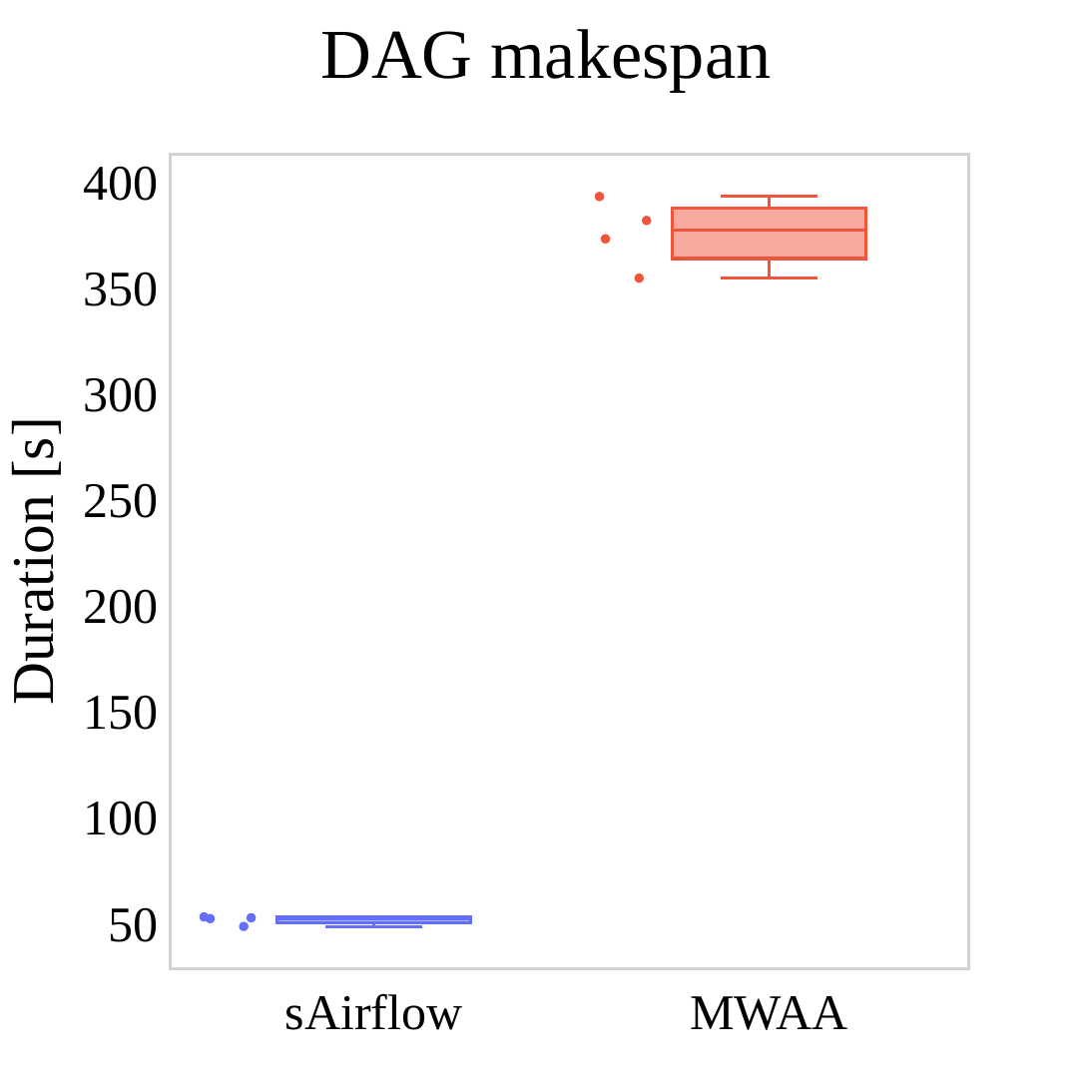}
    \includegraphics[width=0.3\textwidth]{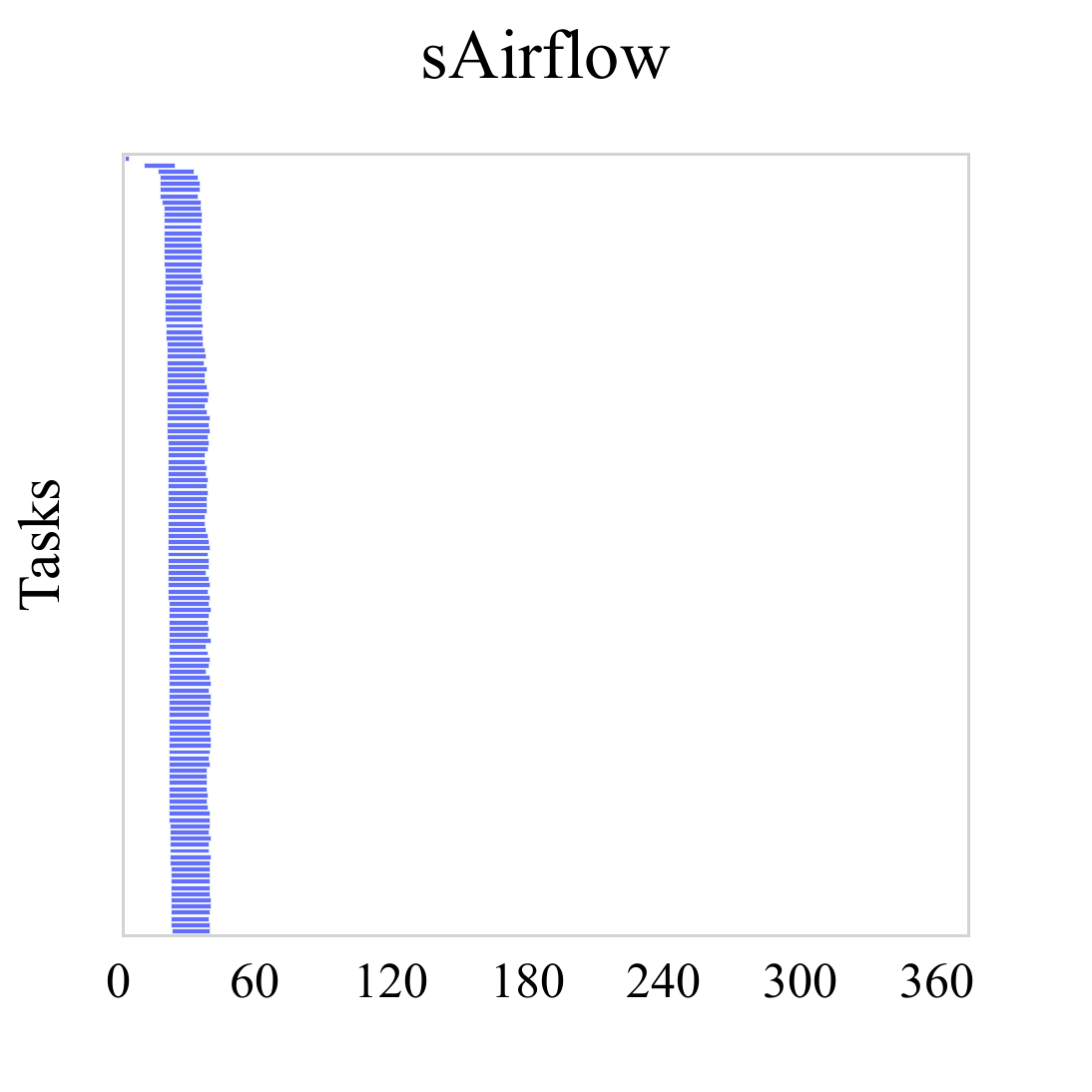}
    \includegraphics[width=0.3\textwidth]{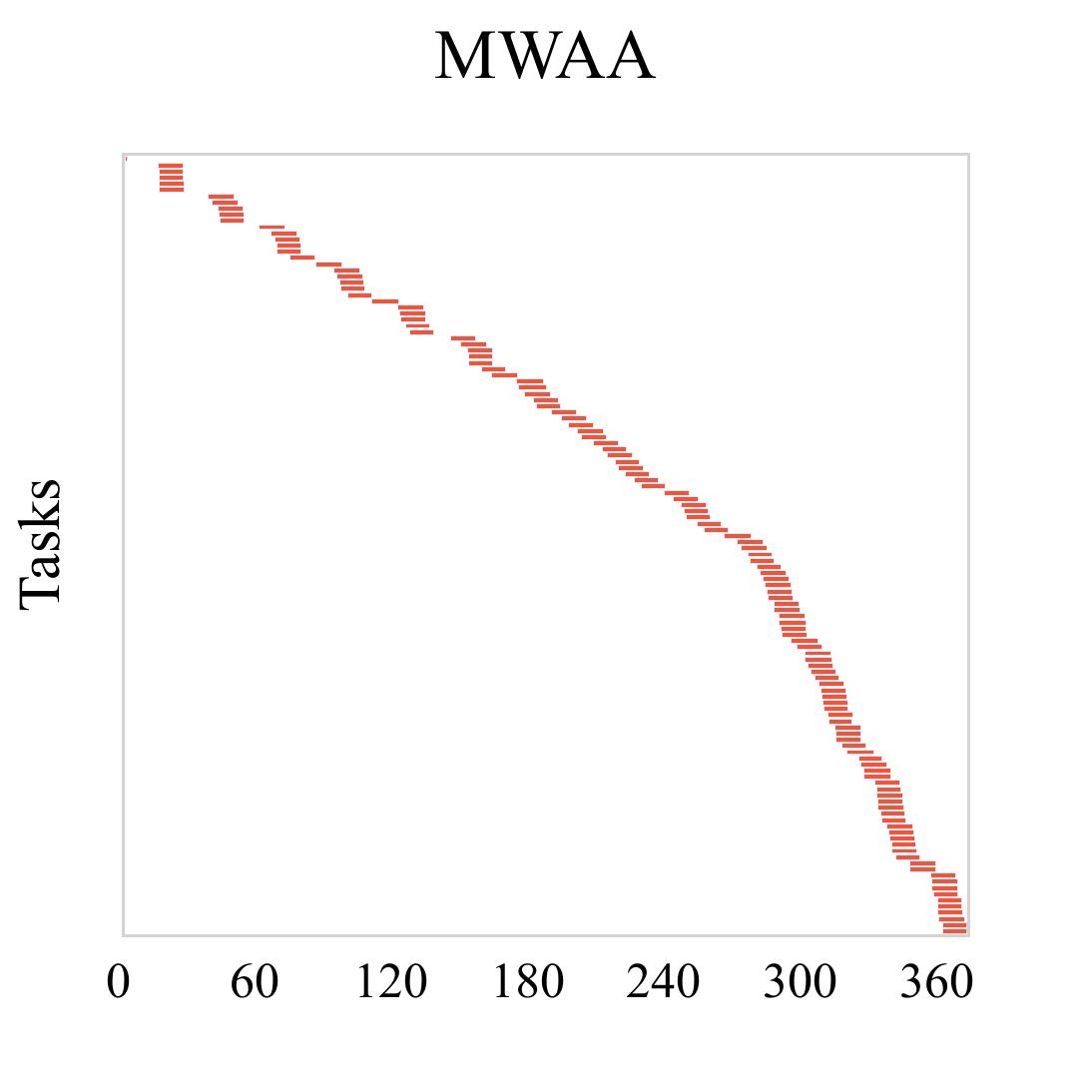}
  \vspace{-1em}
  \caption{Parallel DAGs, $p=10$, $T=30$, $n=125$ (cold starts). sAirflow shortens the makespan by 7.2x (left). Gantt charts (middle, right) show a single run.
  }
  \label{fig:parallel_cold}
\end{figure}

\subsection{Function executor and cold starts}\label{section:experiments:faas:cold}
To measure how the systems handle sporadic load, we run Parallel DAGs with $p=10$ and $n \in \{ 16, 32, 64, 125 \}$. Due to space constraints, Fig.~\ref{fig:parallel_cold} shows the largest $n=125$ (full results in the Appendix).
We run parallel DAGs to focus on workload with enough work. Thus, any inefficiencies will be directly caused by scaling problems.
MWAA is configured to start with one worker and horizontally scale out to up to 25 (supporting 125 concurrent invocations).
sAirflow is similarly limited to 125 concurrent FaaS invocations.
The 30~minute interval between runs ($T=30$) ensures that both systems de-provision resources between consecutive runs.

sAirflow is much faster in scaling out to match the demand, exposing Lambda's horizontal scaling with minimal overheads. The managed version of Airflow needs up to \emph{5 minutes} to add a new worker node (Fig.~\ref{fig:parallel_cold}, right), whereas sAirflow starts all workers almost immediately, thus completing the whole workload in less than a minute (Fig.~\ref{fig:parallel_cold}, first column). Due to MWAA's long horizontal scaling time, 
sAirflow reduces the makespan by, on average: 1.9 times ($n=16$), 3.7 times ($n=32$), 6.13 times ($n=64$) and 7.2 times~($n=125$). 

In sAirflow, the recorded task durations ($c_i-s_i$) increase when more tasks try to start at the same time: a 10-second-long task takes 12~s to complete when $n=64$ and 17s when $n=128$. 
In these settings, the transactional nature of the internal Airflow's code becomes a bottleneck.

\subsection{Function executor and warm starts}\label{section:experiments:faas:warm}
\begin{figure}[tbp]\vspace{-1em}
   \subfloat[chain, $n=5$\label{fig:5task_line}]{
    \includegraphics[width=0.3\textwidth]{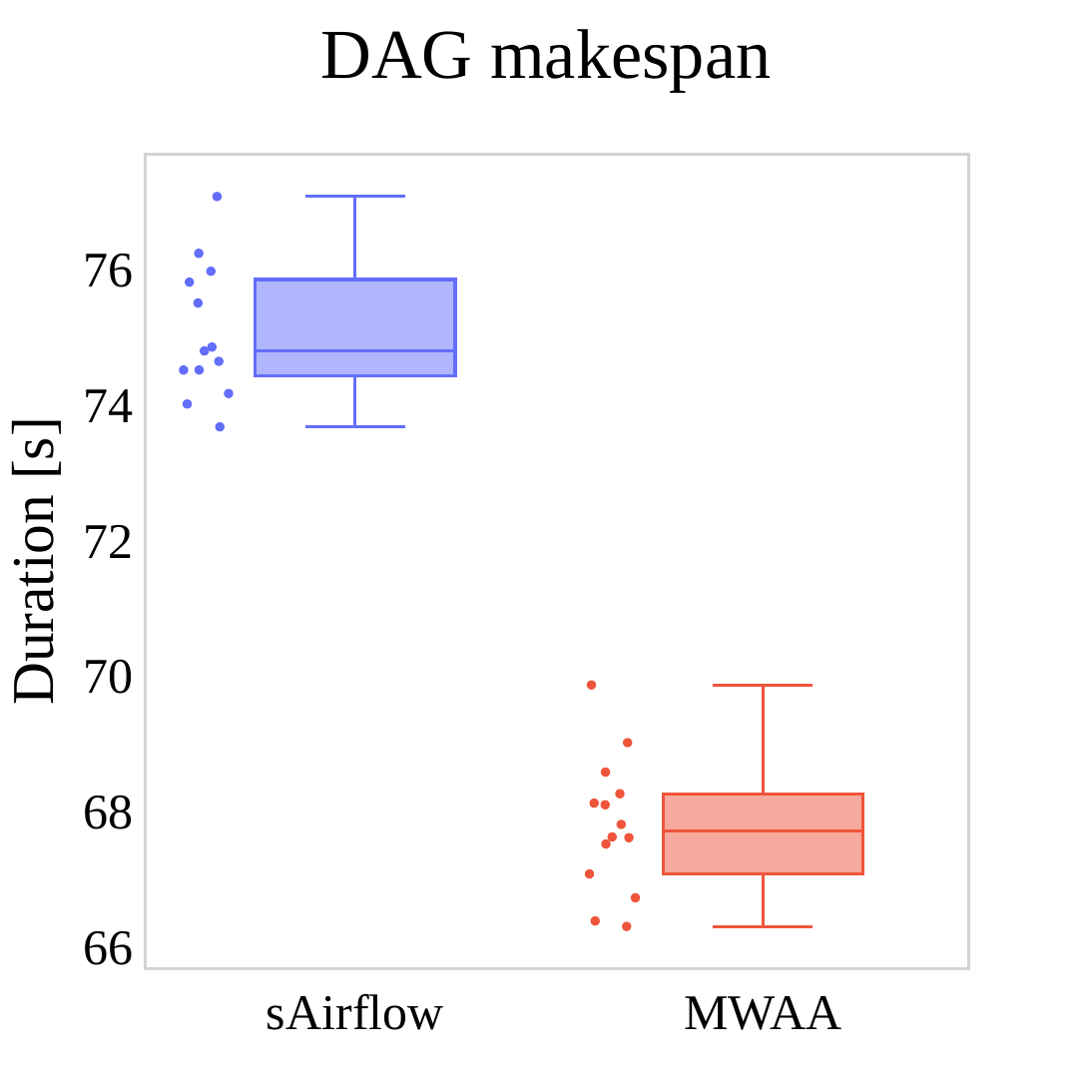}
   }
   \subfloat[parallel, $n=16$\label{fig:16task_parallel}]{
    \includegraphics[width=0.3\textwidth]{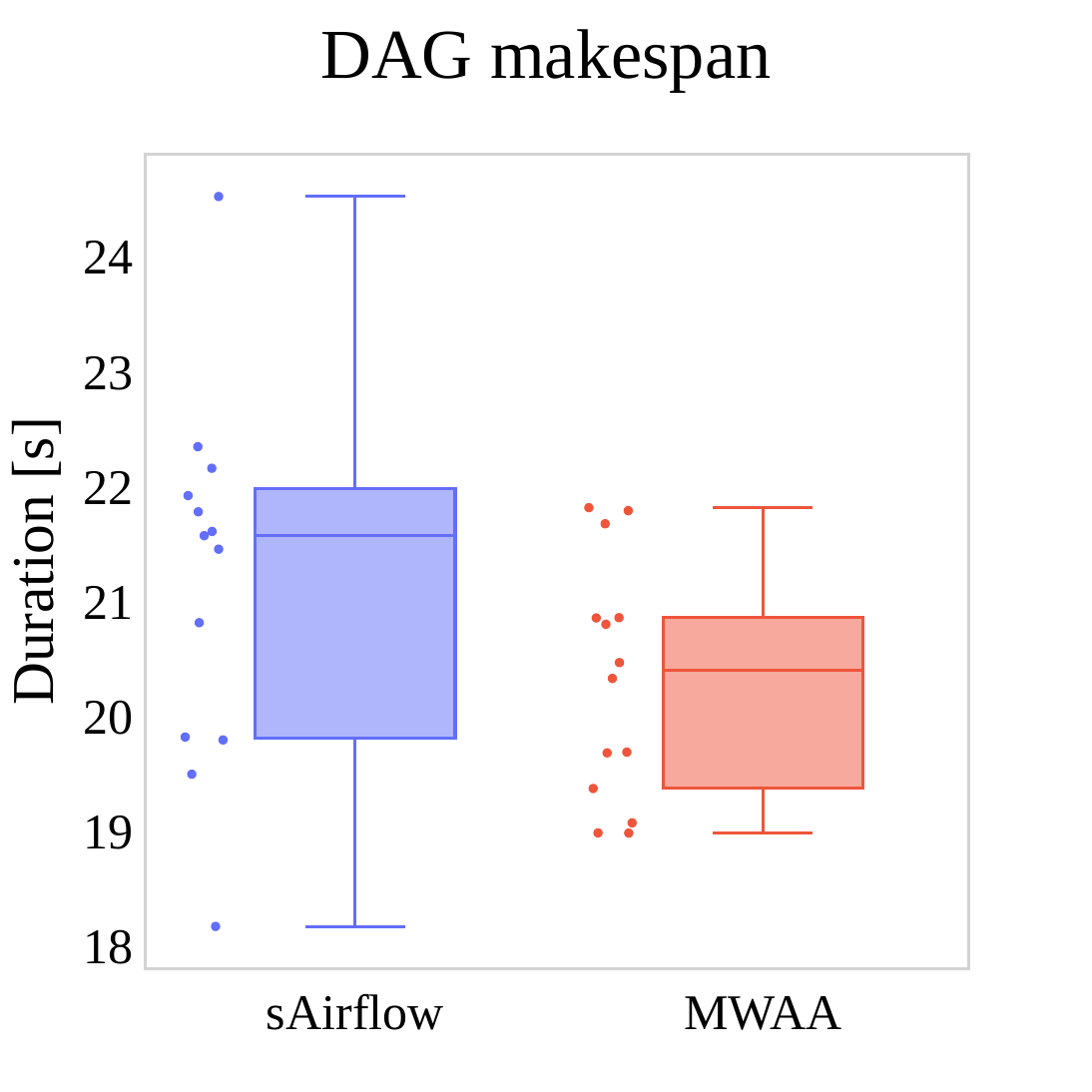}
   }
   \subfloat[parallel, $n=125$\label{fig:125task_parallel}]{
    \includegraphics[width=0.3\textwidth]{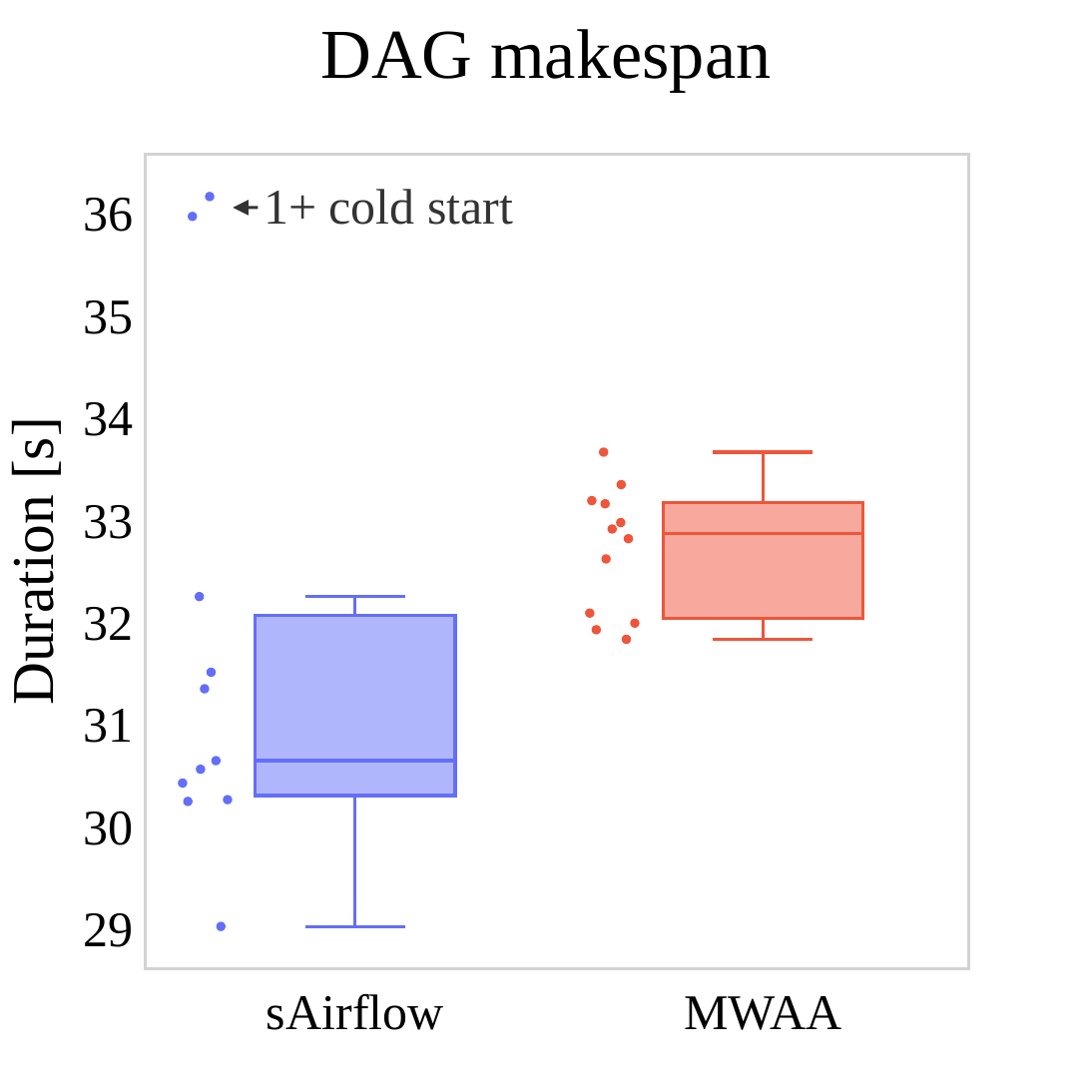}
   }
  \vspace{-1em}
  \caption{Warm system, $p=10$, $T=5$. The first DAG run is not reported.}
\end{figure}

To measure the performance under constant load, we now focus on warm invocations.
To ensure MWAA is run warm, we disable horizontal scaling by equating the minimum and the maximum number of workers (equal to 25, thus 125 parallel tasks).
In sAirflow, we pre-warm the system with a single invocation of the DAG: this invocation warms the lambdas executing the control plane code, as well as the workers.
To isolate this effect, we used a one-task DAG and we measured the cold start increasing the waiting time to almost 12~s (vs a median of 2.5~s on a warm system).
To focus on warm performance, we now exclude the first DAG invocation from the results unless explicitly stated.
However, there is no guarantee that the remaining invocations are all warm --- occasional cold starts do happen, and we will take them into account when describing results.

\emph{Chain} DAGs emphasize the per-task overheads (Fig.~\ref{fig:5task_line}): sAirflow is on the average 0.8s slower than MWAA when launching a task, a consequence of the lack of real-time CDC data streaming on AWS. As multiple events have to be sent in sequence to execute the next task: the previous task's completion triggers the scheduler, which marks the next task as queued, which in turn triggers the push of another event. This sequence results in a higher latency due AWS DMS overheads: it might take up to 1 second to push the event out of the database through Kinesis to EventBridge.

\emph{Parallel} DAGs: When fewer tasks run in parallel ($n=16$, Fig.~\ref{fig:16task_parallel}, and $n=32$), MWAA's and sAirflow's DAG execution times are comparable. MWAA is marginally faster for $n=16$ (by 1.2~s); and similar to sAirflow for $n=32$. The task wait time for sAirflow is shorter and less variable due to sAirflow's event-driven architecture, in contrast with MWAA's polling executor. sAirflow is faster than MWAA on larger DAGs with more parallelism, $n=64$ and $n=125$, Fig.~\ref{fig:125task_parallel}. Each of the two outliers in Fig.~\ref{fig:125task_parallel} can be traced to a cold start of a FaaS worker. 

\begin{figure}[tbp]
  \centering
  \subfloat[30 DAGs \label{fig:alibaba-makespans}]
  {\includegraphics[width=0.24\textwidth]{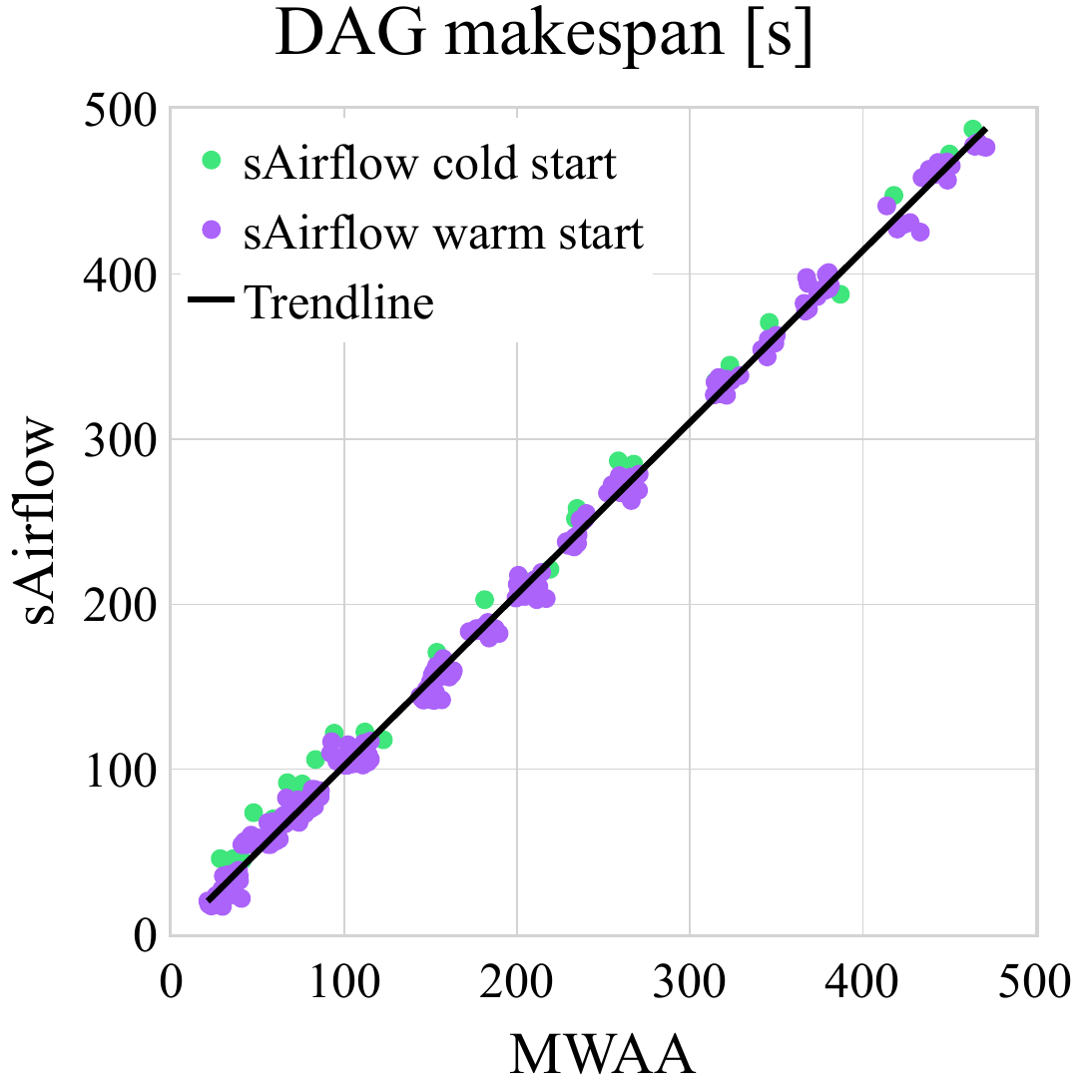}}
  \subfloat[job 3441830 \label{fig:j_3441830_stat}]{\includegraphics[width=0.24\textwidth]{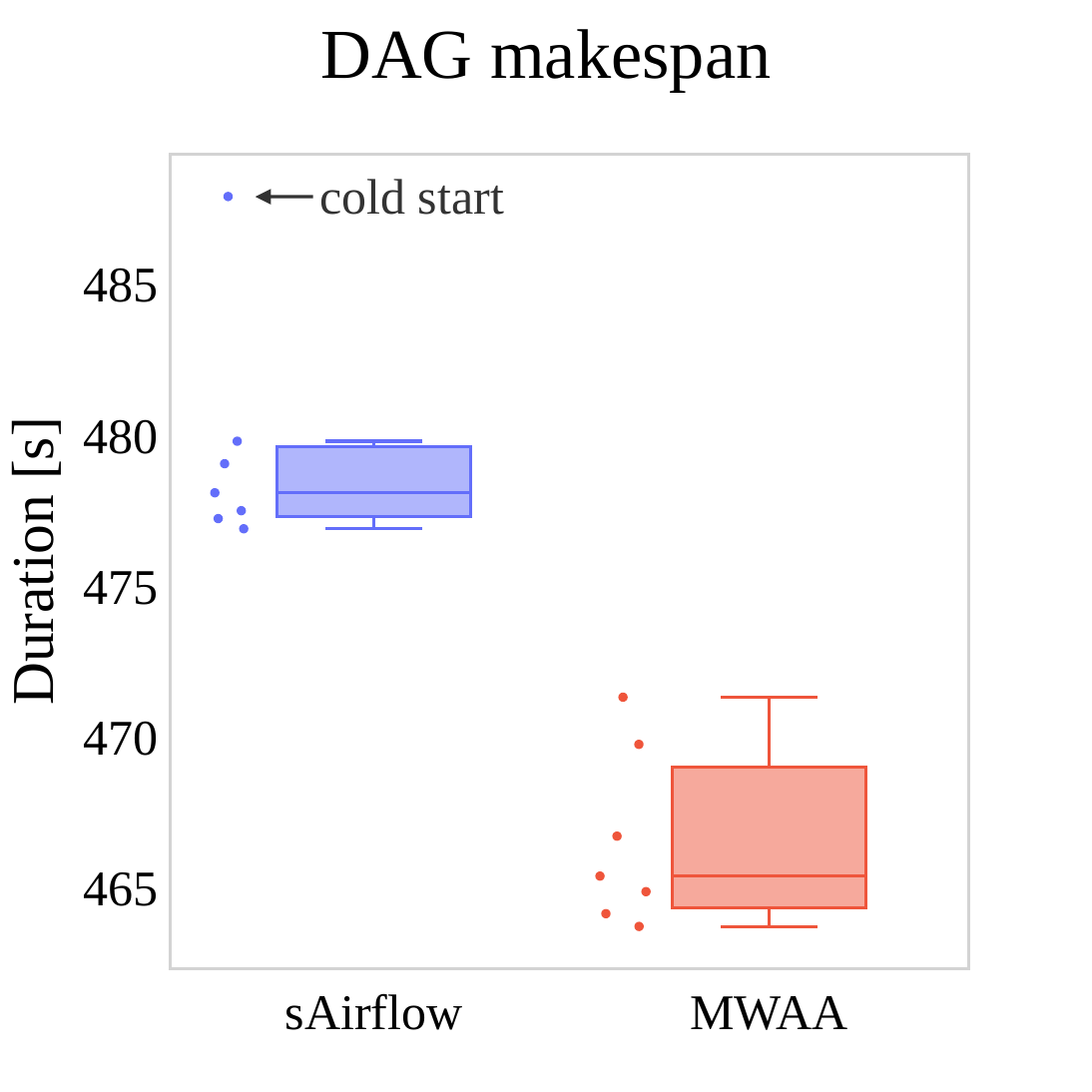}}
  \subfloat[job 581851 \label{fig:j_581851_stat}]{\includegraphics[width=0.24\textwidth]{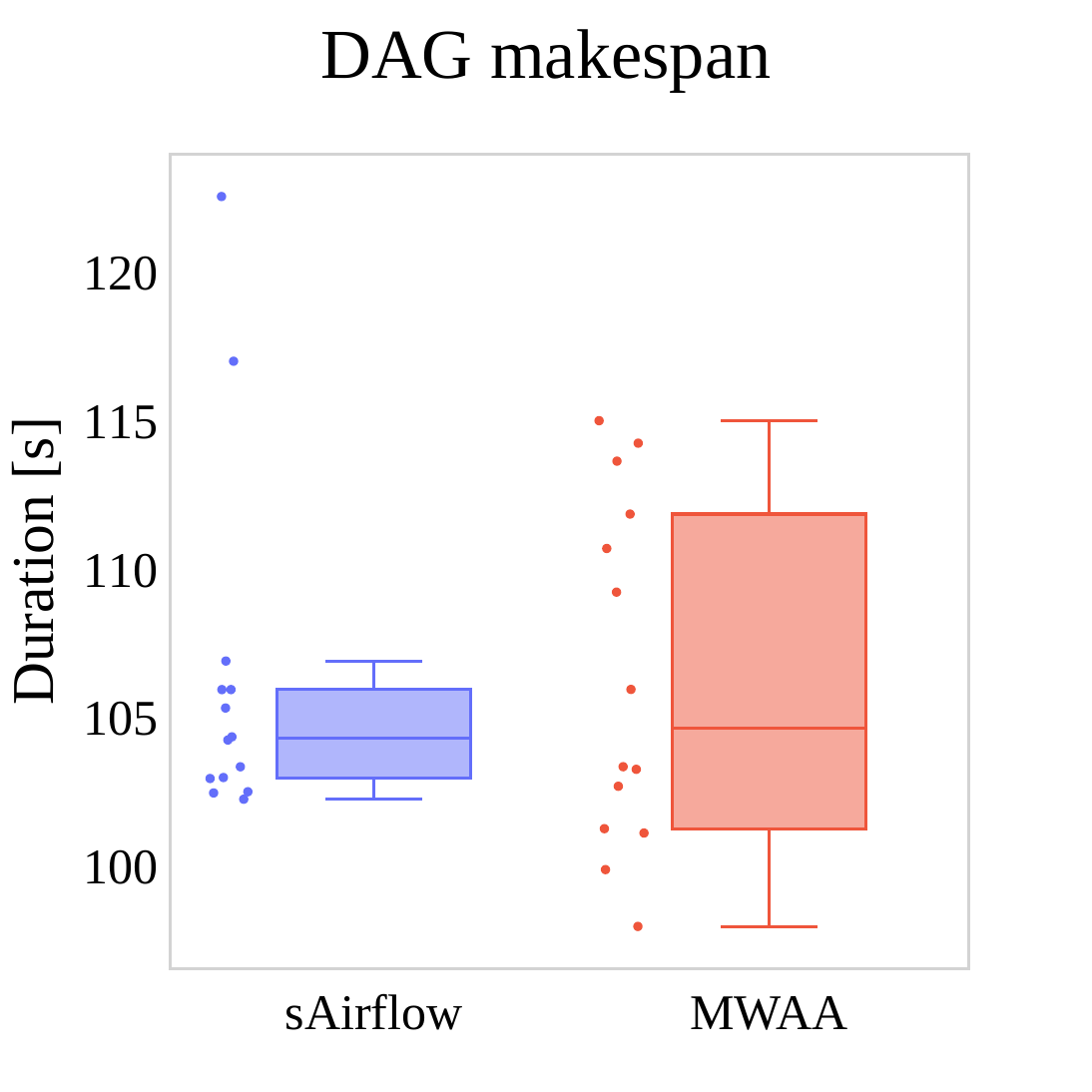}}
  \subfloat[job 3302772 \label{fig:j_3302772_stat}]{\includegraphics[width=0.24\textwidth]{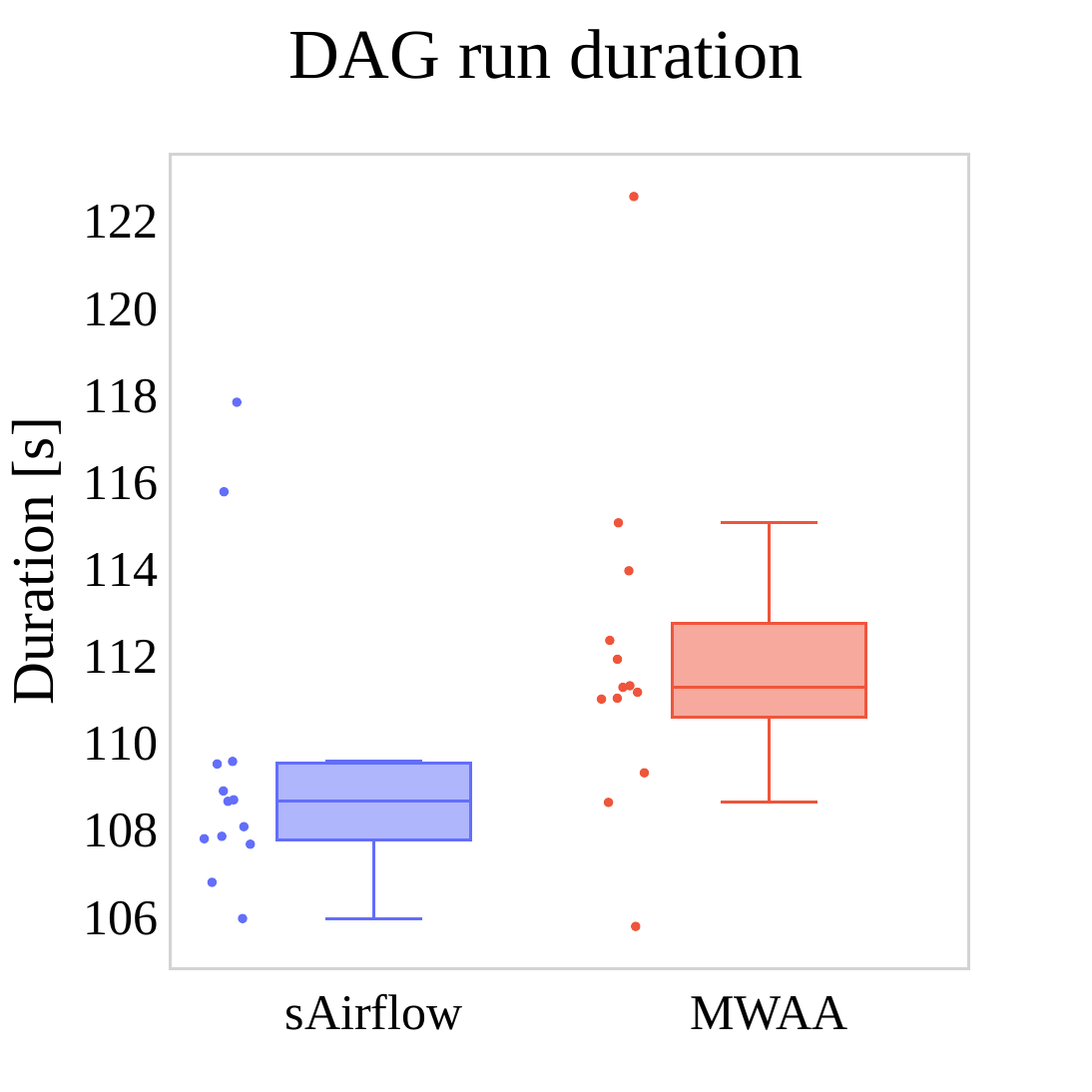}}
  \vspace{-1em}\caption{Alibaba DAGs: makespans on all DAGs (left) and a detailed analysis on the three DAGs from Fig.~\ref{fig:example_dags}}
 \end{figure}

\emph{Alibaba traces: comparable performance}: DAGs derived from the Alibaba traces show that performance on realistic, industrial workloads confirms the observed trends, with sAirflow outperforming MWAA when parallelism is sufficient. In this analysis, we include the first cold-start execution for sAirflow. Overall, makespans are similar (as emphasized by the trend line in the scatterplot, Fig.~\ref{fig:alibaba-makespans}). A detailed analysis of three example DAGs confirms the earlier trends. On chain-like DAGs, such as the one in Fig.~\ref{fig:alibaba_example}, sAirflow's makespan is minimally worse than MWAA (Fig.~\ref{fig:j_3441830_stat}): 478~s for sAirflow vs. 465~s for MWAA. The task wait time is the same in both systems, while the per-task overheads are higher in sAirflow; thus, the overall makespan is longer (478~s for sAirflow vs. 465~s for MWAA). There are 8 nodes on the critical path for the DAG; thus, the 13~s increase can be attributed to the per-task overheads. The DAG from Fig.~\ref{fig:alibaba_j_581851} shows a workload where both systems perform similarly (Fig.~\ref{fig:j_581851_stat}). Where sAirflow loses in terms of the task duration overhead, it gains on better performance concerning the task wait time. 
Finally, the DAG in Fig.~\ref{fig:alibaba_j_3302772} is similar to our parallel DAG, with over 70 tasks that can be executed in parallel; as expected, sAirflow completes the DAG slightly faster than MWAA.

\subsection{Container executor}\label{section:experiments:containers}
We measure sAirflow using container workers (AWS Batch with AWS Fargate). Due to space constraints, we only report the key metrics; full results are available in the Appendix. Launching even a single task DAG on a container worker increases the waiting time to 100.5s (from 2.5s with FaaS worker). Yet, beyond that delay, sAirflow with container workers still efficiently scales horizontally: a parallel DAG with $n=32$ tasks of $p=10$ seconds completes in approx. 140s (compared with approx. 160s needed by cold-starting MWAA).

\subsection{Monetary Cost Estimates}\label{sec:cost-comparison}
While the cloud pricing changes, we believe that the 
qualitative price difference between on-demand VMs and transient serverless should remain relatively stable, rendering the cost comparison based on current prices at least qualitatively correct in the longer term. 
We assume systems run continuously over 24 hours; we analyze costs with four types of workload (heavy, sporadic parallelizable, sporadic light, and constant light, see Appendix for the exact definitions and results).
We lower-bound MWAA costs by assuming that MWAA'a autoscaling bugs are resolved. 
We upper-bound sAirflow costs by excluding the free tier, and assuming the database and the CDC are always available (while with a sporadic load, CDC might be switched off).

Overall, sAirflow halves the fixed cost. The total cost of sAirflow is lower by 17--48\%.
As serverless products eliminate paying for idle resources, sAirflow is cheaper on sporadic and unpredictable workloads. sAirflow, due to the platform's elasticity, also eliminates the need to account for the worst-case load when deploying the service (in contrast to MWAA, that cannot reliably downscale).
In both systems, the costs are driven principally by the size of the database.

\section{Conclusions}
We show how to adapt an existing, complex application, Apache Airflow, to the serverless cloud.
Our prototype uses FaaS for Airflow's control plane, message queues populated by change data capture (CDC) for internal messaging, and FaaS and CaaS for workers.
Through micro-benchmarks and 30 real-world DAGs derived from the Alibaba Cloud traces, we compare the performance of our system with a commercially-maintained version on AWS (MWAA).
Our results show that sAirflow with FaaS workers scales notably better than MWAA. 
When a workflow has enough parallelism, a cold system scales in seconds to 125 workers, reducing completion times by 2x-7x. 
Conversely, sequential workflows, particularly with long tasks requiring containers, highlight increased latencies stemming from propagating CDC events (approx. 2~s); and launching containers through a queuing system (approx. 90~s).

Adopting a comprehensive, legacy system like Apache Airflow to serverless illustrates the difficulties and effort involved. Our extensive experiments show the performance penalty directly resulting from gaps in the current serverless offerings: the SQL database and the CDC process. Ideally, these two capabilities should be integrated into a single cloud-native serverless service.

\section*{Acknowledgements}
This research was supported by Polish National Science Center grant Opus (UMO-2017/25/B/ST6/00116), and by the Department of Energy under grant DE-SC0024387.
\printbibliography

@article{li2022funcx,
  title={FuncX: Federated function as a service for science},
  author={Li, Zhuozhao and Chard, Ryan and Babuji, Yadu and Galewsky, Ben and Skluzacek, Tyler J and Nagaitsev, Kirill and Woodard, Anna and Blaiszik, Ben and Bryan, Josh and Katz, Daniel S and others},
  journal={IEEE Transactions on Parallel and Distributed Systems},
  volume={33},
  number={12},
  pages={4948--4963},
  year={2022},
  publisher={IEEE}
}

@inproceedings{chard2020funcx,
  title={Funcx: A federated function serving fabric for science},
  author={Chard, Ryan and Babuji, Yadu and Li, Zhuozhao and Skluzacek, Tyler and Woodard, Anna and Blaiszik, Ben and Foster, Ian and Chard, Kyle},
  booktitle={Proceedings of the 29th International symposium on high-performance parallel and distributed computing},
  pages={65--76},
  year={2020}
}

@book{harenslak2021data,
  title={Data pipelines with apache airflow},
  author={Harenslak, Bas P and de Ruiter, Julian},
  year={2021},
  publisher={Simon and Schuster}
}

@book{gilbert2021software,
	title = {Software Architecture Patterns for Serverless Systems: Architecting for innovation with events, autonomous services, and micro frontends},
	isbn = {978-1-80020-073-9},
	OPTurl = {https://books.google.pl/books?id=gUs2EAAAQBAJ},
	publisher = {Packt Publishing},
	author = {Gilbert, J. and Price, E.},
	date = {2021},
}

@inproceedings{DBLP:conf/sc/PrzybylskiPZLMR22,
  author       = {Bartlomiej Przybylski and
                  Maciej Pawlik and
                  Pawel Zuk and
                  Bartlomiej Lagosz and
                  Maciej Malawski and
                  Krzysztof Rzadca},
  title        = {Using Unused: Non-Invasive Dynamic FaaS Infrastructure with HPC-Whisk},
  booktitle    = {Supercomputing},
  OPTpages        = {40:1--40:15},
  publisher    = {{IEEE}},
  year         = {2022},
  OPTurl          = {https://doi.org/10.1109/SC41404.2022.00045},
  OPTdoi          = {10.1109/SC41404.2022.00045},
  }

@incollection{hamza2024JourneyServerlessMigration,
	title = {The Journey to Serverless Migration: An Empirical Analysis of Intentions, Strategies, and Challenges},
	volume = {14483},
	isbn = {978-3-031-49265-5 978-3-031-49266-2},
	shorttitle = {The Journey to Serverless Migration},
	OPTpages = {100--115},
	booktitle = {Product-Focused Software Process Improvement},
	publisher = {Springer},
	author = {Hamza, Muhammad and Akbar, Muhammad Azeem and Smolander, Kari},
	OPTeditor = {Kadgien, Regine and Jedlitschka, Andreas and Janes, Andrea and Lenarduzzi, Valentina and Li, Xiaozhou},
	date = {2024},
	OPTdoi = {10.1007/978-3-031-49266-2_7},
	series = {LNCS},
	OPTfile = {Available Version (via Google Scholar):/Users/krz/Zotero/storage/6PNTPLUB/Hamza et al. - 2024 - The Journey to Serverless Migration An Empirical .pdf:application/pdf},
}

@ARTICLE{fairbanks19ignore-refactor-rewrite,
  author={Fairbanks, George},
  journal={IEEE Software}, 
  title={Ignore, Refactor, or Rewrite}, 
  year={2019},
  volume={36},
  number={2},
  pages={133-136},
  OPTdoi={10.1109/MS.2018.2880662}
  }

@inproceedings{jonas2017occupy,
  title={Occupy the cloud: Distributed computing for the 99\%},
  author={Jonas, Eric and Pu, Qifan and Venkataraman, Shivaram and Stoica, Ion and Recht, Benjamin},
  booktitle={SoCC, Proc.},
  pages={445--451},
  year={2017}
}

@inproceedings{maheo2021serverless,
  title={The serverless shell},
  author={Mah{\'e}o, Aur{\`e}le and Sutra, Pierre and Tarrant, Tristan},
  booktitle={Middleware: Industrial Track},
  pages={9--15},
  year={2021}
}

@article{gimenez2019framework,
  title={A framework and a performance assessment for serverless MapReduce on AWS Lambda},
  author={Gim{\'e}nez-Alventosa, Vicent and Molt{\'o}, Germ{\'a}n and Caballer, Miguel},
  journal={FGCS},
  volume={97},
  pages={259--274},
  year={2019},
  publisher={Elsevier}
}

@misc{copik22faaskeeper,
  author = {Copik, Marcin and Calotoiu, Alexandru and Taranov, Konstantin and Hoefler, Torsten},
  title = {{FaasKeeper}: a Blueprint for Serverless Services},
  publisher = {arXiv},
        eprint={2202.05711},
      archivePrefix={arXiv},
      primaryClass={cs.DC},
  year = {2022},
}

@misc{https://doi.org/10.48550/arxiv.1905.10270,
      eprint={1905.10270},
      archivePrefix={arXiv},
      primaryClass={cs.DC},
  author = {Ilyushkin, Alexey and Bauer, André and Papadopoulos, Alessandro V. and Deelman, Ewa and Iosup, Alexandru},
  title = {Performance-Feedback Autoscaling with Budget Constraints for Cloud-based Workloads of Workflows},
  year = {2019},
}

@online{aws_mwaa_autoscaling_downscaling_flaw,
  author = {Thom Bedford},
  title = {{Diagnosing Airflow’s Auto-Scaling Flaw in AWS MWAA}},
  year = 2022,
  url = {https://technical.thombedford.com/267},
  urldate = {2023-01-22}
}

@online{airflow_native_aws_executors,
  author = {Ahmed Elzeiny},
  title = {Apache {Airflow}: Native {AWS} Executors},
  url = {https://github.com/aelzeiny/airflow-aws-executors},
  OPTurldate = {2023-01-22}
}

@online{aws-batch-fargate,
  author = {{Amazon Web Services}},
  title = {{AWS Batch on AWS Fargate}},
  url = {https://docs.aws.amazon.com/batch/latest/userguide/fargate.html},
  urldate = {2023-01-22}
}

@online{aws_mwaa_pricing,
  author = {{Amazon Web Services}},
  title = {{Amazon Managed Workflows for {Apache Airflow} Pricing}},
  url = {https://aws.amazon.com/managed-workflows-for-apache-airflow/pricing/},
  urldate = {2023-01-22}
}

@online{alibaba-clusterdata,
  author = {Alibaba},
  title = "Clusterdata: Public trace data sets of production clusters",
  url = "https://github.com/alibaba/clusterdata/tree/master/cluster-trace-v2018",
}

@online{aws-step-functions-pricing,
  author = {{Amazon Web Services}},
  title = {{AWS {Step Functions} Pricing}},
  url = {https://aws.amazon.com/step-functions/pricing/},
  urldate = {2023-04-13}
}

@online{aws-s3-pricing,
  author = {{Amazon Web Services}},
  title = {Amazon {S3} Pricing},
  url = "https://aws.amazon.com/s3/pricing/?p=pm&c=s3&z=4",
  urldate = {2023-04-13}
}

@online{aws-eventbridge-pricing,
  author = {{Amazon Web Services}},
  title = {Amazon {EventBridge} Pricing},
  url = "https://aws.amazon.com/eventbridge/pricing/",
  urldate = {2023-04-13}
}

@online{aws-sqs-pricing,
  author = {{Amazon Web Services}},
  title = {Amazon {SQS} Pricing},
  url = "https://aws.amazon.com/sqs/pricing/",
  urldate = {2023-04-23}
}

@online{aws-sqs-short-long-polling,
  author = {{Amazon Web Services}},
  title = {Amazon {SQS} Short and Long Polling},
  url = "https://docs.aws.amazon.com/AWSSimpleQueueService/latest/SQSDeveloperGuide/sqs-short-and-long-polling.html#sqs-short-long-polling-differences",
  urldate = {2023-04-23}
}

@INPROCEEDINGS{9582324,
  author={Burkat, Krzysztof and Pawlik, Maciej and Balis, Bartosz and Malawski, Maciej and Vahi, Karan and Rynge, Mats and da Silva, Rafael Ferreira and Deelman, Ewa},
  booktitle={eScience},
  title={Serverless Containers – Rising Viable Approach to Scientific Workflows}, 
  year={2021},
  volume={},
  number={},
  OPTpages={40-49},
  OPTdoi={10.1109/eScience51609.2021.00014}}

@article{eismann2021state,
  title={The state of serverless applications: Collection, characterization, and community consensus},
  author={Eismann, Simon and Scheuner, Joel and Van Eyk, Erwin and Schwinger, Maximilian and Grohmann, Johannes and Herbst, Nikolas and Abad, Cristina L and Iosup, Alexandru},
  journal={IEEE TSE},
  volume={48},
  number={10},
  pages={4152--4166},
  year={2021},
  publisher={IEEE}
}

@misc{lin2022global,
      title={Global Optimization of Data Pipelines in Heterogeneous Cloud Environments}, 
      author={Erica Lin and Luna Xu and Suraj Bramhavar and Marco Montes de Oca and Sean Gorsky and Lingyun Yi and Arianna Groetsema and Jeffrey Chou},
      year={2022},
      eprint={2202.05711},
      archivePrefix={arXiv},
      primaryClass={cs.DC}
}

@article{10.1145/3368454,
    author = {Castro, Paul and Ishakian, Vatche and Muthusamy, Vinod and Slominski, Aleksander},
    title = {The Rise of Serverless Computing},
    year = {2019},
    issue_date = {December 2019},
    publisher = {ACM},
    volume = {62},
    number = {12},
    issn = {0001-0782},
    OPTurl = {https://doi.org/10.1145/3368454},
    OPTdoi = {10.1145/3368454},
    journal = {Commun. ACM},
    
    pages = {44–54},
    numpages = {11}
}

@ARTICLE{9393895,
  author={Ahmad, Zulfiqar and Jehangiri, Ali Imran and Ala'anzy, Mohammed Alaa and Othman, Mohamed and Latip, Rohaya and Zaman, Sardar Khaliq Uz and Umar, Arif Iqbal},
  journal={IEEE Access}, 
  title={Scientific Workflows Management and Scheduling in Cloud Computing: Taxonomy, Prospects, and Challenges}, 
  year={2021},
  volume={9},
  number={},
  OPTpages={53491-53508},
  OPTdoi={10.1109/ACCESS.2021.3070785}}

@ARTICLE{9837056,
  author={Ahmad, Zulfiqar and Jehangiri, Ali Imran and Mohamed, Nader and Othman, Mohamed and Umar, Arif Iqbal},
  journal={IEEE Access}, 
  title={Fault Tolerant and Data Oriented Scientific Workflows Management and Scheduling System in Cloud Computing}, 
  year={2022},
  volume={10},
  number={},
  OPTpages={77614-77632},
  OPTdoi={10.1109/ACCESS.2022.3193151}}

@ARTICLE{9812609,
  author={Kamran, Ali and Farooq, Umar and Rabbi, Ihsan and Zia, Kashif and Assam, Muhammad and Alsolai, Hadeel and Al-Wesabi, Fahd N.},
  journal={IEEE Access}, 
  title={A Unified Mechanism for Cloud Scheduling of Scientific Workflows}, 
  year={2022},
  volume={10},
  number={},
  OPTpages={71233-71246},
  OPTdoi={10.1109/ACCESS.2022.3187704}}

@ARTICLE{9066946,
  author={Versluis, Laurens and Mathá, Roland and Talluri, Sacheendra and Hegeman, Tim and Prodan, Radu and Deelman, Ewa and Iosup, Alexandru},
  journal={TPDS},
  title={The Workflow Trace Archive: Open-Access Data From Public and Private Computing Infrastructures}, 
  year={2020},
  volume={31},
  number={9},
  OPTpages={2170-2184},
  OPTdoi={10.1109/TPDS.2020.2984821}
  }

@ARTICLE{8013738,
  author={Pham, Thanh-Phuong and Durillo, Juan J. and Fahringer, Thomas},
  journal={IEEE Transactions on Cloud Computing}, 
  title={Predicting Workflow Task Execution Time in the Cloud Using A Two-Stage Machine Learning Approach}, 
  year={2020},
  volume={8},
  number={1},
  pages={256-268},
  doi={10.1109/TCC.2017.2732344}}

@ARTICLE{8976136,
  author={Farid, Mazen and Latip, Rohaya and Hussin, Masnida and Abdul Hamid, Nor Asilah Wati},
  journal={IEEE Access}, 
  title={Scheduling Scientific Workflow Using Multi-Objective Algorithm With Fuzzy Resource Utilization in Multi-Cloud Environment}, 
  year={2020},
  volume={8},
  number={},
  pages={24309-24322},
  doi={10.1109/ACCESS.2020.2970475}}

@ARTICLE{7839905,
  author={Cai, Zhicheng and Li, Xiaoping and Ruiz, Rubén},
  journal={IEEE TCC}, 
  title={Resource Provisioning for Task-Batch Based Workflows with Deadlines in Public Clouds}, 
  year={2019},
  volume={7},
  number={3},
  pages={814-826},
}

@INPROCEEDINGS{9005494,
  author={Mitchell, Ryan and Pottier, Loїc and Jacobs, Steve and Silva, Rafael Ferreira da and Rynge, Mats and Vahi, Karan and Deelman, Ewa},
  booktitle={Big Data},
  title={Exploration of Workflow Management Systems Emerging Features from Users Perspectives}, 
  year={2019},
  volume={},
  number={},
  pages={4537-4544},
  OPTdoi={10.1109/BigData47090.2019.9005494}
  }

@inproceedings{10.1145/3366623.3368137,
author = {Barcelona-Pons, Daniel and Garcia-Lopez, Pedro and Ruiz Alvaro and Gomez-Gomez, Amanda and Paris, Gerard and Sanchez-Artigas, Marc},
title = {Faa{S} Orchestration of Parallel Workloads},
year = {2019},
isbn = {9781450370387},
OPTpublisher = {ACM},
booktitle={WOSC},
OPTpages = {25–30},
}

@inproceedings{10.1145/3415958.3433082,
author = {Dessalk, Yared Dejene and Nikolov, Nikolay and Matskin, Mihhail and Soylu, Ahmet and Roman, Dumitru},
title = {Scalable Execution of Big Data Workflows Using Software Containers},
year = {2020},
isbn = {9781450381154},
publisher = {ACM},
OPTdoi = {10.1145/3415958.3433082},
booktitle = {MEDES},
OPTpages = {76–83},
}

@inproceedings{10.1145/3401025.3401731,
author={Lopez, Pedro Garcia and Arjona, Aitor and Sampe, Josep and Slominski, Aleksander and Villard, Lionel},
title = {Triggerflow: Trigger-Based Orchestration of Serverless Workflows},
year = {2020},
isbn = {9781450380287},
publisher = {ACM},
OPTdoi = {10.1145/3401025.3401731},
OPTabstract = {As more applications are being moved to the Cloud thanks to serverless computing, it is increasingly necessary to support native life cycle execution of those applications in the data center.But existing systems either focus on short-running workflows (like IBM Composer or Amazon Express Workflows) or impose considerable overheads for synchronizing massively parallel jobs (Azure Durable Functions, Amazon Step Functions, Google Cloud Composer). None of them are open systems enabling extensible interception and optimization of custom workflows.We present Triggerflow: an extensible Trigger-based Orchestration architecture for serverless workflows built on top of Knative Eventing and Kubernetes technologies. We demonstrate that Triggerflow is a novel serverless building block capable of constructing different reactive schedulers (State Machines, Directed Acyclic Graphs, Workflow as code). We also validate that it can support high-volume event processing workloads, auto-scale on demand and transparently optimize scientific workflows.},
booktitle = {DEBS},
OPTpages = {3–14},
numpages = {12},
}

@article{deelman2019evolution,
  title={The evolution of the pegasus workflow management software},
  author={Deelman, Ewa and Vahi, Karan and Rynge, Mats and Mayani, Rajiv and da Silva, Rafael Ferreira and Papadimitriou, George and Livny, Miron},
  journal={Computing in Science \& Engineering},
  volume={21},
  number={4},
  pages={22--36},
  year={2019},
  publisher={IEEE}
}

@article{novella2019container,
  title={Container-based bioinformatics with Pachyderm},
  author={Novella, Jon Ander and Emami Khoonsari, Payam and Herman, Stephanie and Whitenack, Daniel and Capuccini, Marco and Burman, Joachim and Kultima, Kim and Spjuth, Ola},
  journal={Bioinformatics},
  volume={35},
  number={5},
  pages={839--846},
  year={2019},
  publisher={Oxford University Press}
}

@inproceedings{manner2018cold,
  title={Cold start influencing factors in function as a service},
  author={Manner, Johannes and Endress, Martin and Heckel, Tobias and Wirtz, Guido},
  booktitle={UCC Proc.},
  pages={181--188},
  year={2018},
  publisher={IEEE}
}

@article{shahrad2020characterization,
  title={Characterization and Optimization of the Serverless Workload at a Large Cloud Provider},
  author={Shahrad, Mohammad and Fonseca, Rodrigo and Goiri, I{\~n}igo and Chaudhry, Gohar and Bianchini, Ricardo},
  journal={USENIX},
  year={2020}
}

@inproceedings{zuk2022call,
  title={Call Scheduling to Reduce Response Time of a FaaS System},
  author={Zuk, Pawe{\l} and Przybylski, Bart{\l}omiej and Rzadca, Krzysztof},
  booktitle={CLUSTER},
  pages={172--182},
  year={2022},
  publisher={IEEE}
}

@inproceedings{aumala2019beyond,
  title={Beyond load balancing: Package-aware scheduling for serverless platforms},
  author={Aumala, Gabriel and Boza, Edwin and Ortiz-Aviles, Luis and Totoy, Gustavo and Abad, Cristina},
  booktitle={CCGRID, Proc.},
  OPTpages={282--291},
  year={2019},
  publisher={IEEE}
}

@article{burckhardt2022netherite,
  title={Netherite: Efficient execution of serverless workflows},
  author={Burckhardt, Sebastian and Chandramouli, Badrish and Gillum, Chris and Justo, David and Kallas, Konstantinos and McMahon, Connor and Meiklejohn, Christopher S and Zhu, Xiangfeng},
  journal={VLDB},
  volume={15},
  number={8},
  pages={1591--1604},
  year={2022},
  publisher={VLDB Endowment}
}

@inproceedings{jiang2017serverless,
  title={Serverless execution of scientific workflows},
  author={Jiang, Qingye and Lee, Young Choon and Zomaya, Albert Y},
  booktitle={International Conference on Service-Oriented Computing},
  pages={706--721},
  year={2017},
  organization={Springer}
}

@inproceedings{kijak2018challenges,
  title={Challenges for scheduling scientific workflows on cloud functions},
  author={Kijak, Joanna and Martyna, Piotr and Pawlik, Maciej and Balis, Bartosz and Malawski, Maciej},
  booktitle={CLOUD},
  pages={460--467},
  year={2018},
  organization={IEEE}
}

\clearpage
\appendix

\section{Cold starts and Function Executor}
\begin{figure}[!htbp]
  \centering
  \includegraphics[width=0.3\textwidth]{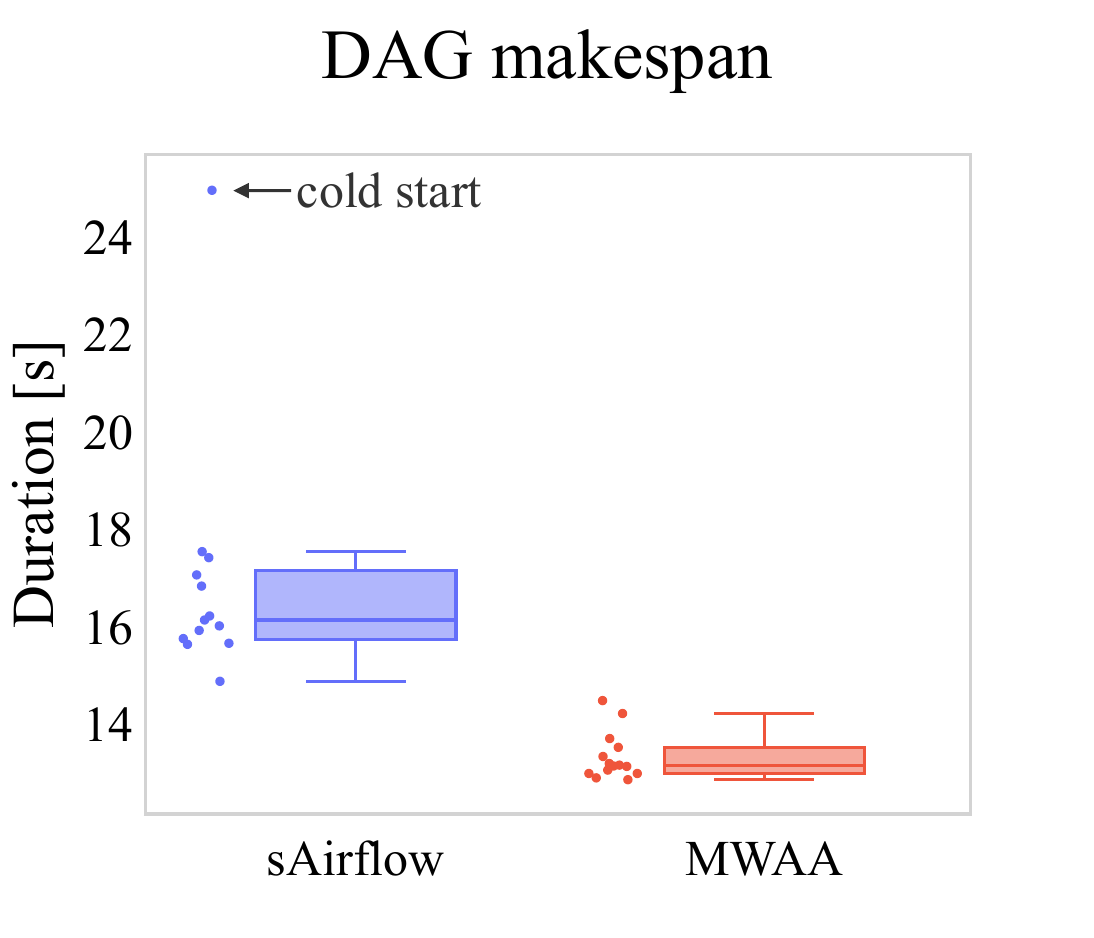}
  \includegraphics[width=0.3\textwidth]{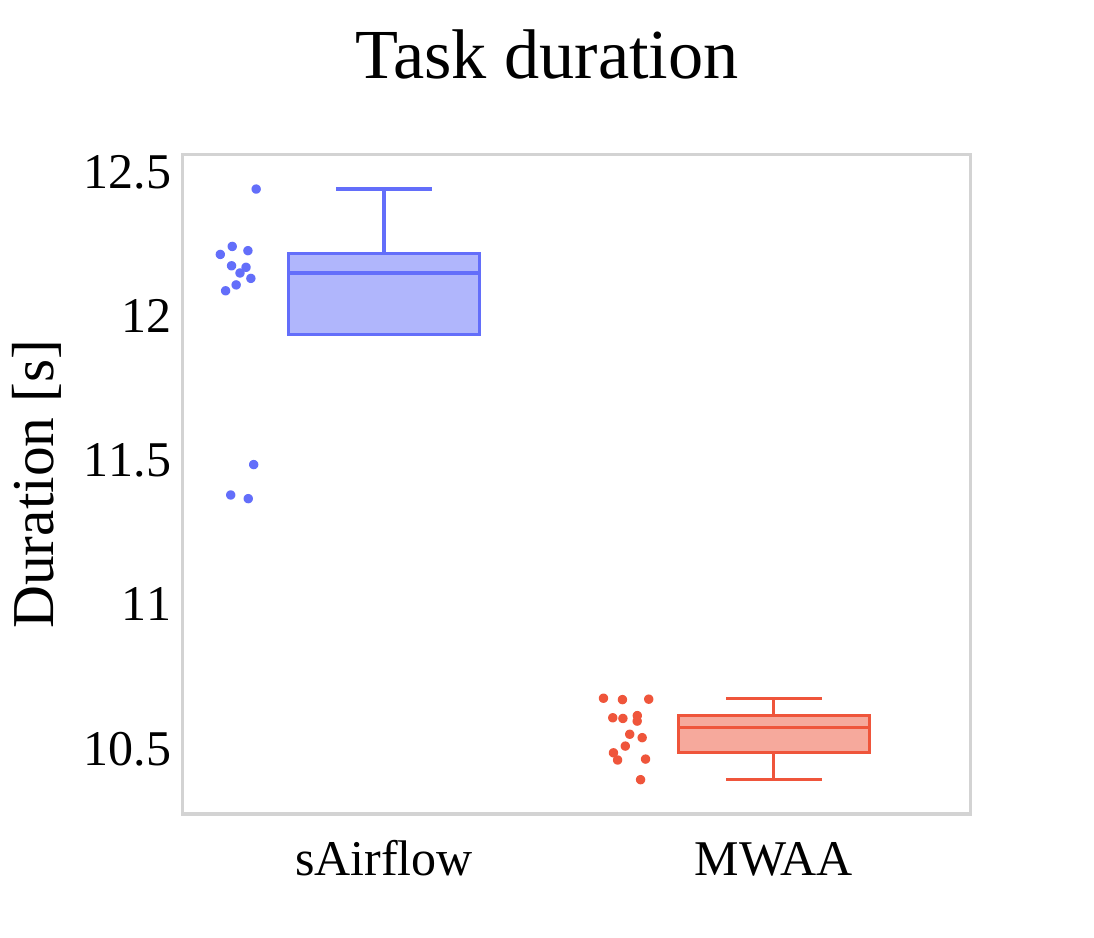}
  \includegraphics[width=0.3\textwidth]{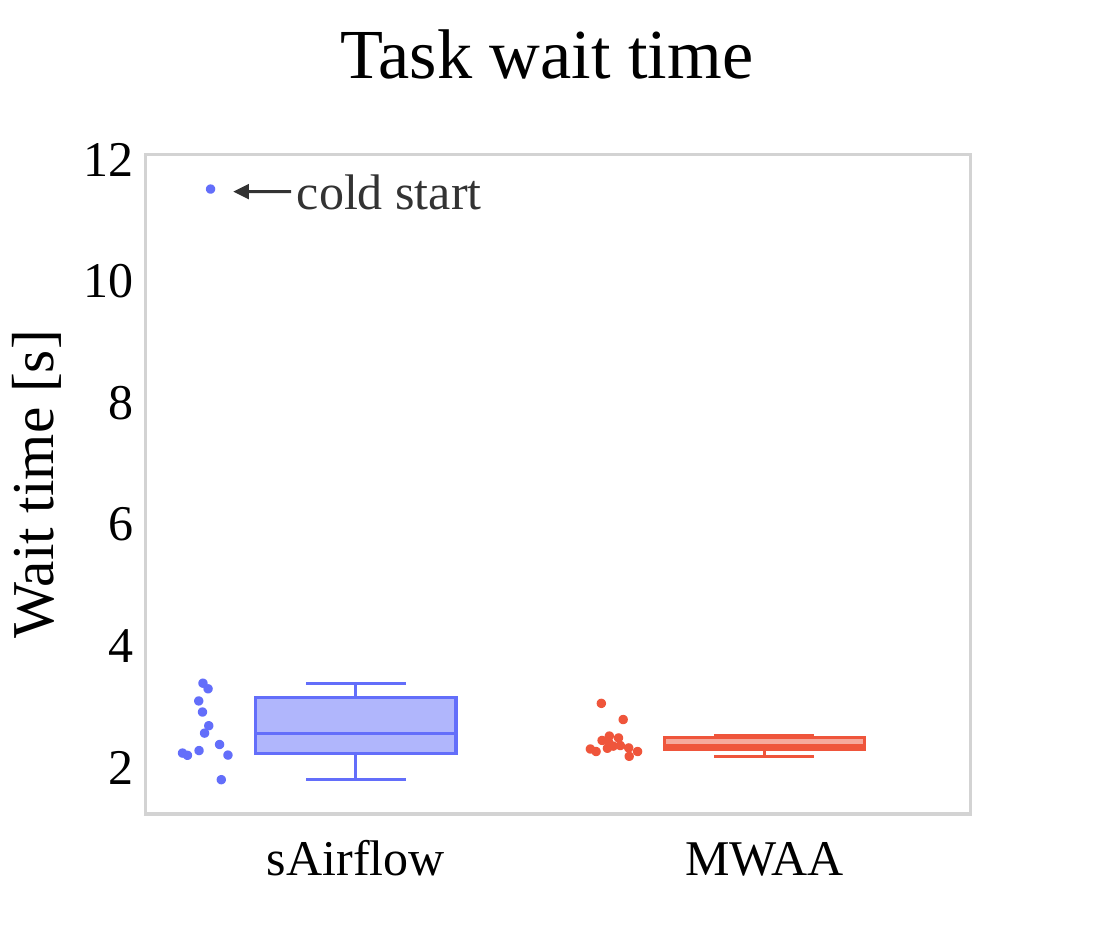}
  \caption{A single-task DAG (a chain with $n=1$), $p=10$, $T=5$. The first DAG run, resulting in a cold start, corresponds to the outlier in the left and right figures.}
  \label{fig:1task_line_cold}
\end{figure}

\begin{figure}[!htbp]
    \centering
    \subfloat[$n=16$\label{fig:16task_parallel_cold}]{
      \includegraphics[width=0.3\textwidth]{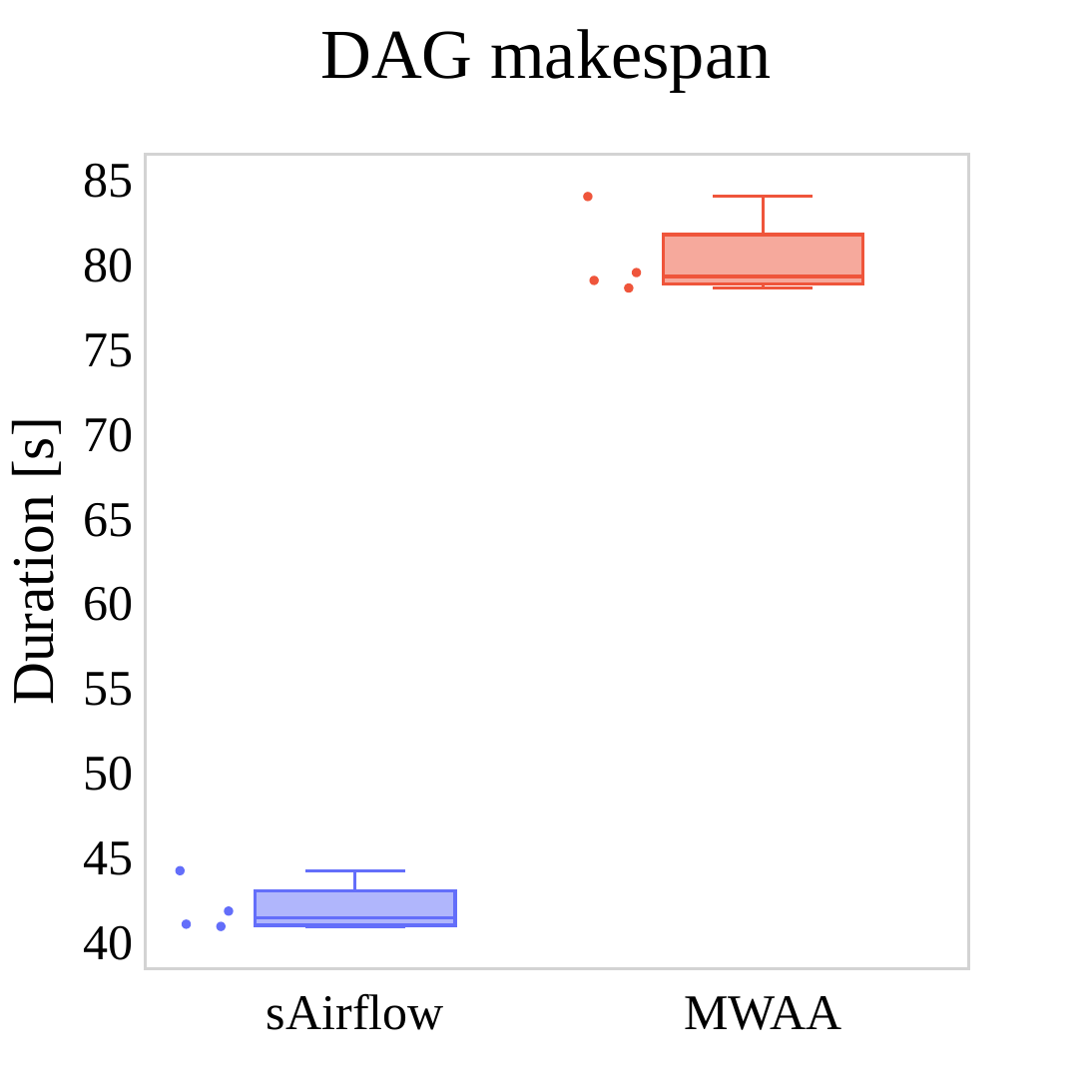}
      \includegraphics[width=0.3\textwidth]{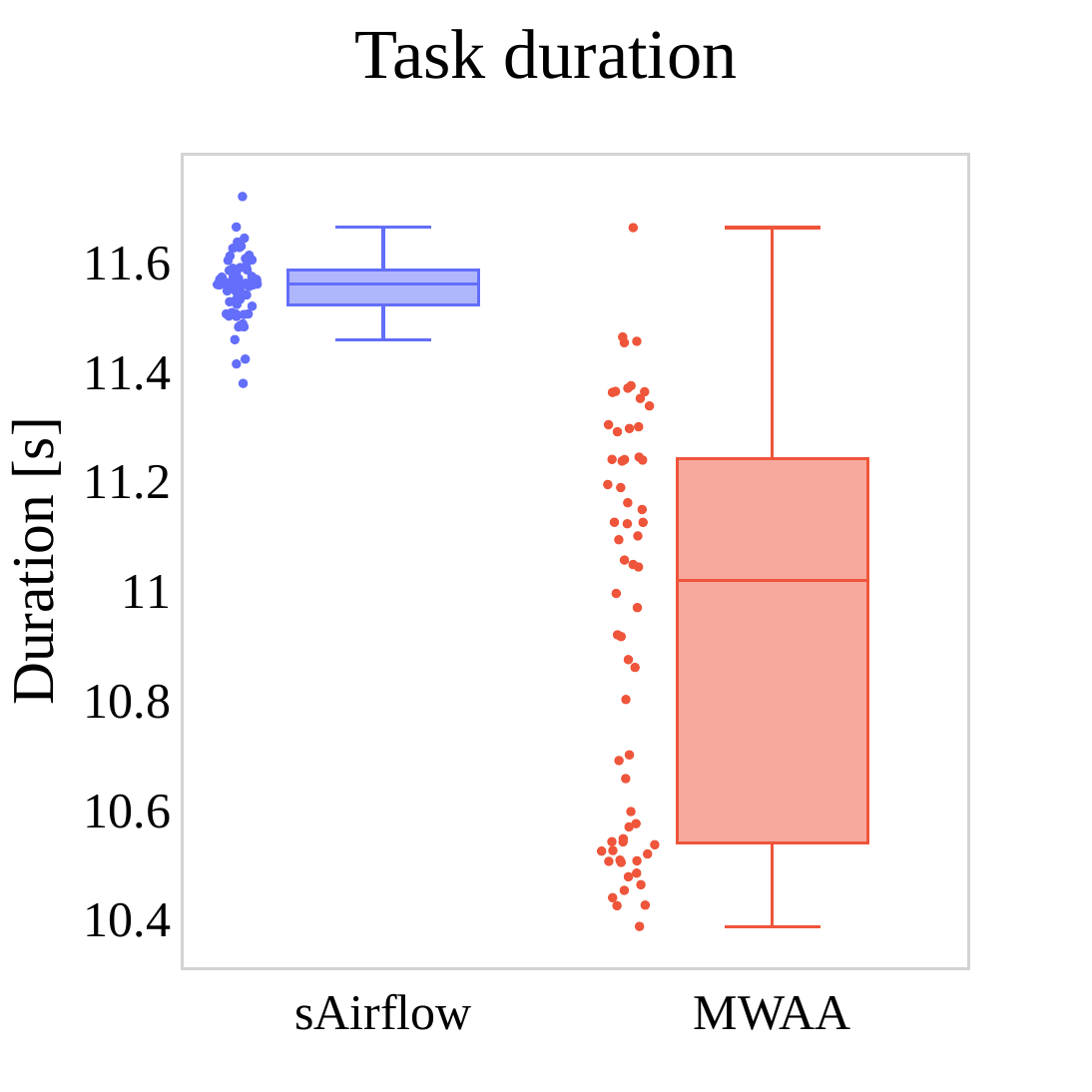}
      \includegraphics[width=0.3\textwidth]{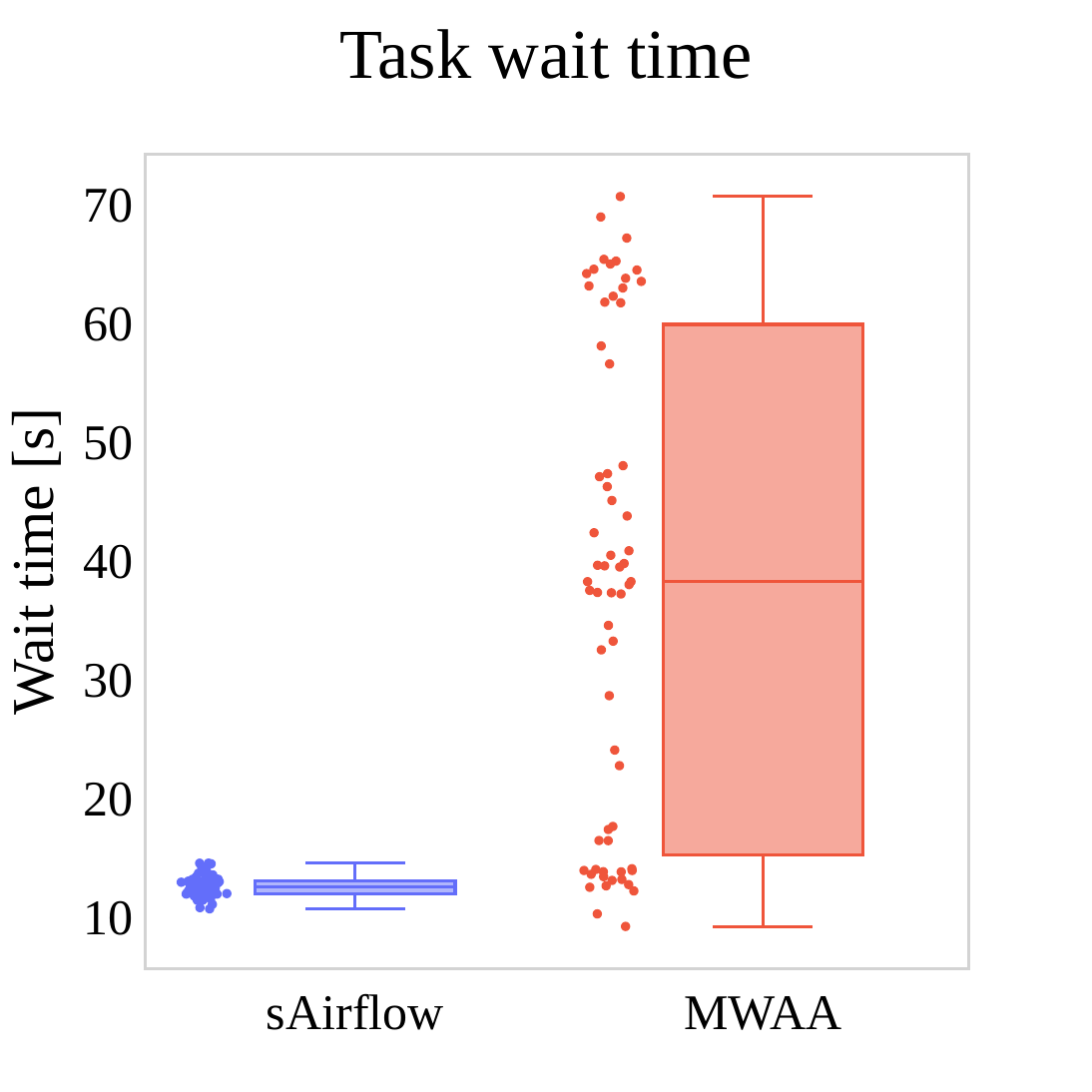}
    }%
    \\
    \subfloat[$n=32$\label{fig:32task_parallel_cold}]{
      \includegraphics[width=0.3\textwidth]{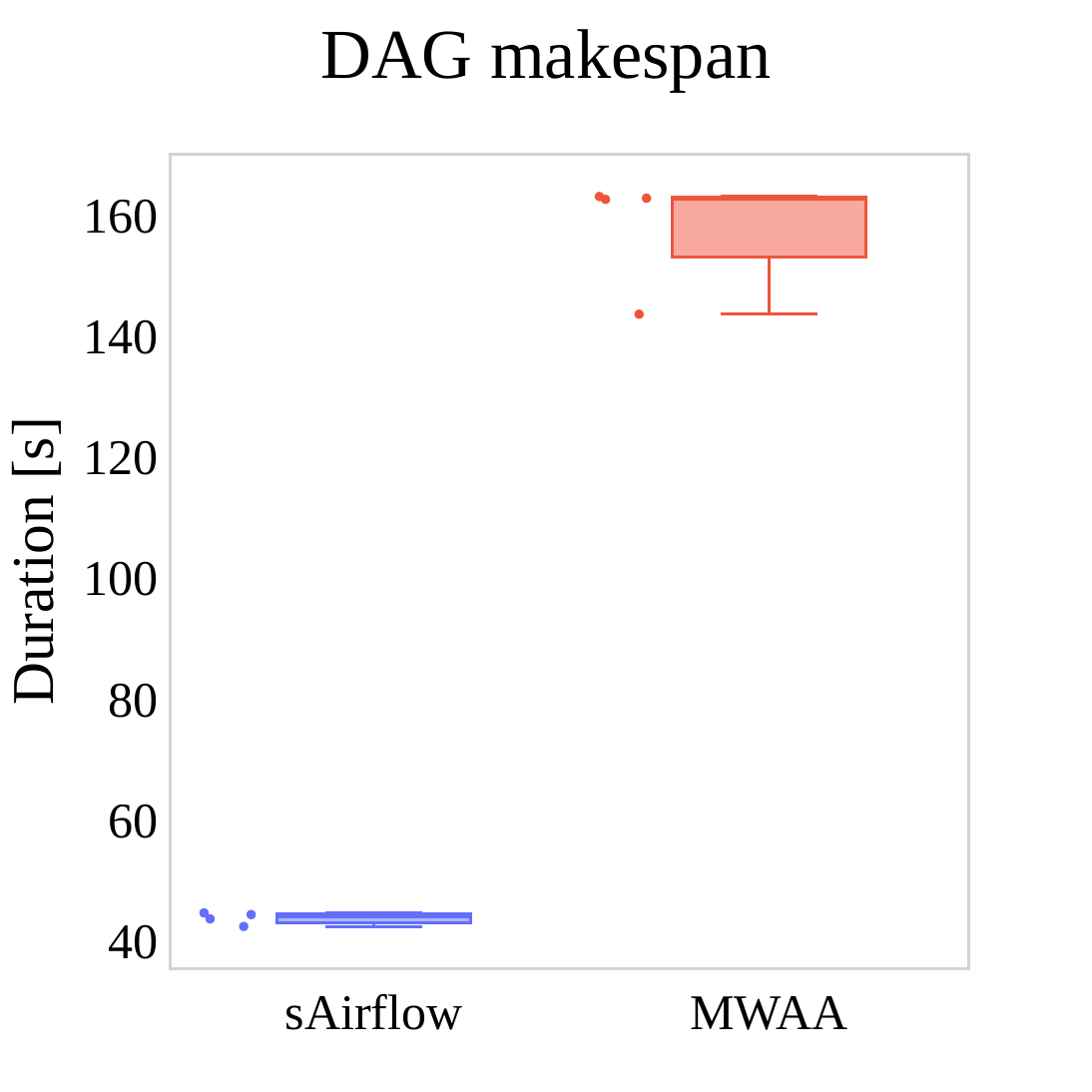}
      \includegraphics[width=0.3\textwidth]{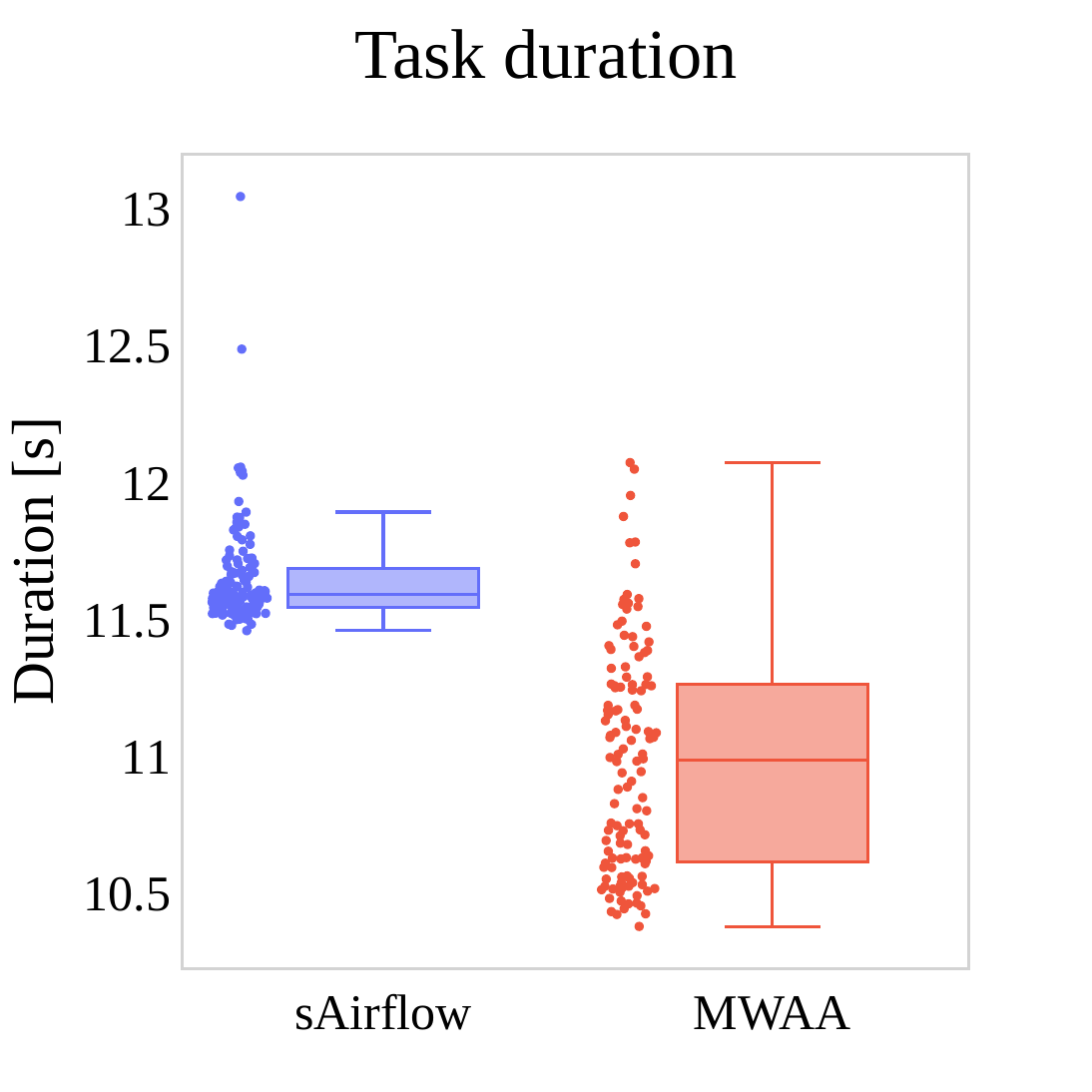}
      \includegraphics[width=0.3\textwidth]{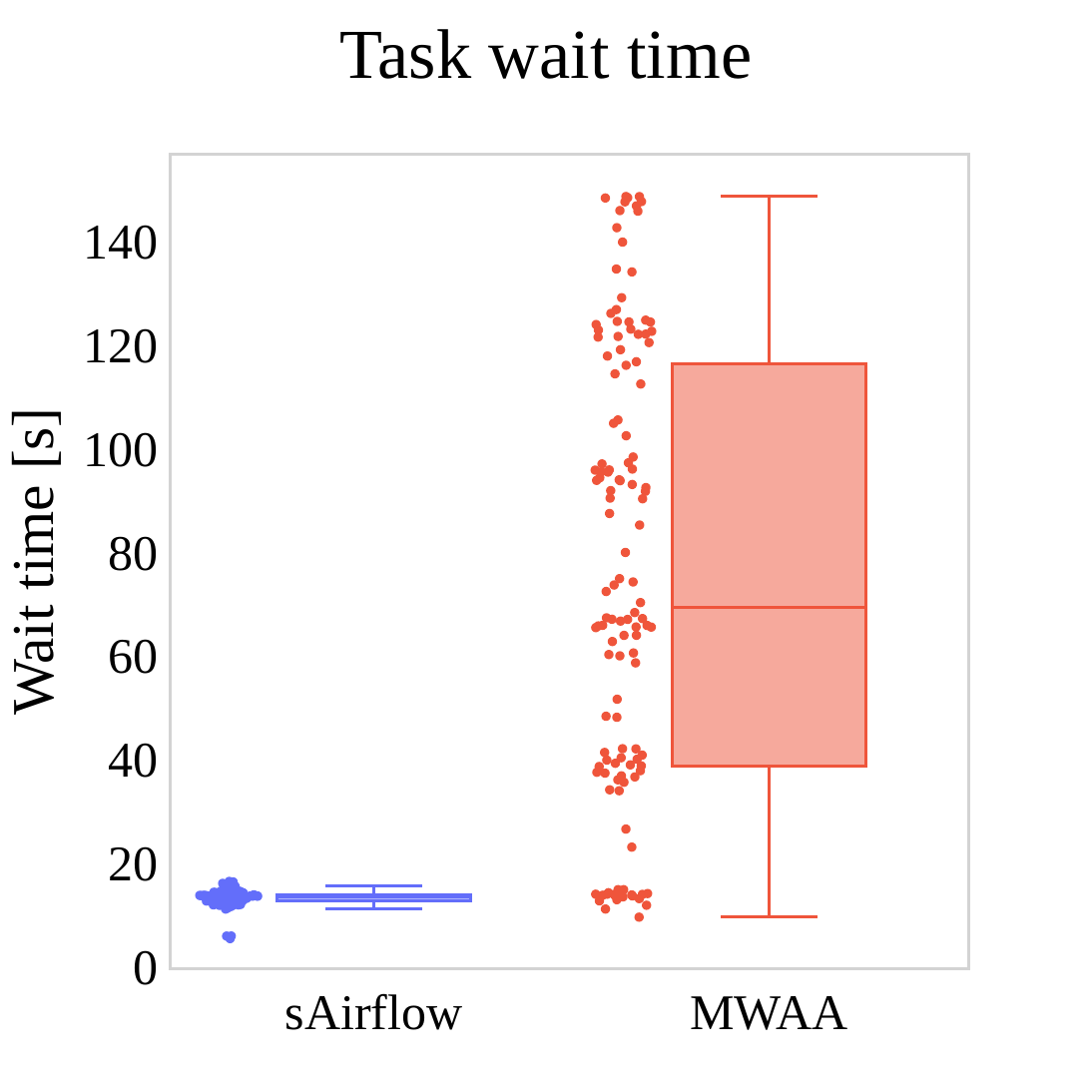}
    }%
    \\
    \subfloat[$n=64$\label{fig:64task_parallel_cold} \label{fig:mwaa_scale_out_64}]{
      \includegraphics[width=0.2\textwidth]{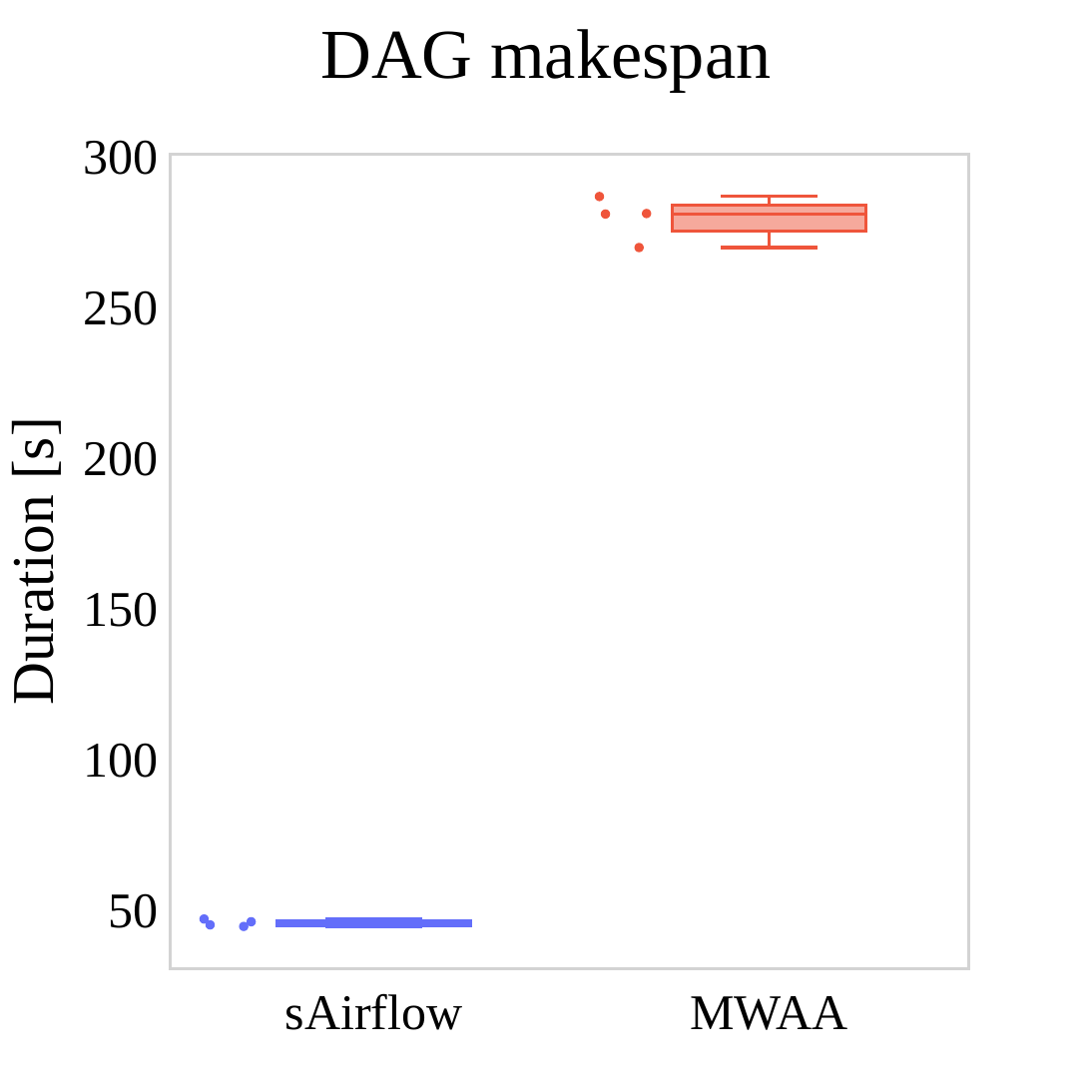}
      \includegraphics[width=0.2\textwidth]{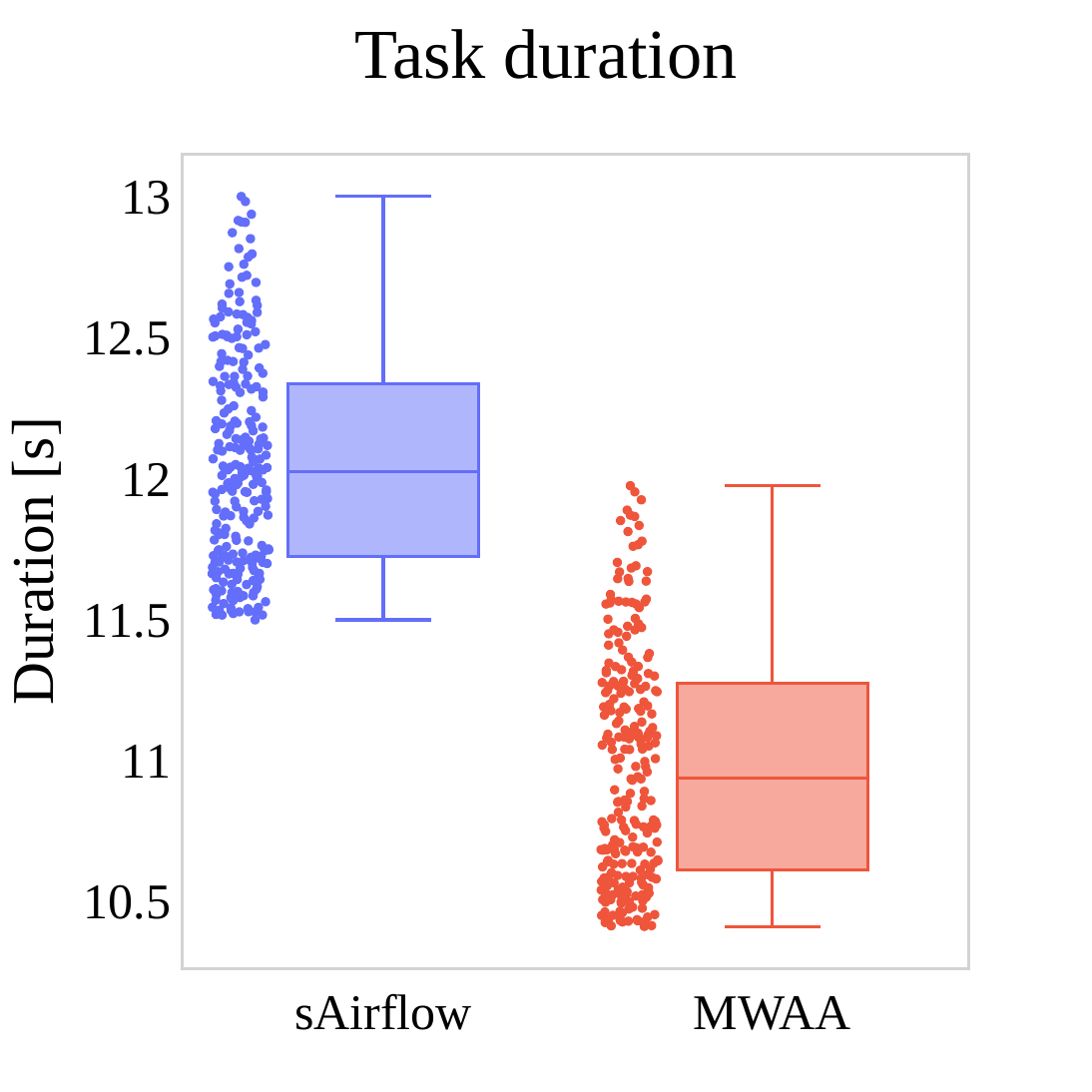}
      \includegraphics[width=0.2\textwidth]{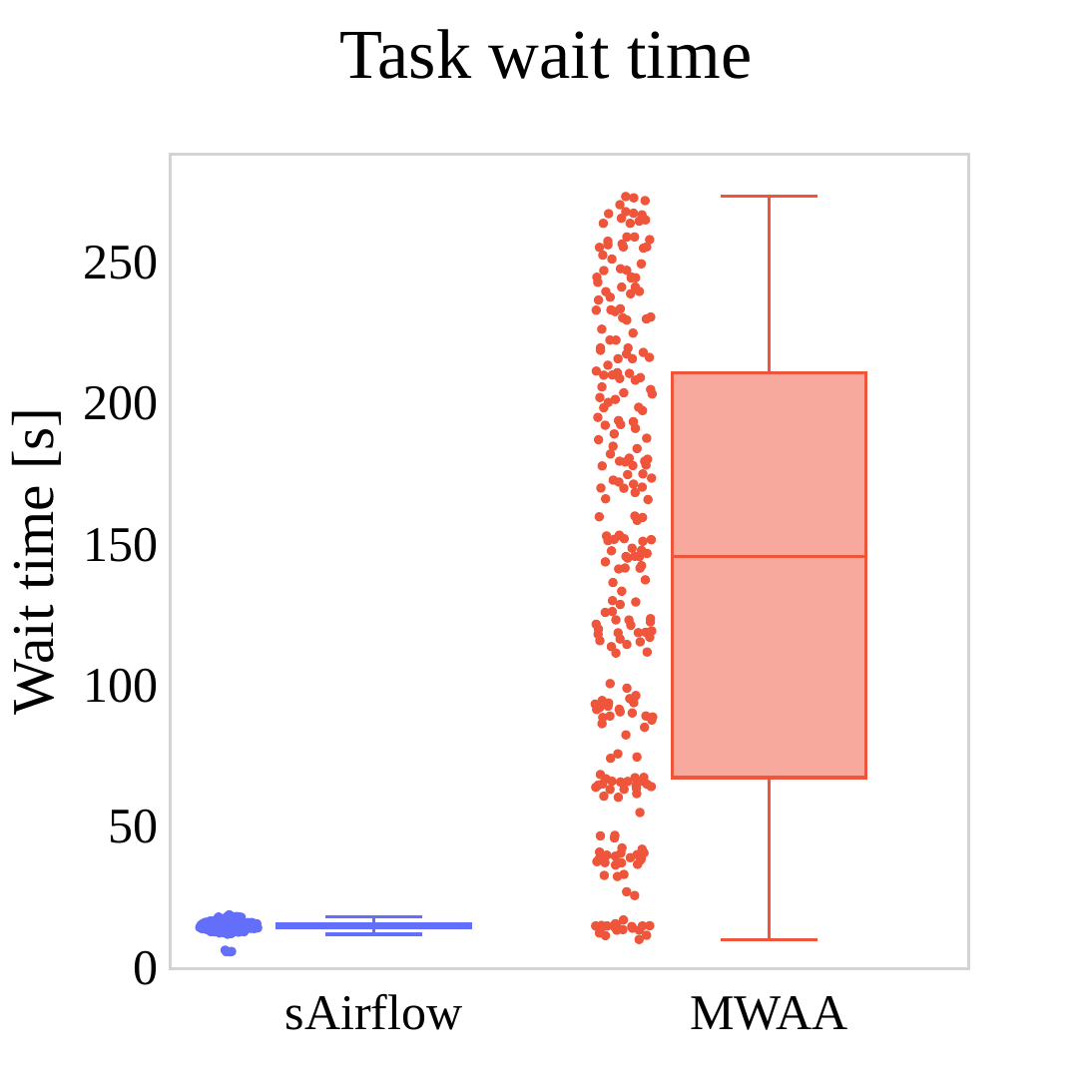}
      \includegraphics[width=0.2\textwidth]{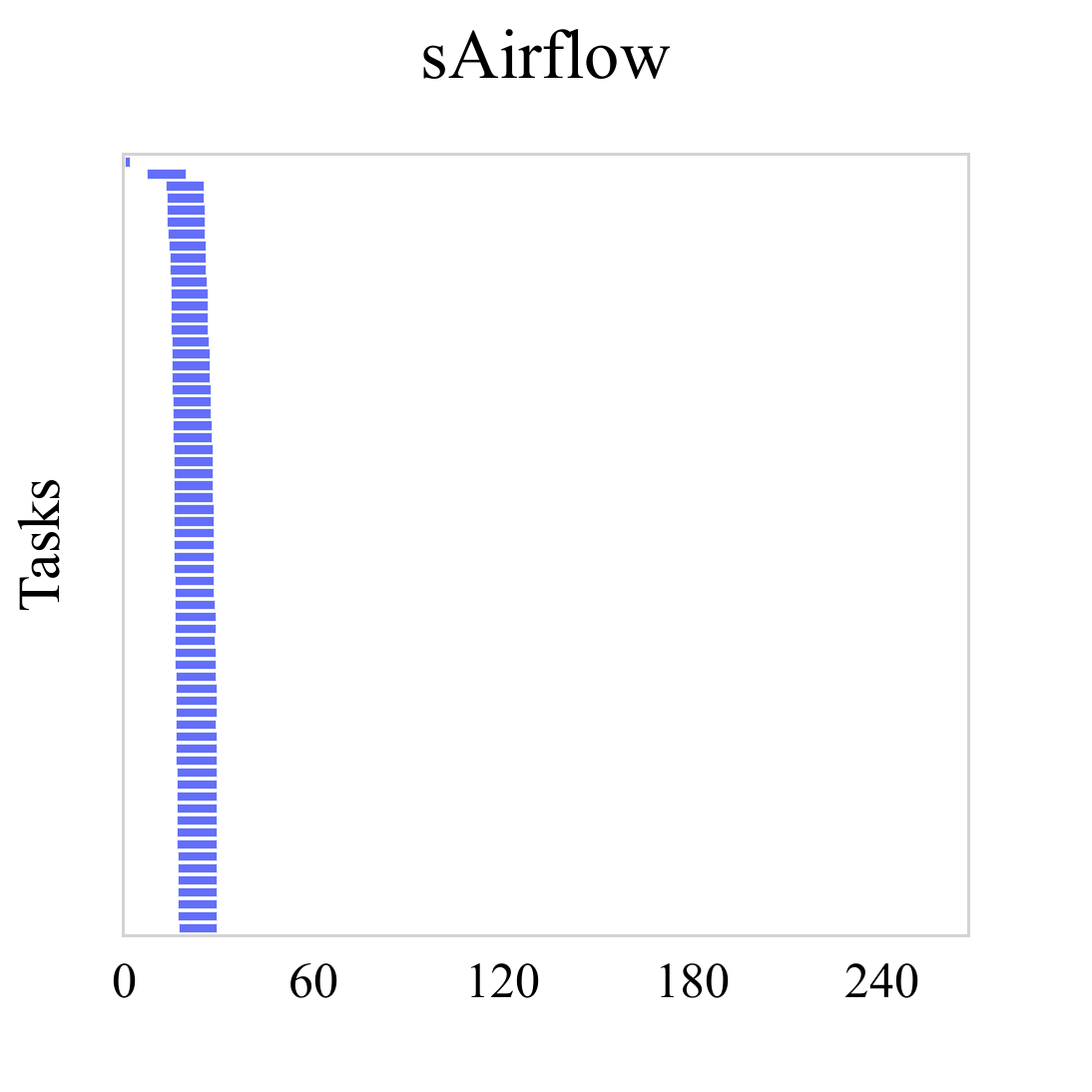}
      \includegraphics[width=0.2\textwidth]{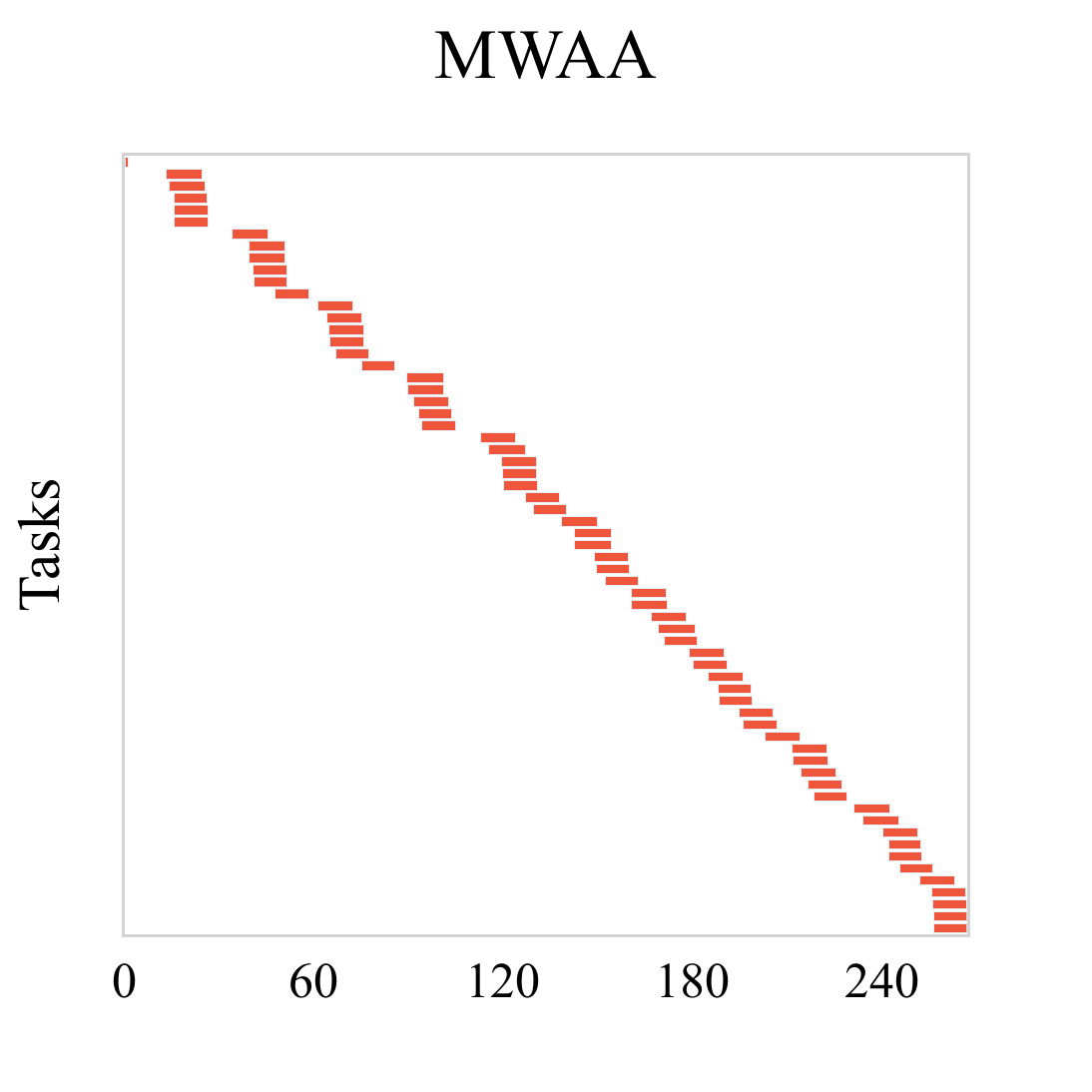}
    }%
    \\
    \subfloat[$n=125$\label{fig:125task_parallel_cold} \label{fig:mwaa_scale_out_125}]{
      \includegraphics[width=0.2\textwidth]{DAG_makespan-125task_parallel_cold.pdf}
      \includegraphics[width=0.2\textwidth]{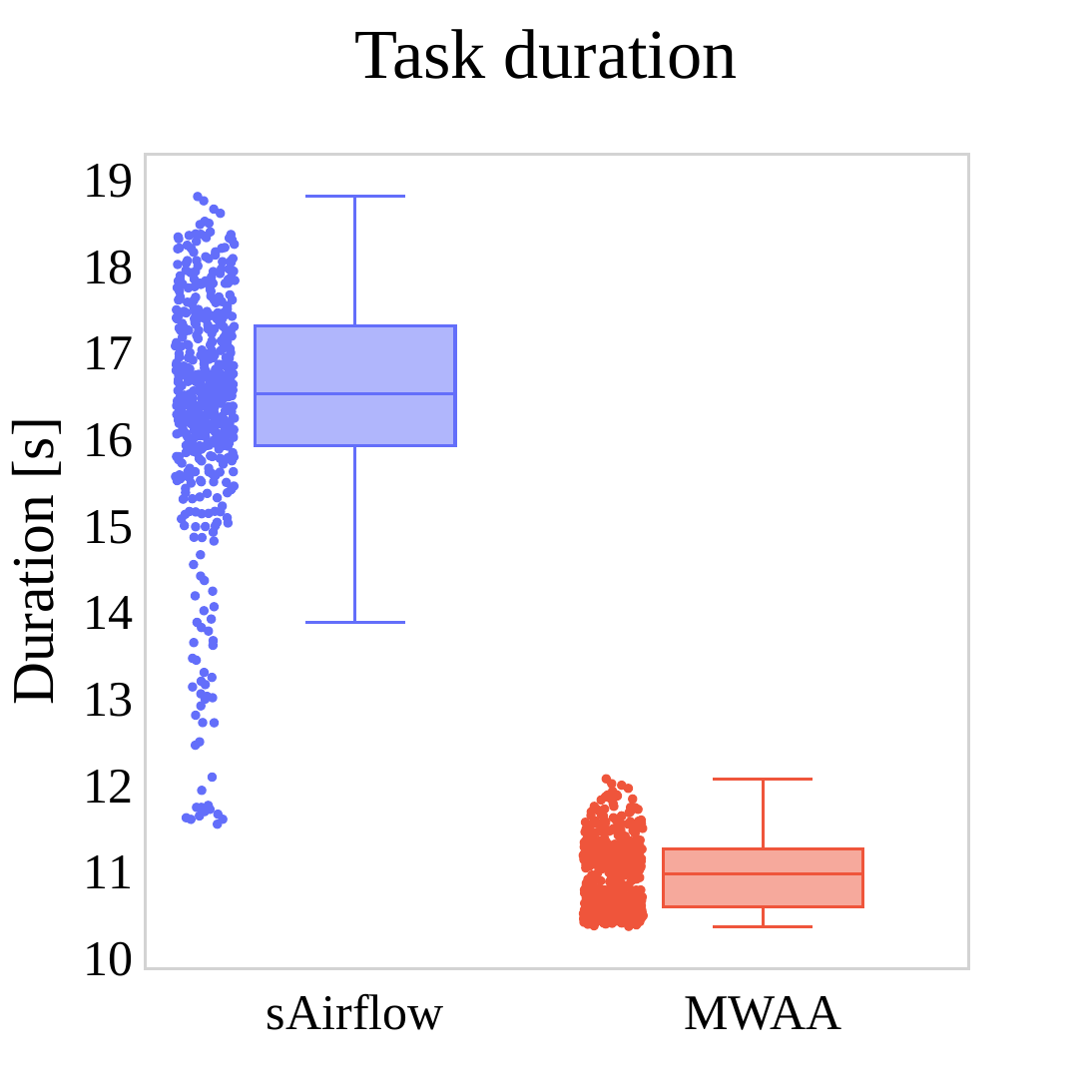}
      \includegraphics[width=0.2\textwidth]{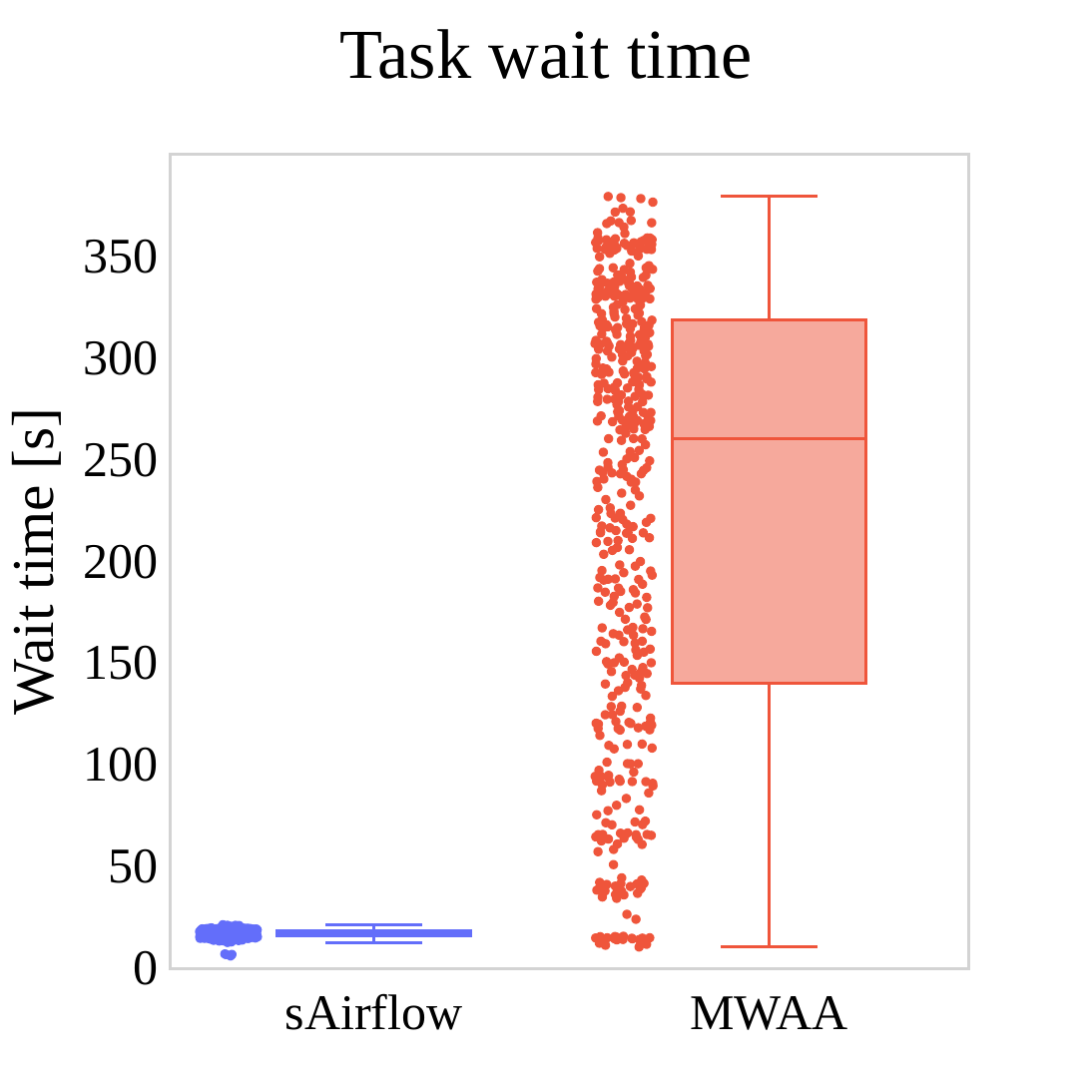}
      \includegraphics[width=0.2\textwidth]{sAirflow-125task_parallel_cold_fmt.pdf}
      \includegraphics[width=0.2\textwidth]{MWAA-125task_parallel_cold_fmt.pdf}
    }
    \caption{Parallel DAGs, function executor, cold starts, $p=10$, $T=30$.
    Gantt charts on the right side correspond to one of the DAG runs.
    }
    \label{fig:app_parallel_cold}
\end{figure}

For transparency, we report full results from the following experiments:
\begin{itemize}
  \item A detailed analysis of the overheads by analyzing a single-task DAG, Fig.~\ref{fig:1task_line_cold}.
  \item Parallel DAGs with $n=16$, $n=32$, $n=64$ and $n=125$, Fig~\ref{fig:app_parallel_cold}.
\end{itemize}
We refer to the main text for the analysis of these results.

\newpage
\section{Warm starts and Function Executor}
\begin{figure}[!htbp]
  \centering
  \subfloat[$n=1$\label{fig:1task_line}]{
    \includegraphics[width=0.3\textwidth]{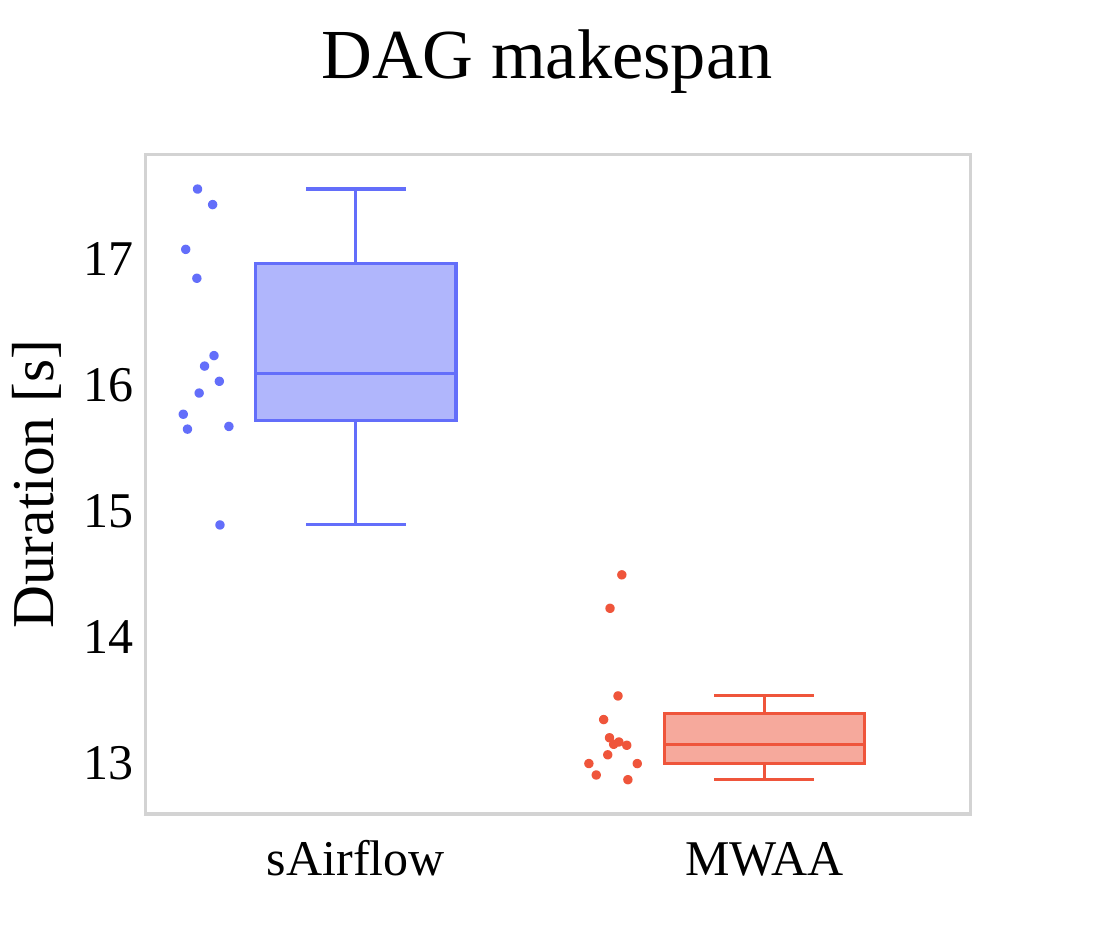}
    \includegraphics[width=0.3\textwidth]{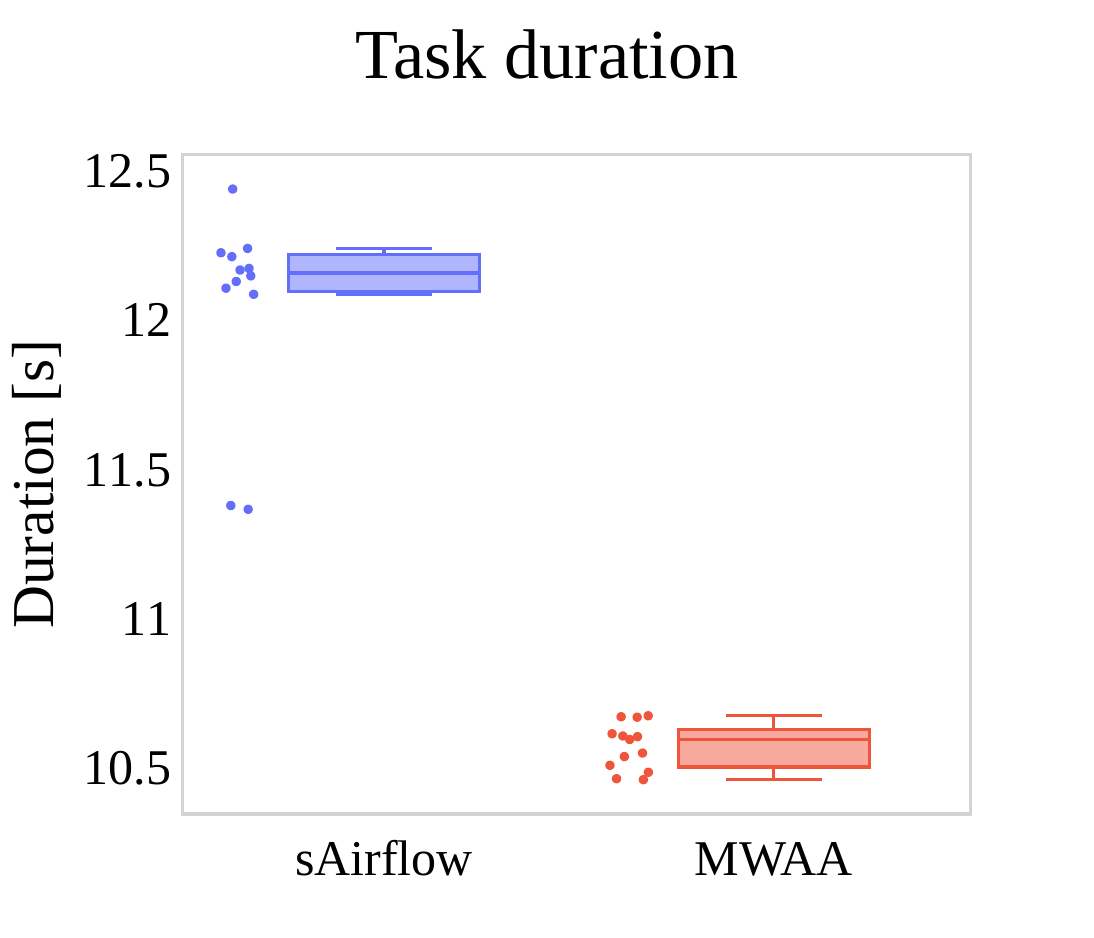}
    \includegraphics[width=0.3\textwidth]{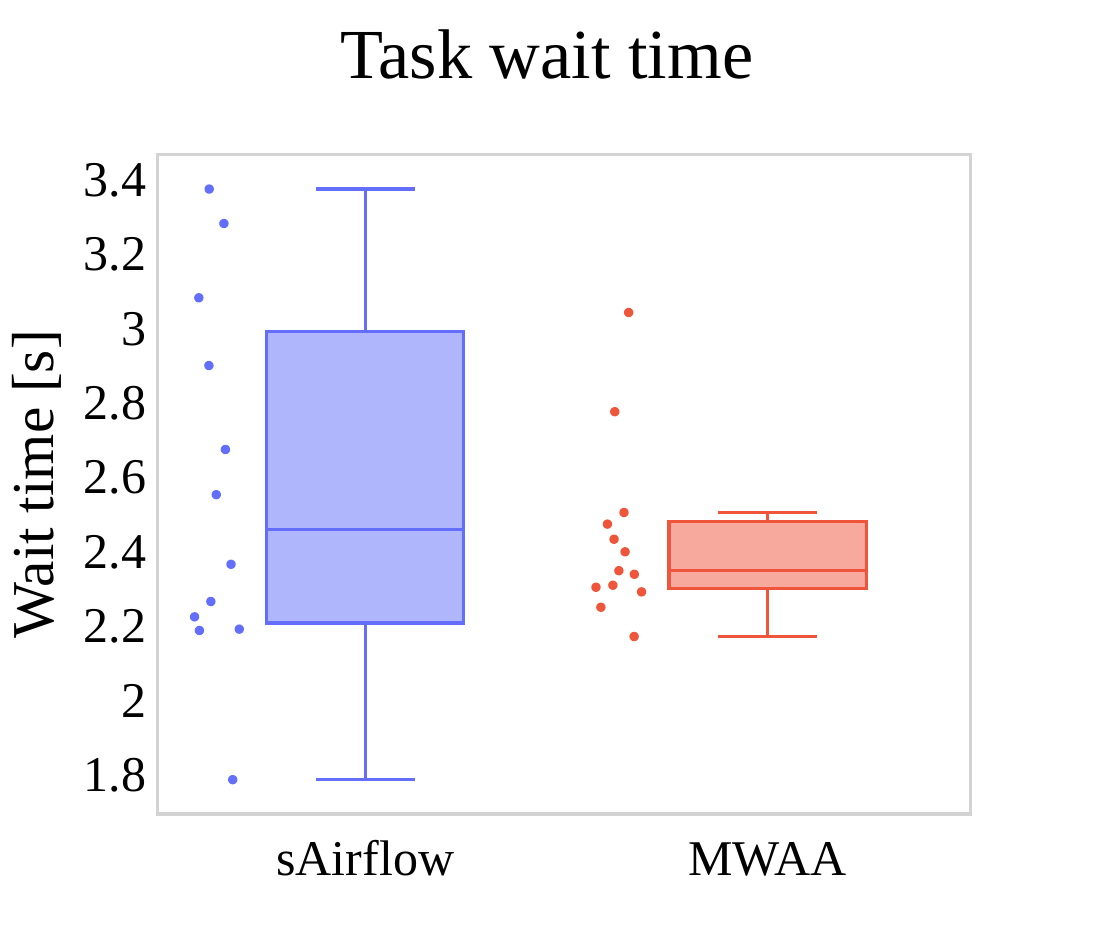}
  }%
  \\
  \subfloat[$n=5$\label{fig:app_5task_line}]{
    \includegraphics[width=0.3\textwidth]{DAG_makespan-5task_line.pdf}
    \includegraphics[width=0.3\textwidth]{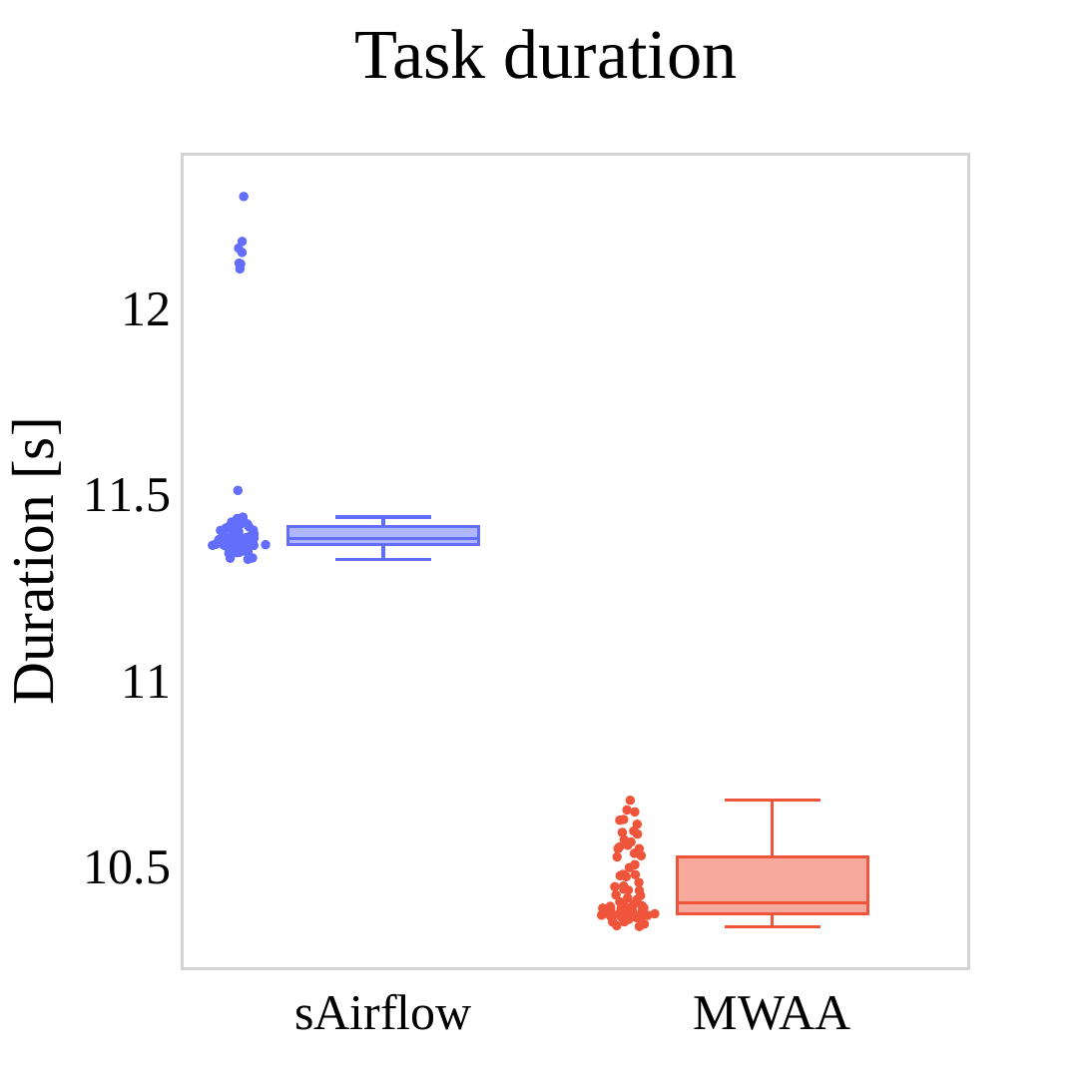}
    \includegraphics[width=0.3\textwidth]{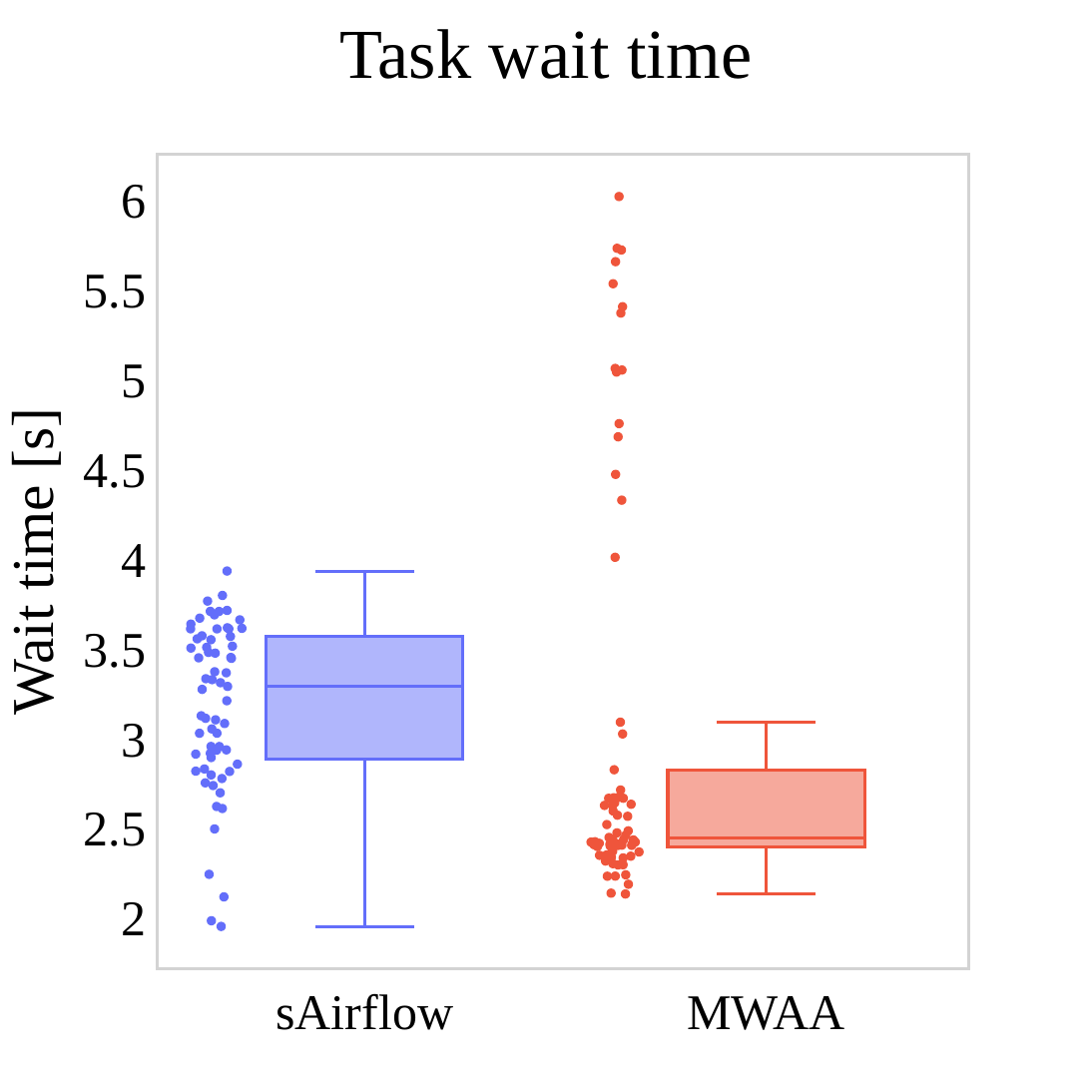}
  }%
  \\
  \subfloat[$n=10$\label{fig:10task_line}]{
    \includegraphics[width=0.3\textwidth]{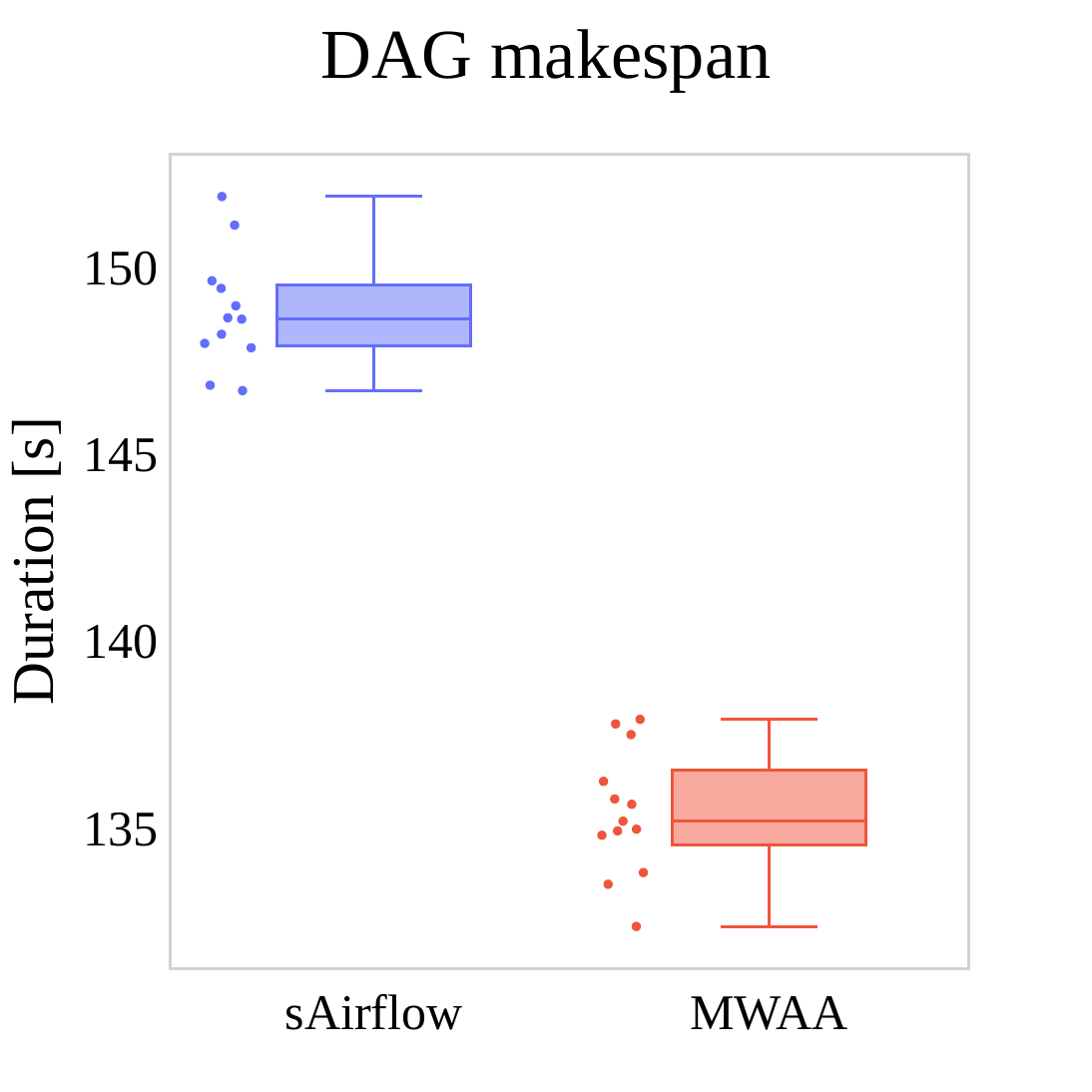}
    \includegraphics[width=0.3\textwidth]{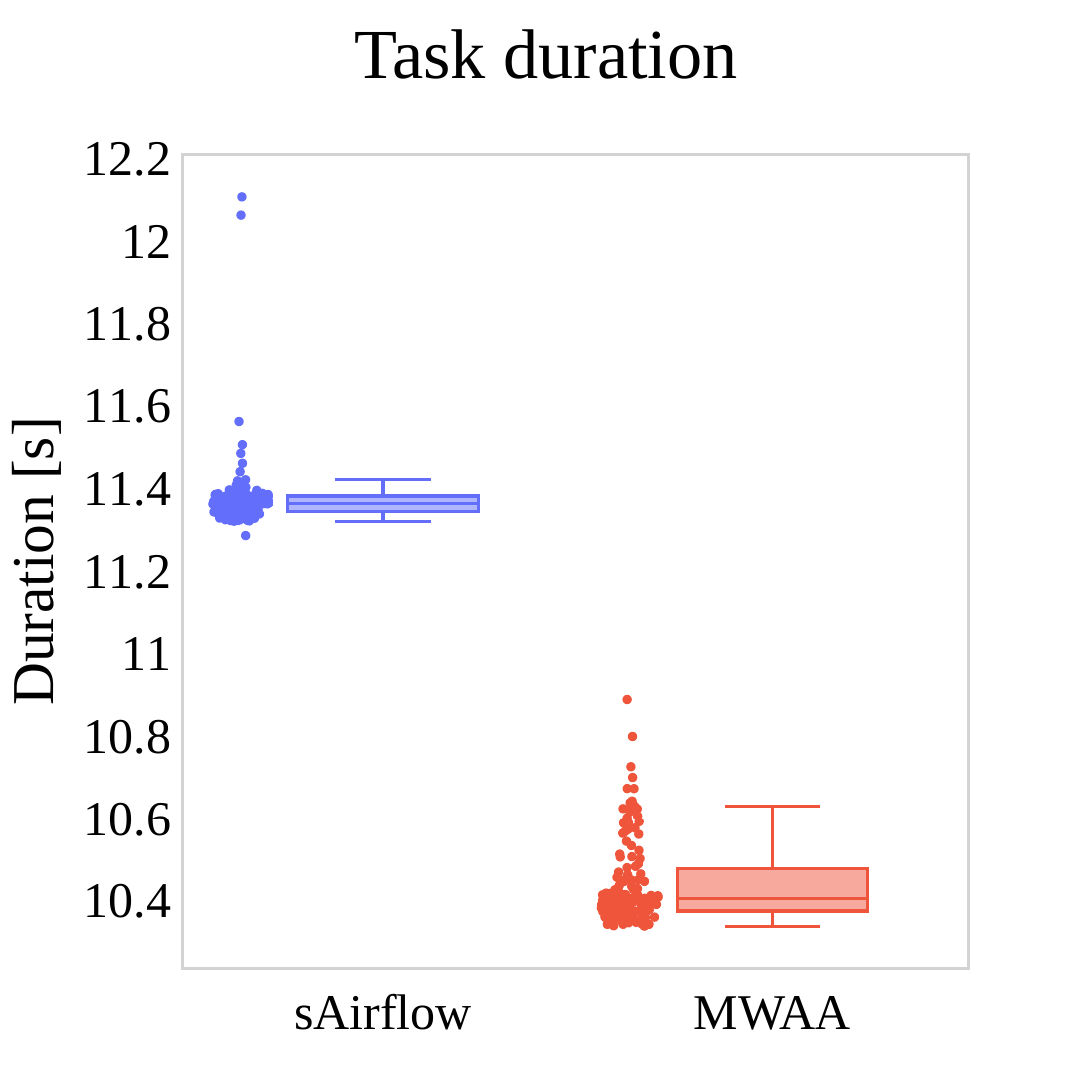}
    \includegraphics[width=0.3\textwidth]{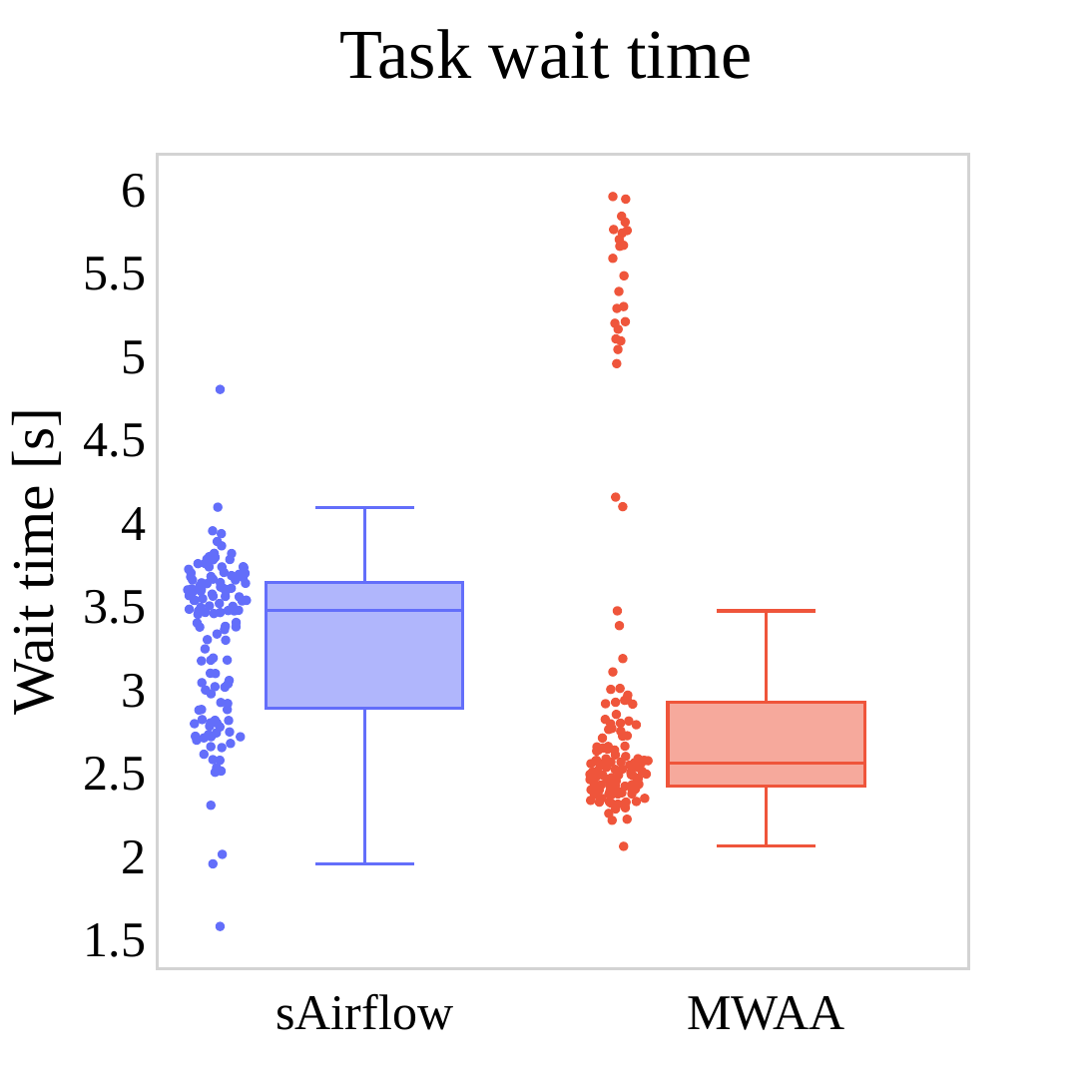}
  }%
  \caption{Chain DAG, function executor, warm starts, $p=10$, $T=5$. The first DAG run is not reported.}\label{fig:chain_warm}
\end{figure}

\begin{figure}[!htbp]
  \centering
  \subfloat[$n=16$]{
    \includegraphics[width=0.3\textwidth]{DAG_makespan-16task_parallel.pdf}
    \includegraphics[width=0.3\textwidth]{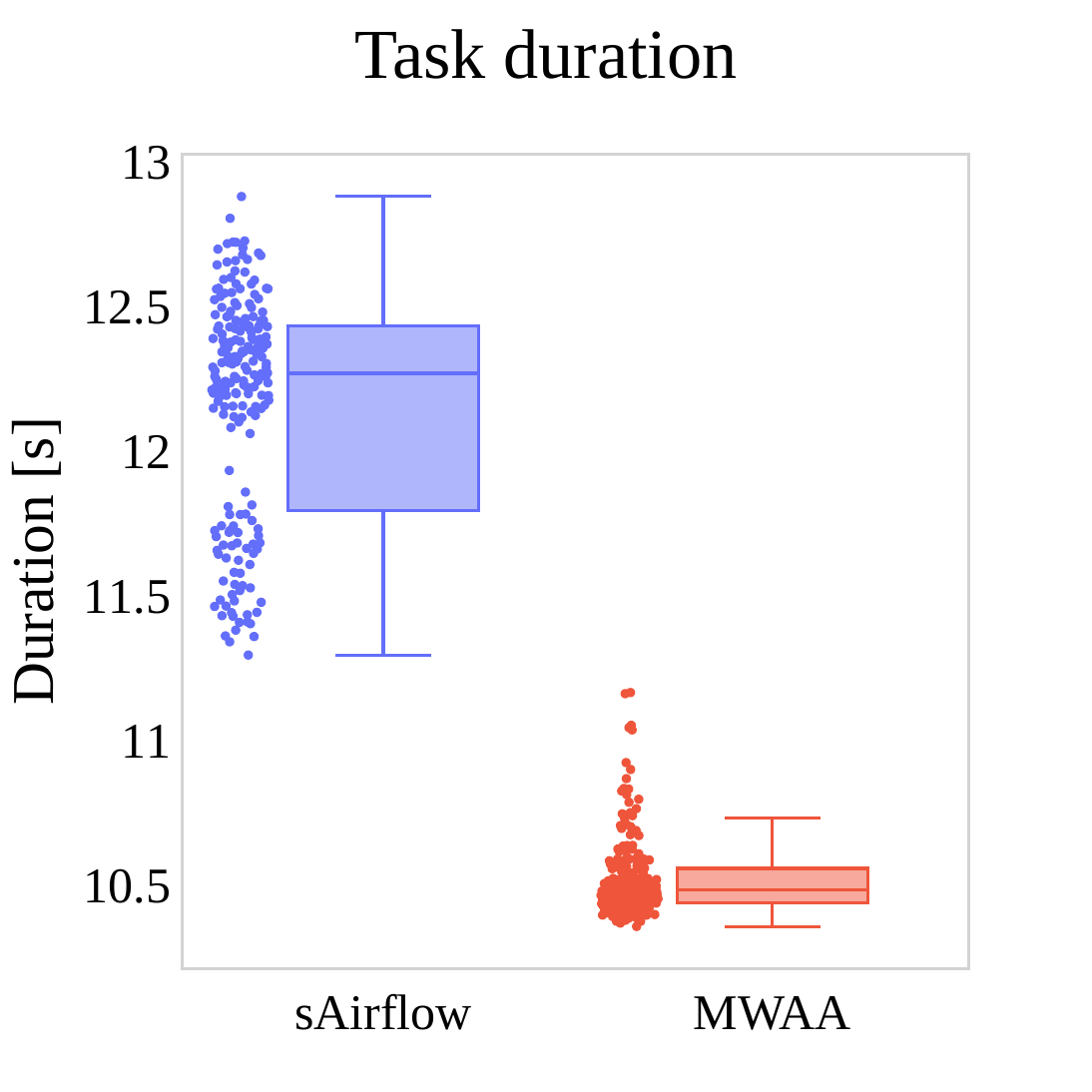}
    \includegraphics[width=0.3\textwidth]{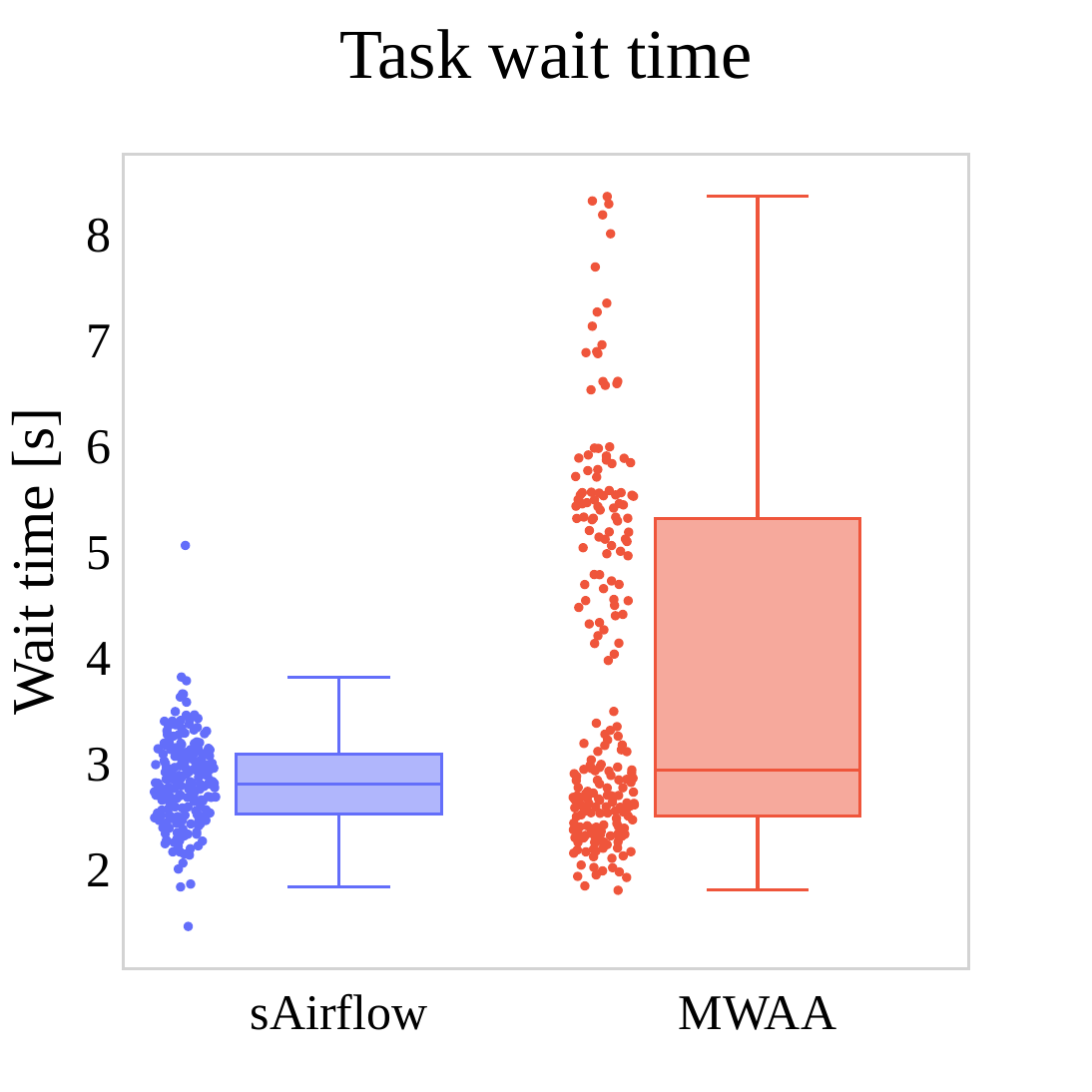}
  }%
  \\
  \subfloat[$n=32$]{
    \includegraphics[width=0.3\textwidth]{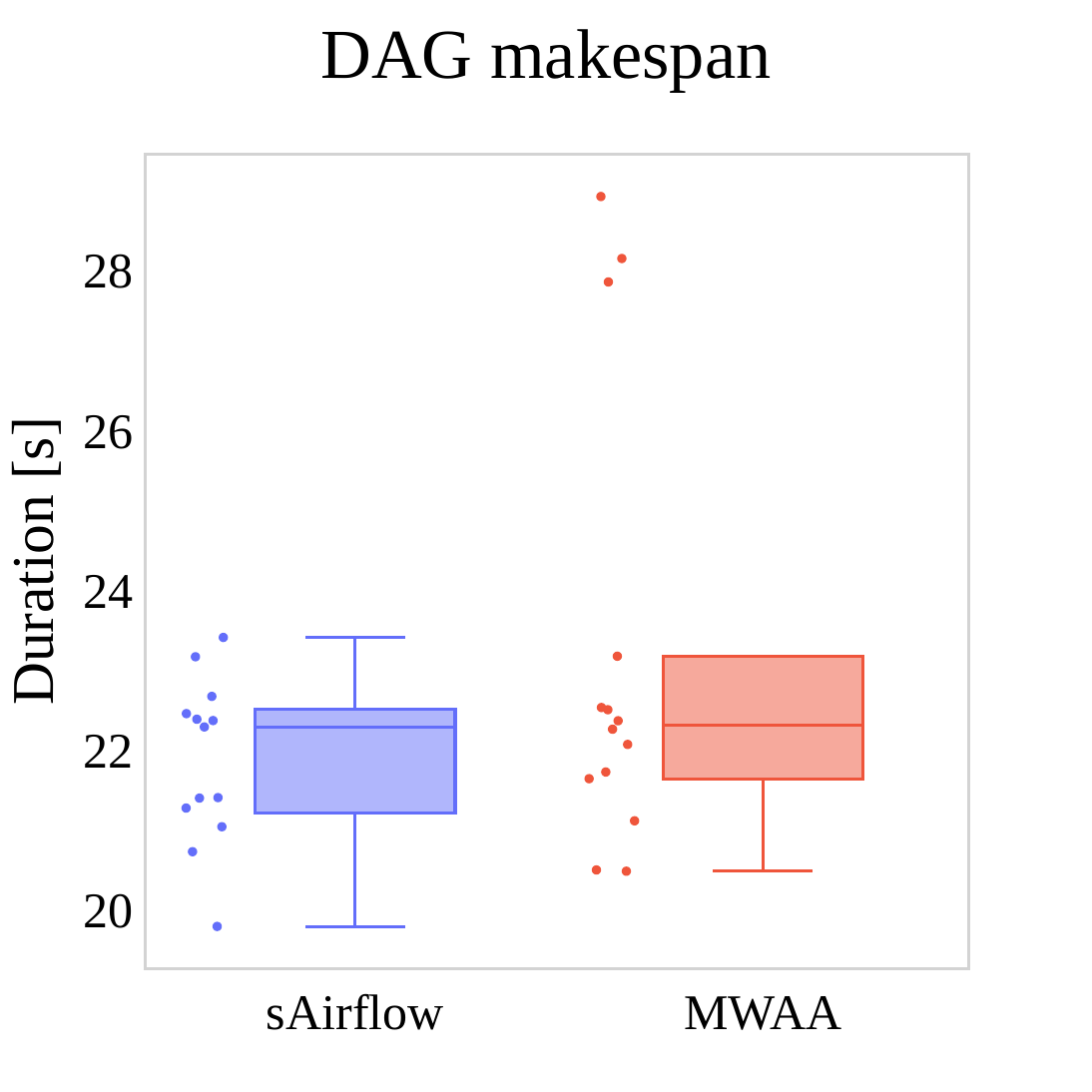}
    \includegraphics[width=0.3\textwidth]{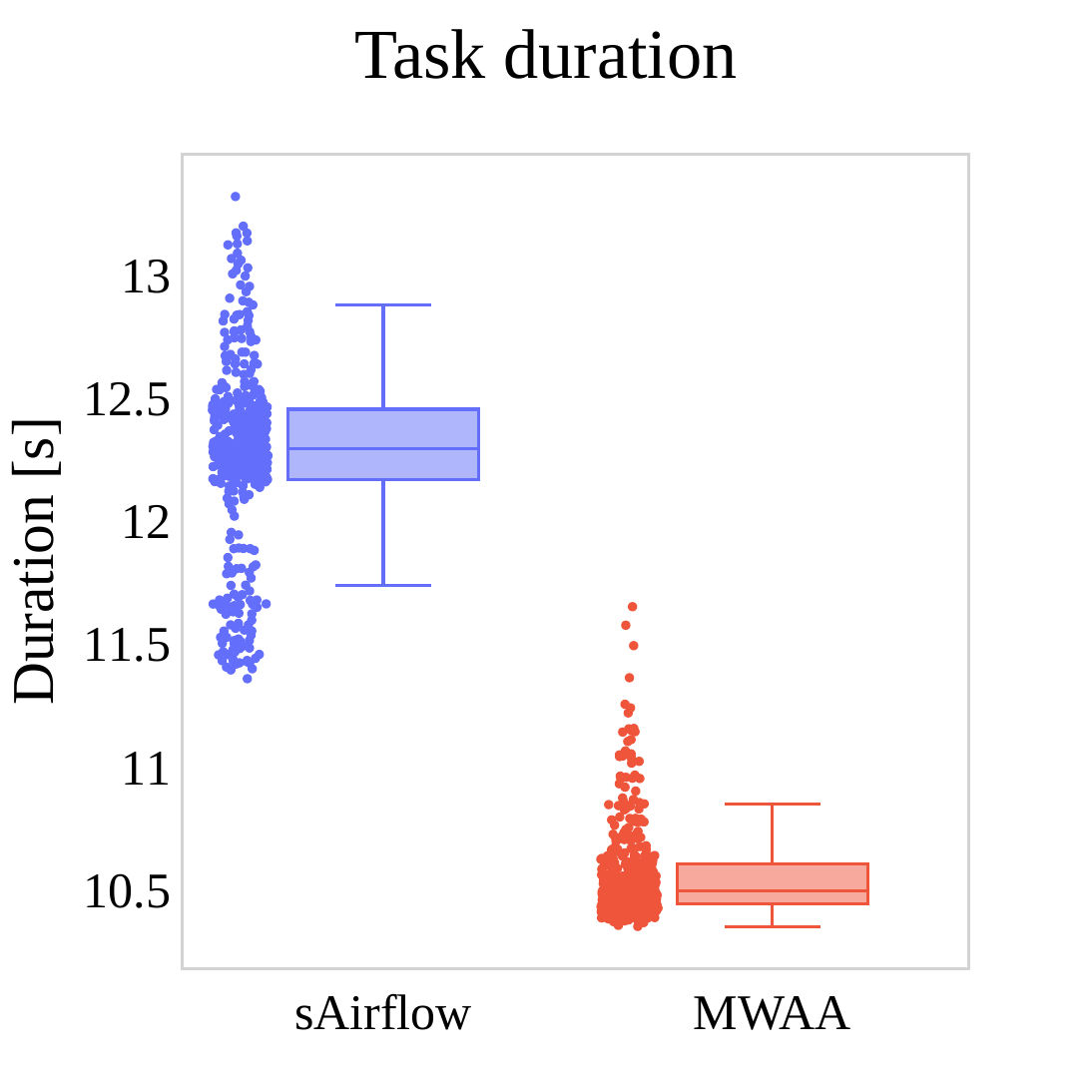}
    \includegraphics[width=0.3\textwidth]{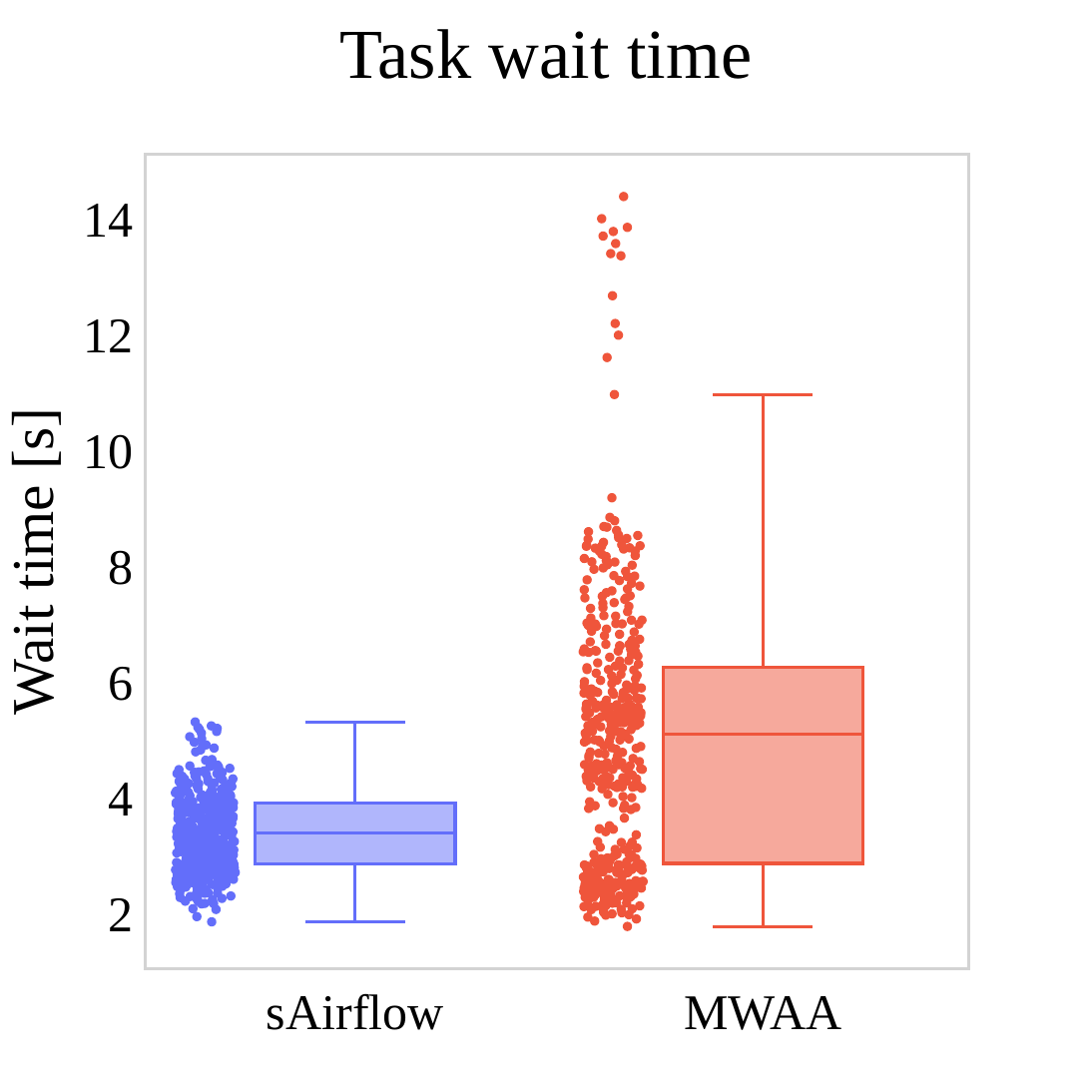}
  }%
  \\
  \subfloat[$n=64$]{
    \includegraphics[width=0.3\textwidth]{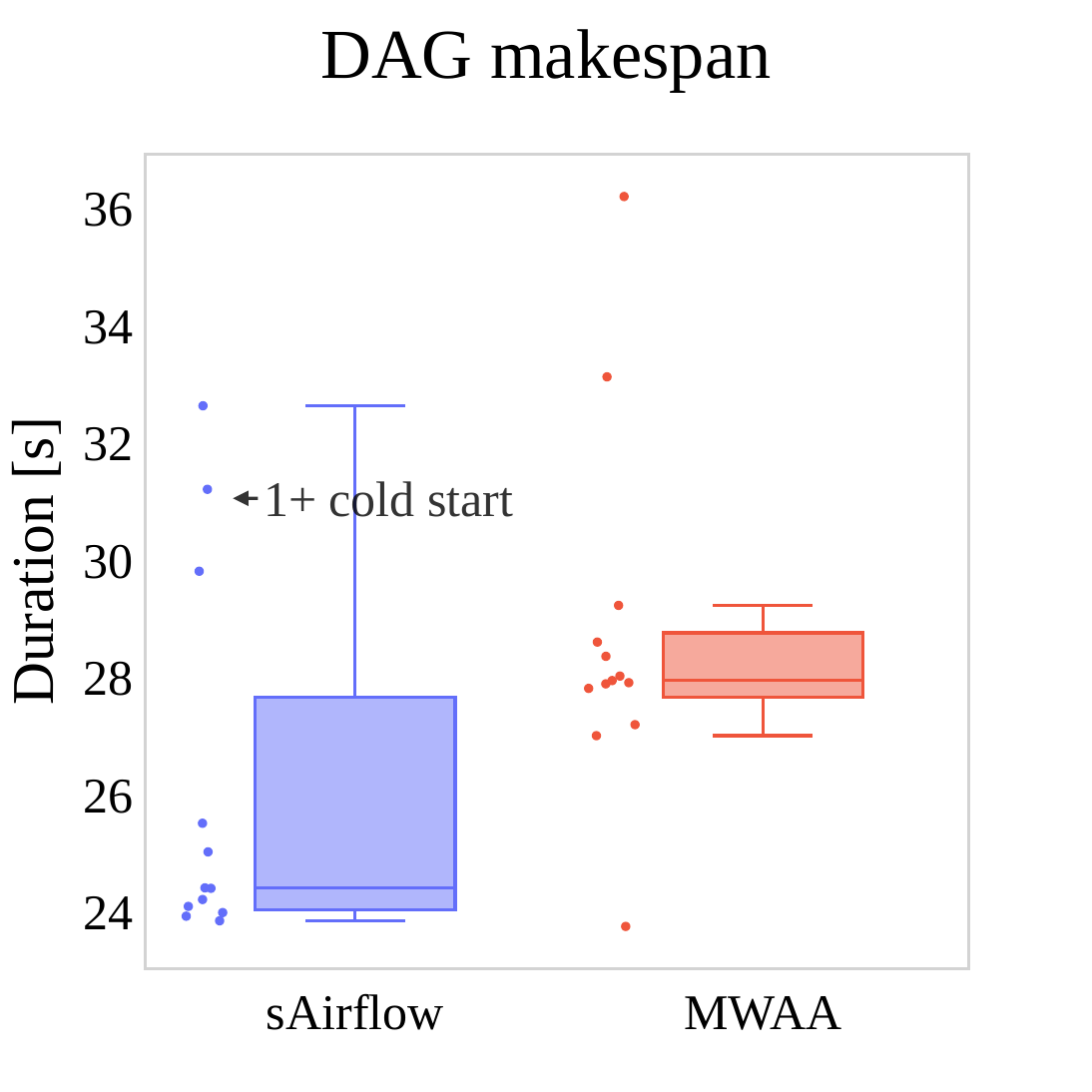}
    \includegraphics[width=0.3\textwidth]{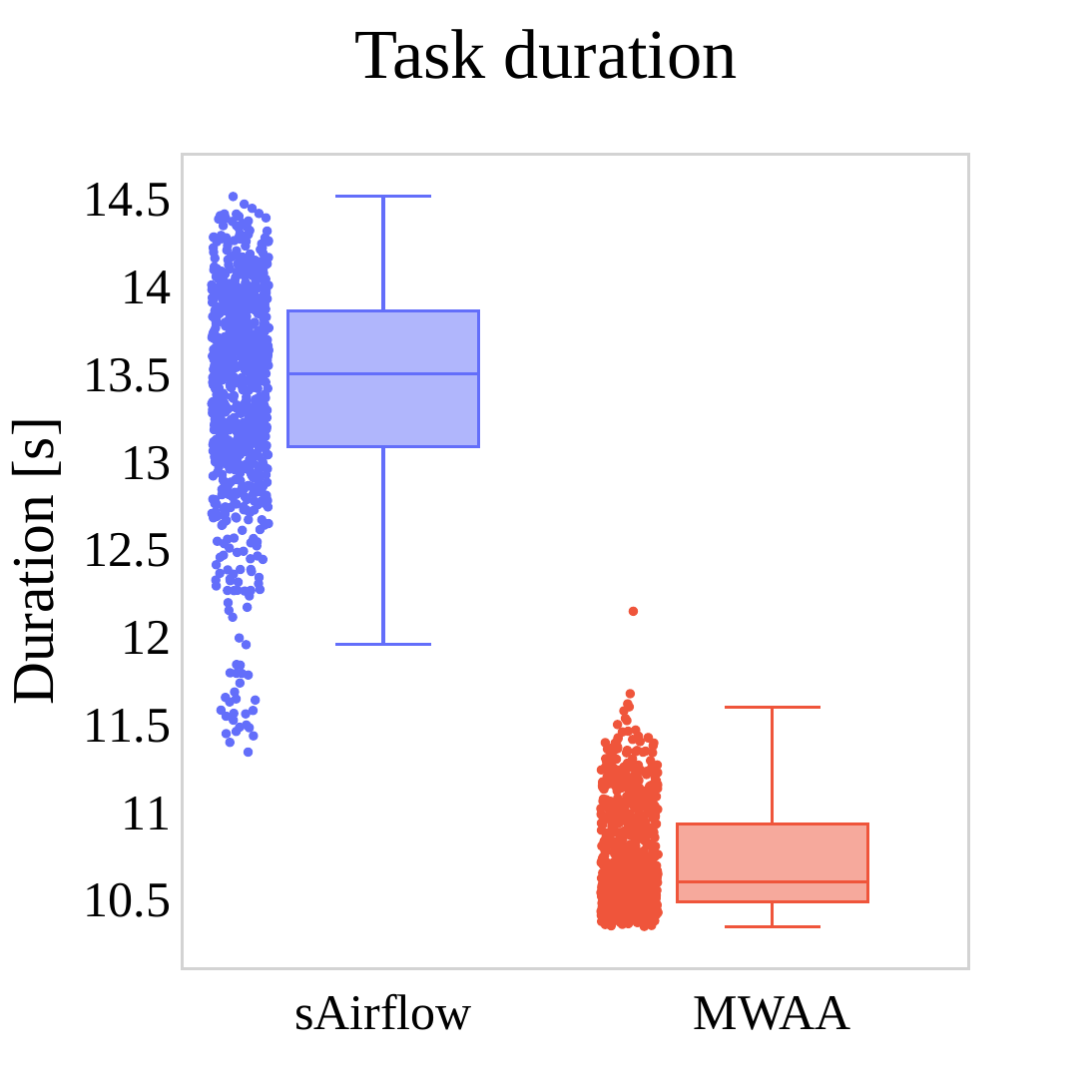}
    \includegraphics[width=0.3\textwidth]{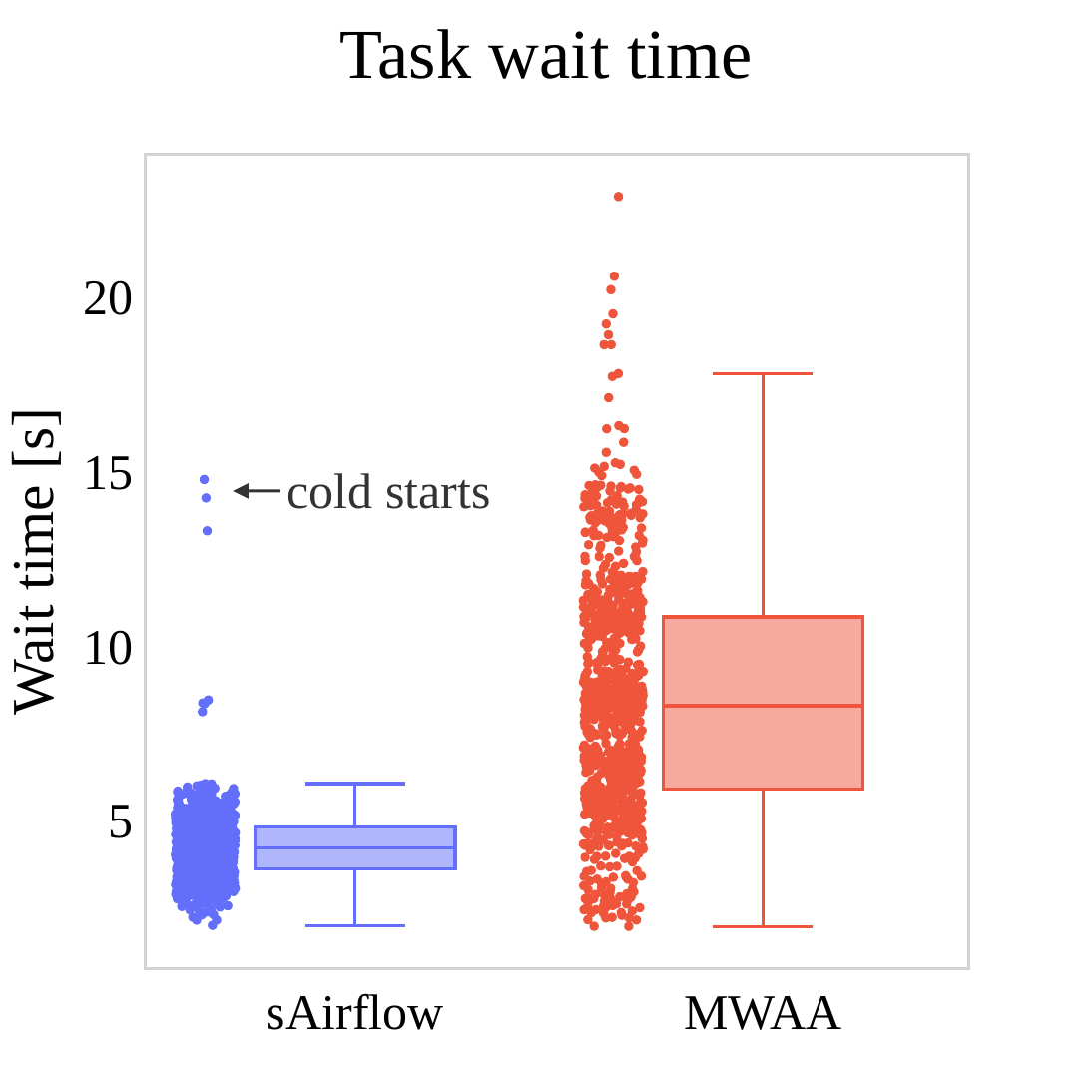}
  }%
  \\
  \subfloat[$n=125$]{
    \includegraphics[width=0.3\textwidth]{DAG_makespan-125task_parallel.pdf}
    \includegraphics[width=0.3\textwidth]{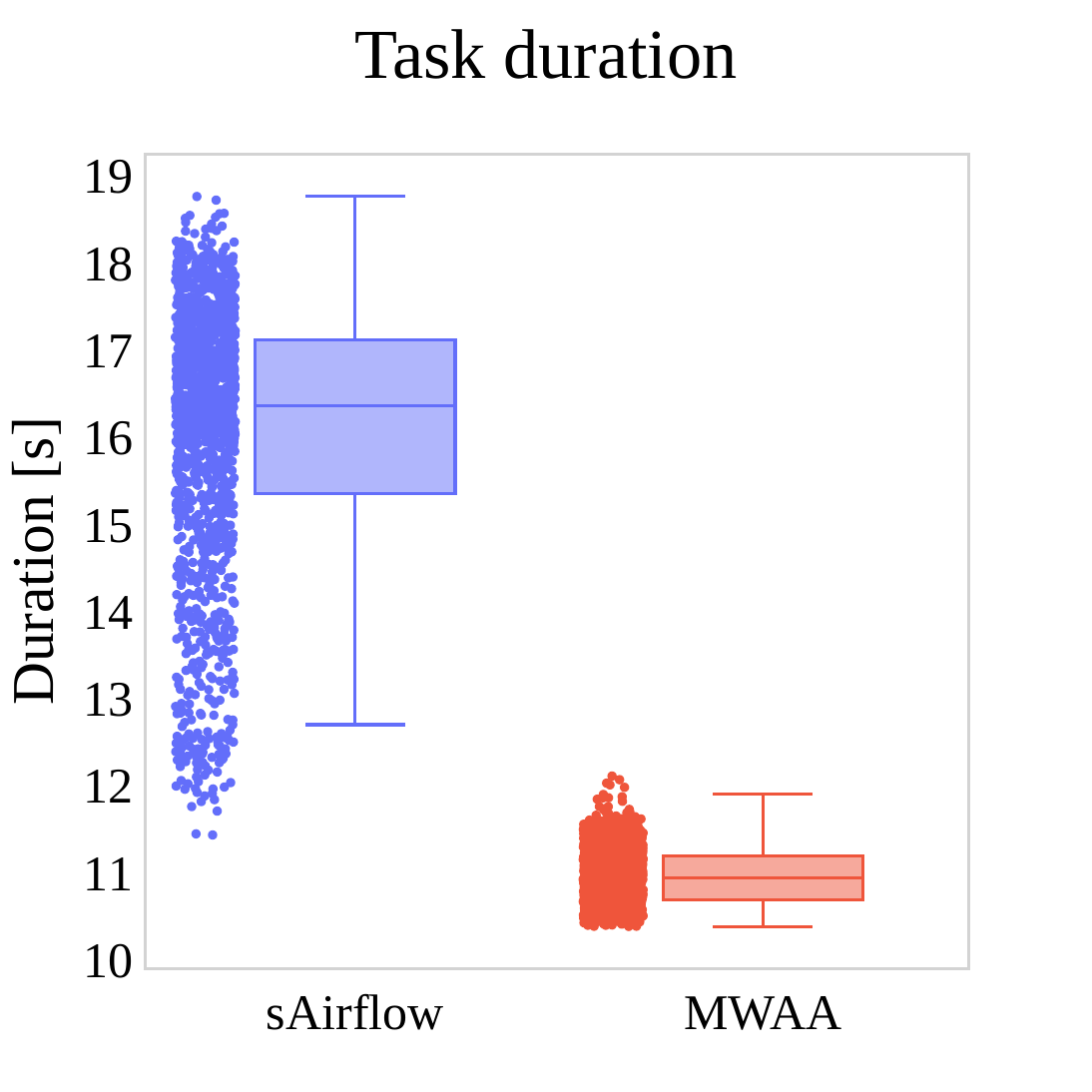}
    \includegraphics[width=0.3\textwidth]{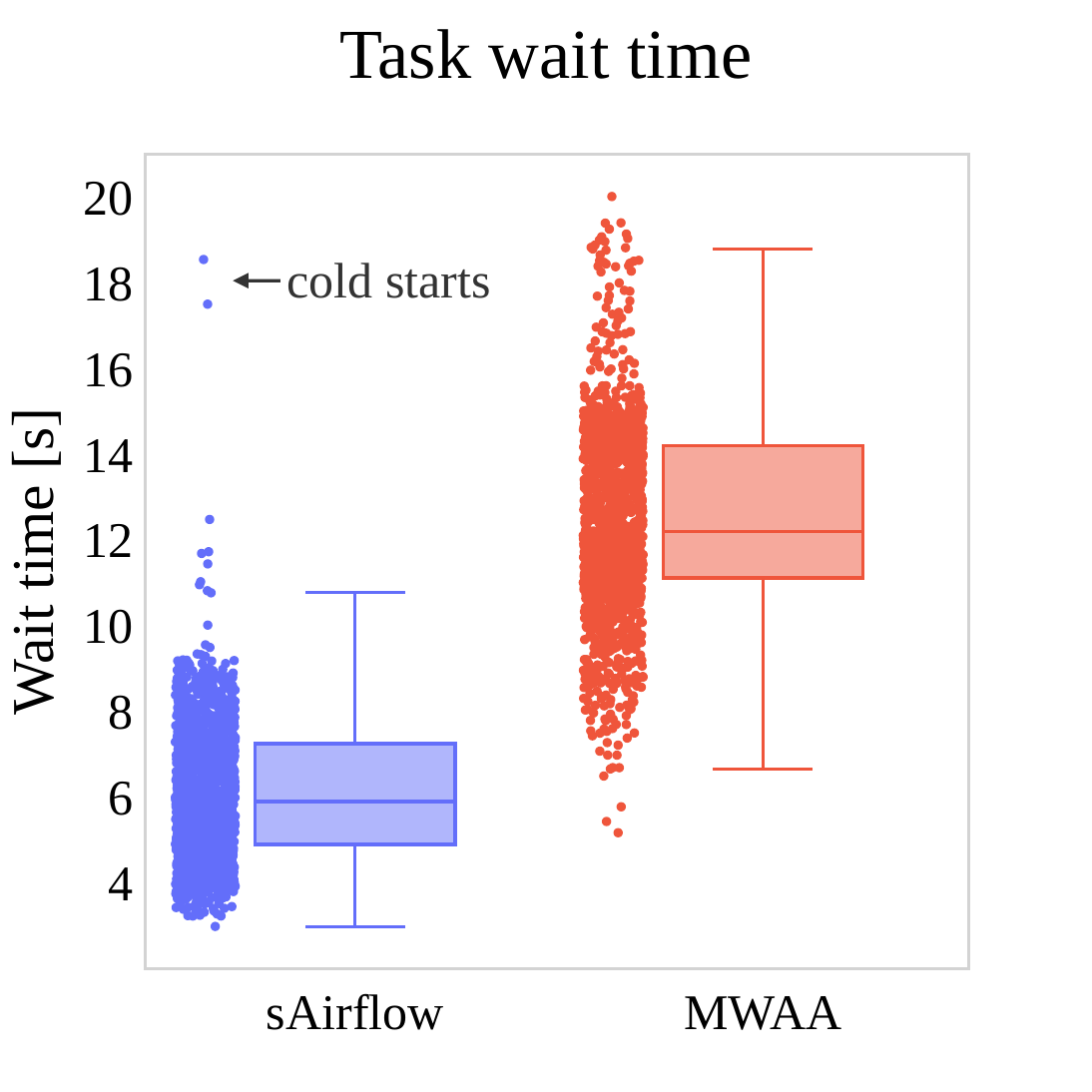}
  }
  \caption{Parallel DAGs, function executor, warm starts, $p=10$, $T=5$.
  Gantt charts on the right side correspond to one of the DAG runs.
  }\label{fig:parallel_warm}
\end{figure}

For transparency, we report the full results from the following experiments:
\begin{itemize}
  \item Chain DAGs with $n=1$, $n=5$ and $n=10$, Fig~\ref{fig:chain_warm}.
  \item Parallel DAGs with $n=16$, $n=32$, $n=64$ and $n=125$, Fig~\ref{fig:parallel_warm}.
\end{itemize}
We refer to the main text for the analysis of these results.

\newpage
\section{Parallel Forest: Both systems perform similarly on multiple DAGs} 

To show that sAirflow works equally well for multiple DAG, we employ an experiment where multiple copies of the same DAG run in parallel. The DAG has the same parameters no matter how many copies of it are run. We use Parallel Forest DAGs with $n=8$, $p=10$, and $k \in \{1, 2, 4, 8\}$. We compare sAirflows overall performance (the first execution of each of the DAGs will warm the system) to MWAA's warm executions.

\begin{figure}[htbp]
\centering
    \includegraphics[width=0.32\textwidth]{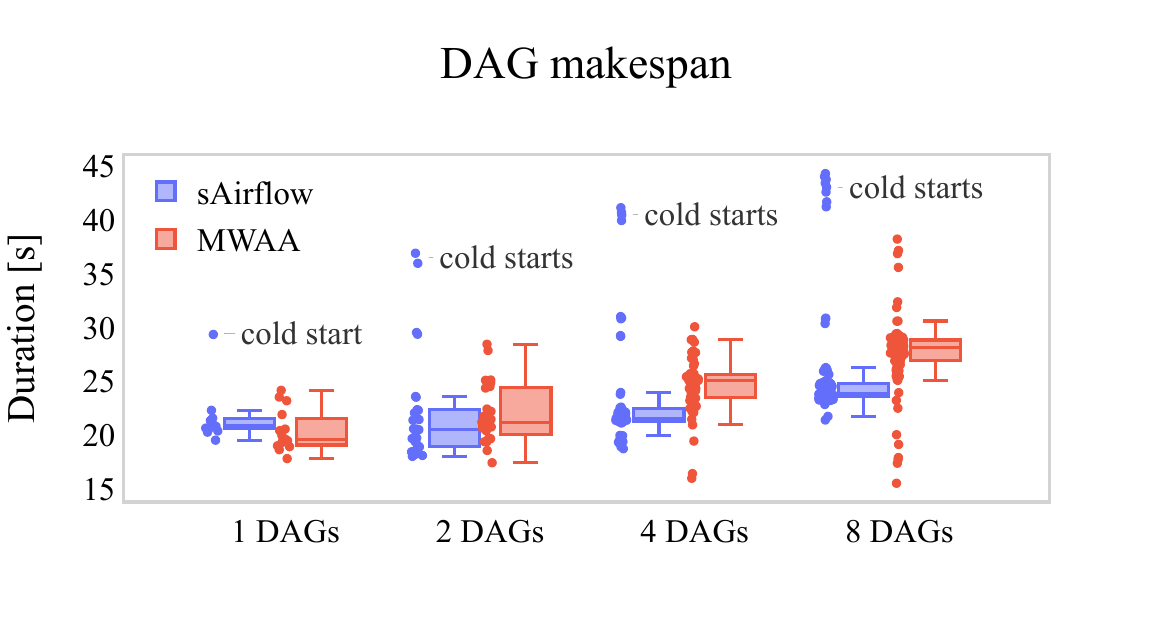}
  \includegraphics[width=0.32\textwidth]{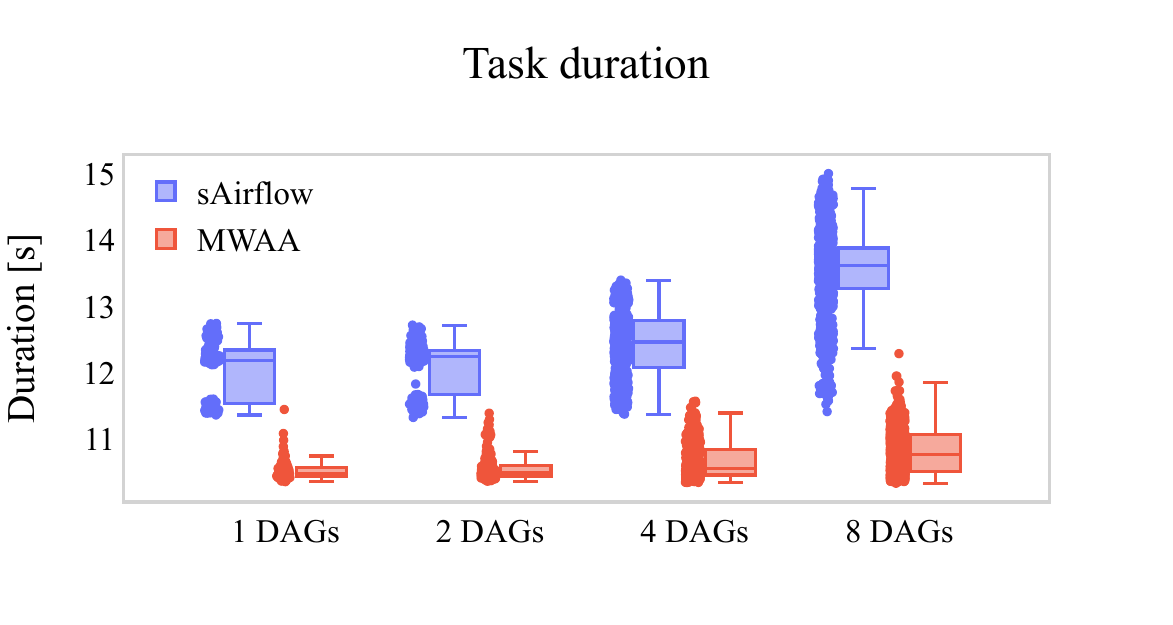}
  \includegraphics[width=0.32\textwidth]{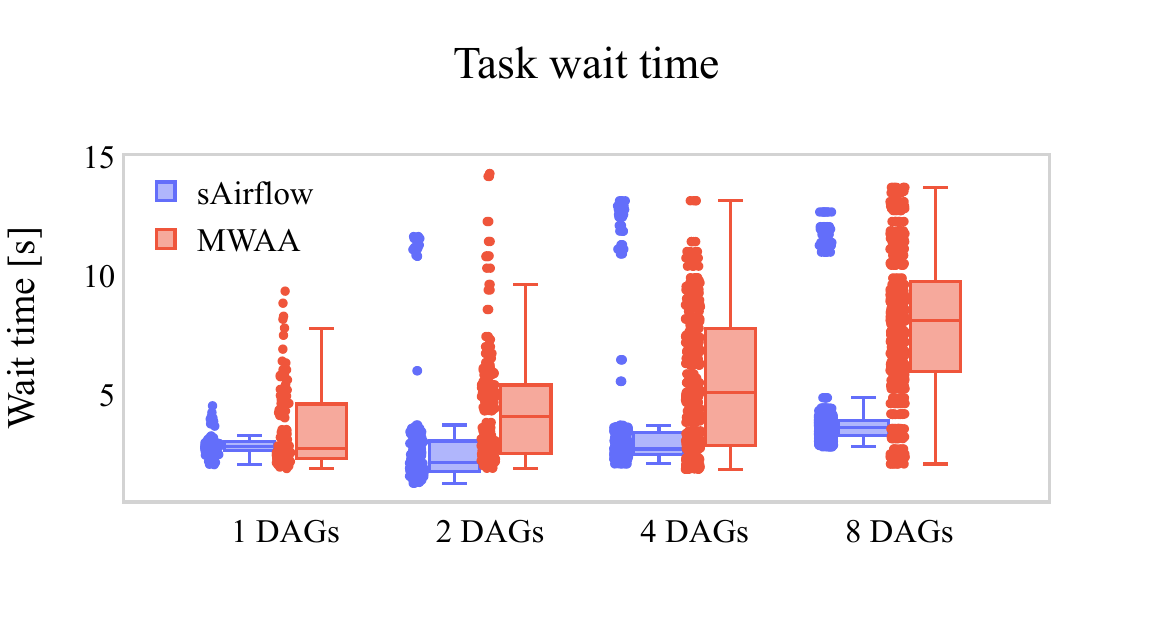}
\caption{Parallel forest DAGs, $n=8$, $p=10$, $T=5$, comparison of the system where the same copies of the DAGs are run in parallel (for $k \in \{1, 2, 4, 8\}$).}
\label{fig:8t_forest}
\end{figure}

The resulting metrics (Figure \ref{fig:8t_forest}) show that the trend across systems is similar. Both systems are almost equally affected by running more copies of the same DAG. As sAirflow is notably better at parallelizing and minimizing the task wait time, the median DAG makespan is not affected as much throughout the experiment. For $k=1$ sAirflow and MWAA yield 20.90 seconds and 19.60 seconds respectively, for $k=8$ it is 28.16 seconds and 23.87 seconds respectively. 

\begin{figure}[htbp]
  \centering
    \includegraphics[width=0.32\textwidth]{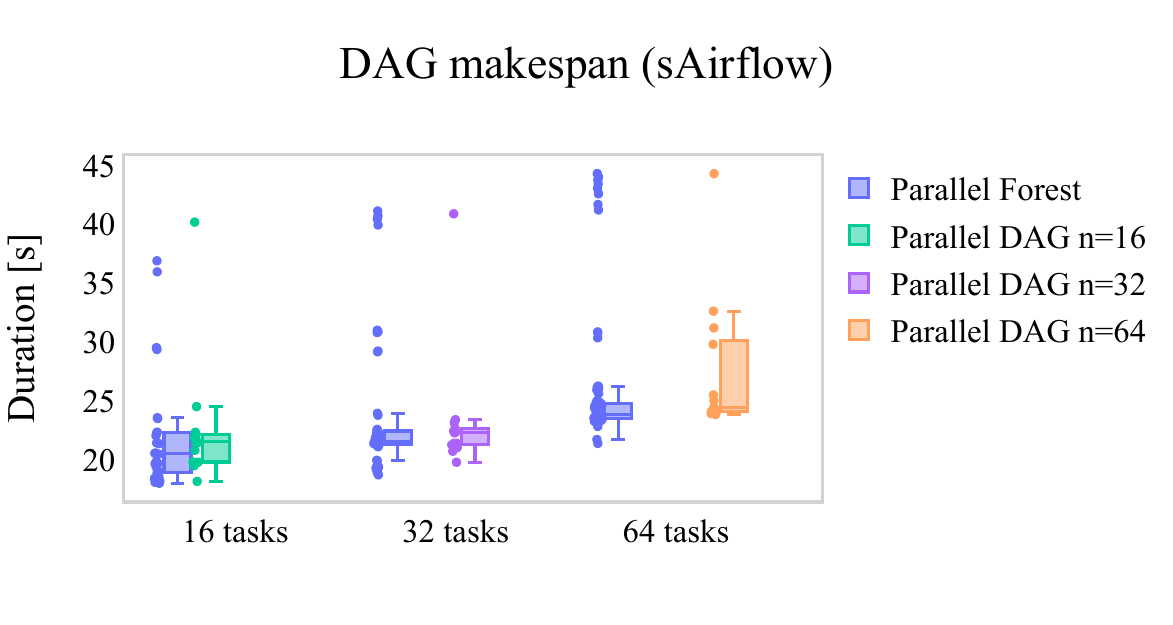}
    \includegraphics[width=0.32\textwidth]{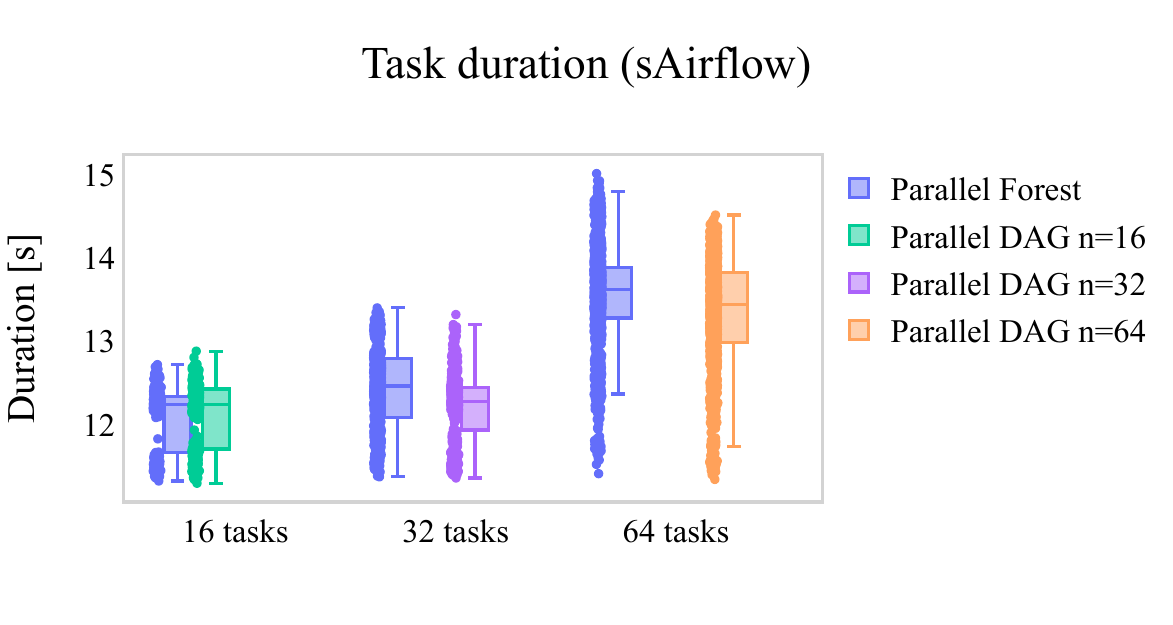}
    \includegraphics[width=0.32\textwidth]{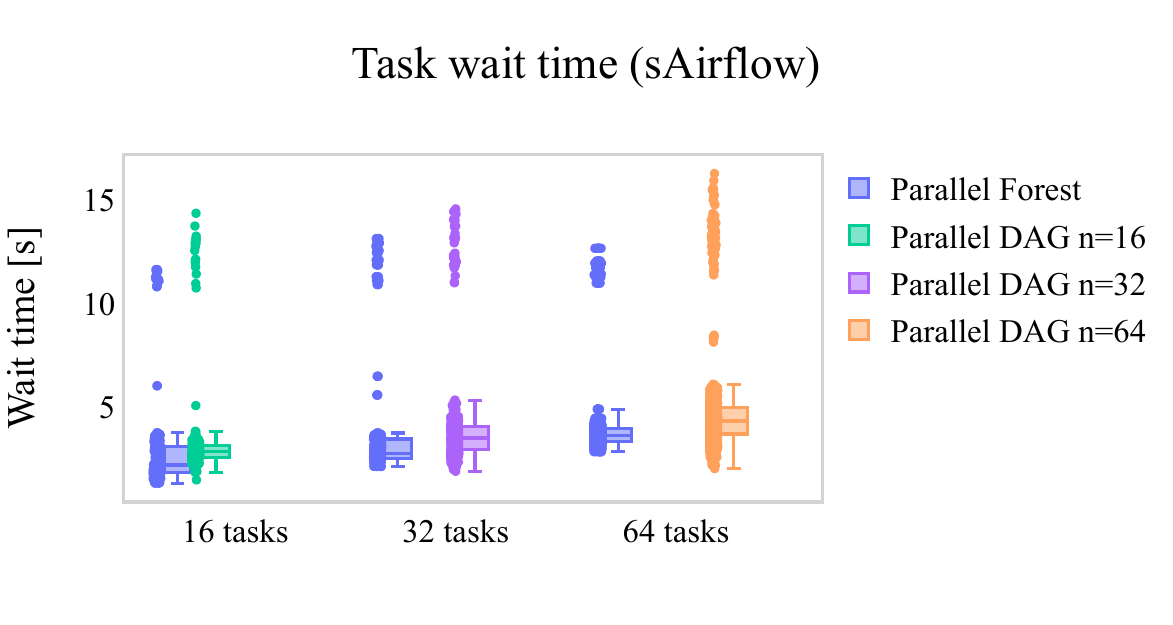}
  \caption{Comparison between experiments on sAirflow for Parallel DAGs and Parallel Forest grouped by the total number of tasks in the DAGs. For instance, for forest with 8 DAGs there were $8*8$ tasks in total, which corresponds to one DAG with $n=64$ tasks. }
  \label{fig:8t_forest_sairflow}
\end{figure}

As the DAG is highly parallel sAirflow faces the same challenges and wins as in the Parallel DAGs experiment. For instance, with $k=8$ there were in total $8*8$ tasks being run across the DAGs. The correlated metrics are close (Figure \ref{fig:8t_forest_sairflow}). Thus, sAirflow works equally well when the workload is split into multiple DAGs. 

\newpage
\section{Function Executor, Alibaba DAGs}
We use the DAGs generated from the Alibaba Cloud traces to show that performance on realistic, industrial workloads confirms the trends observed in earlier experiments --- with sAirflow performing better on more parallel DAGs, while MWAA on more sequential DAGs. In this analysis, we include the first cold-start execution for sAirflow.

In these instances, task durations in a DAG vary. 
The critical path might also be longer than 5~min., the interval we used earlier. Thus, we set $T=5$ for DAGs with a critical path less or equal to 200~s; and $T=10$ for the remaining DAGs. This prevents DAG runs from overlapping, at the risk of getting more cold starts in sAirflow (but not MWAA). 

\begin{figure}
  \centering
  \includegraphics[width=0.3\textwidth]{traces_dag_makespan_raw_comparison.pdf}
  \includegraphics[width=0.3\textwidth]{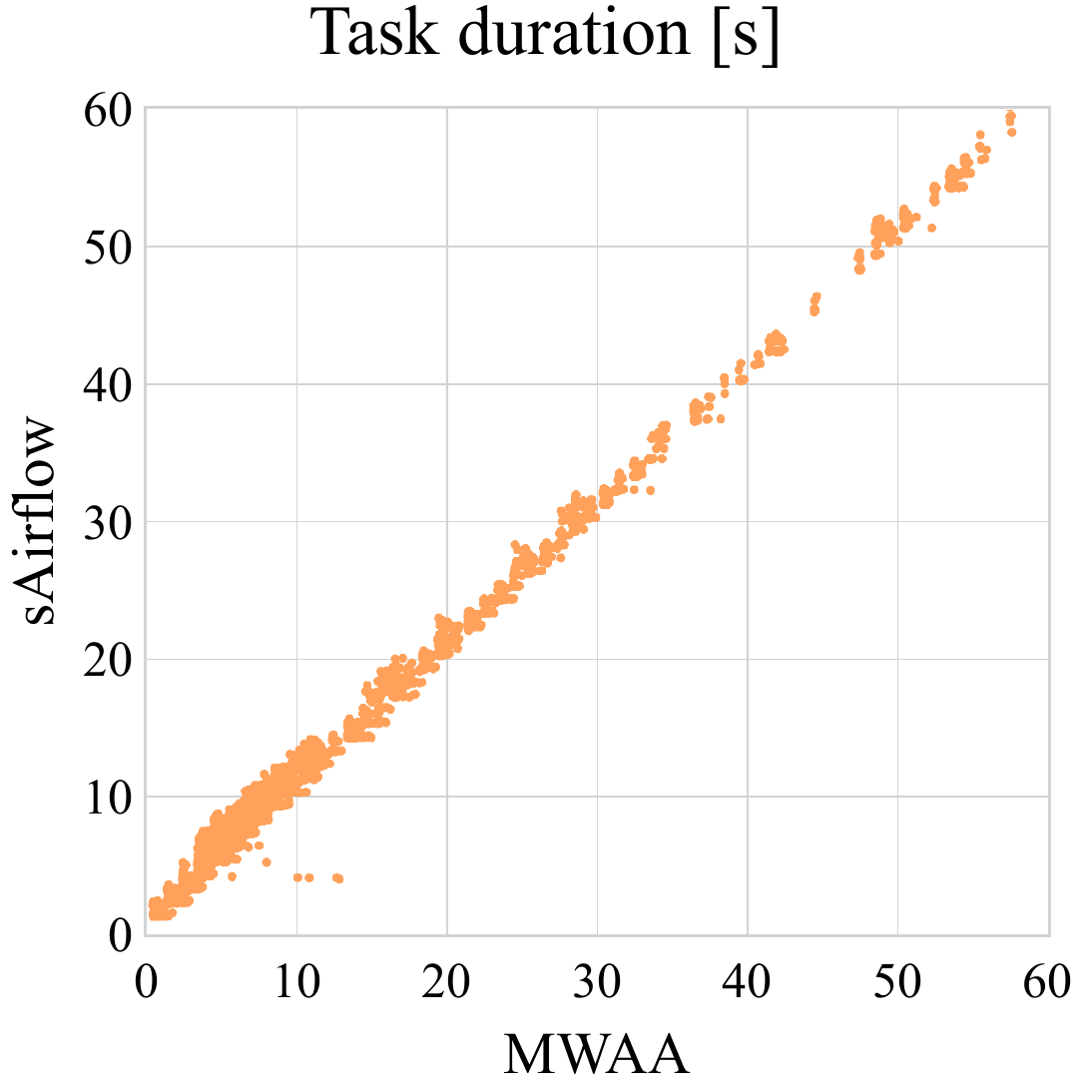}
  \includegraphics[width=0.3\textwidth]{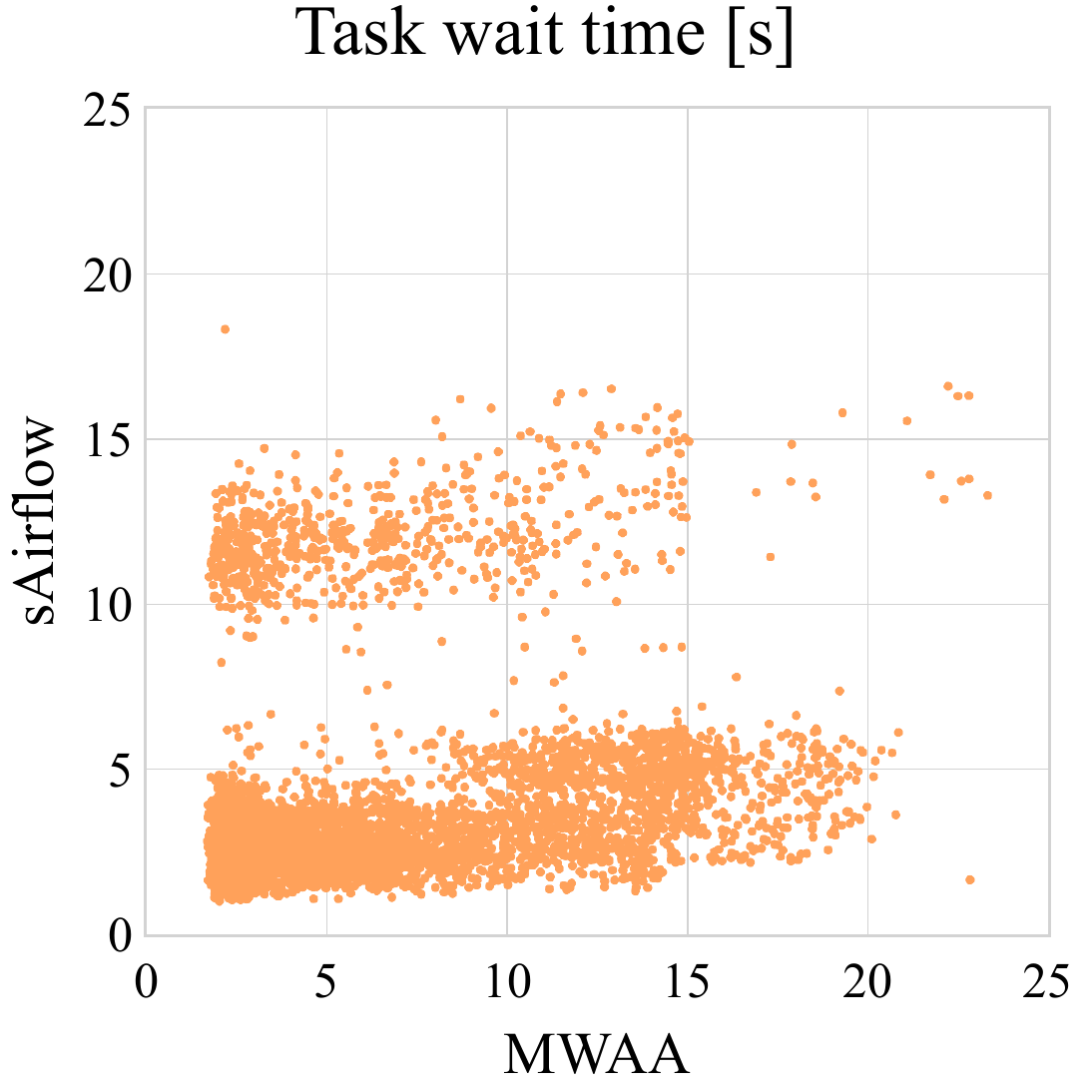}
  \caption{Metrics comparison for MWAA and sAirflow for DAGs generated from the Alibaba Cloud traces.}
  \label{fig:traces_dag_run_raw_comparison}
\end{figure}

\begin{figure}
\centering
  \subfloat[Absolute\label{fig:traces_dag_duration_overhead}]{\includegraphics[width=0.327\textwidth]{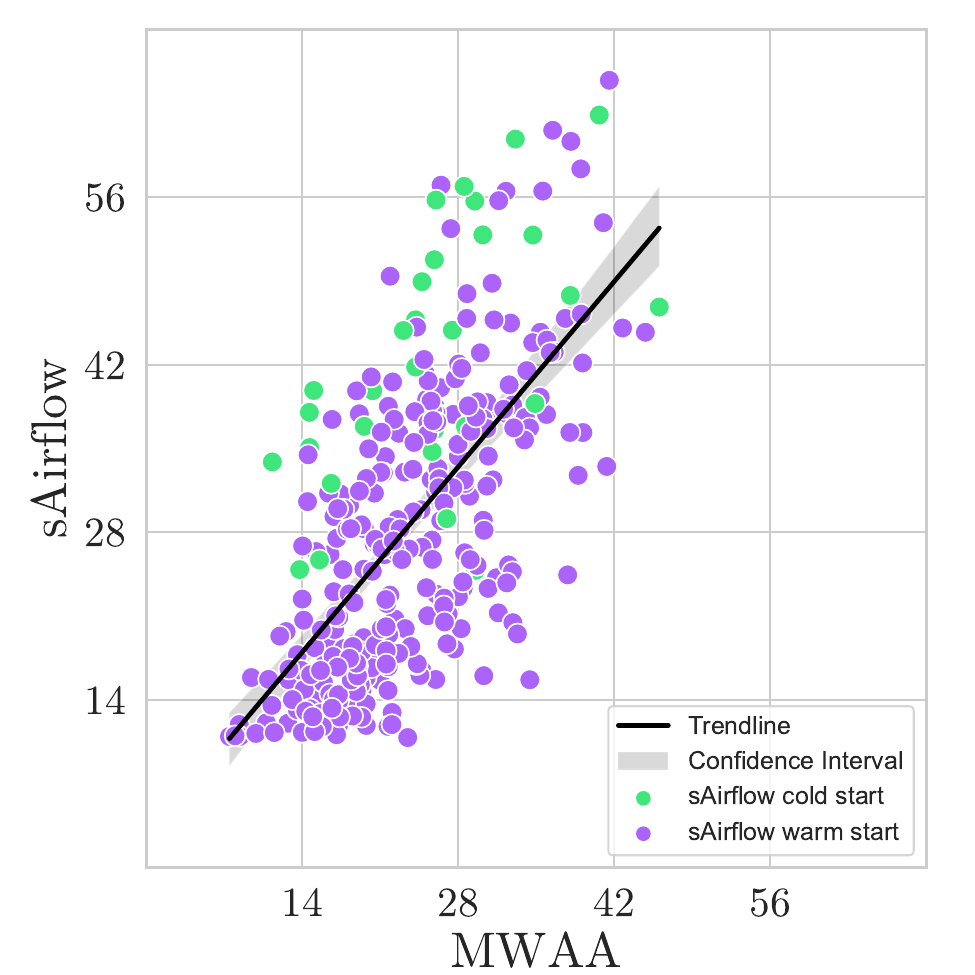}}%
  \subfloat[Per-task\label{fig:traces_dag_duration_overhead_by_tasks_num}]{\includegraphics[width=0.327\textwidth]{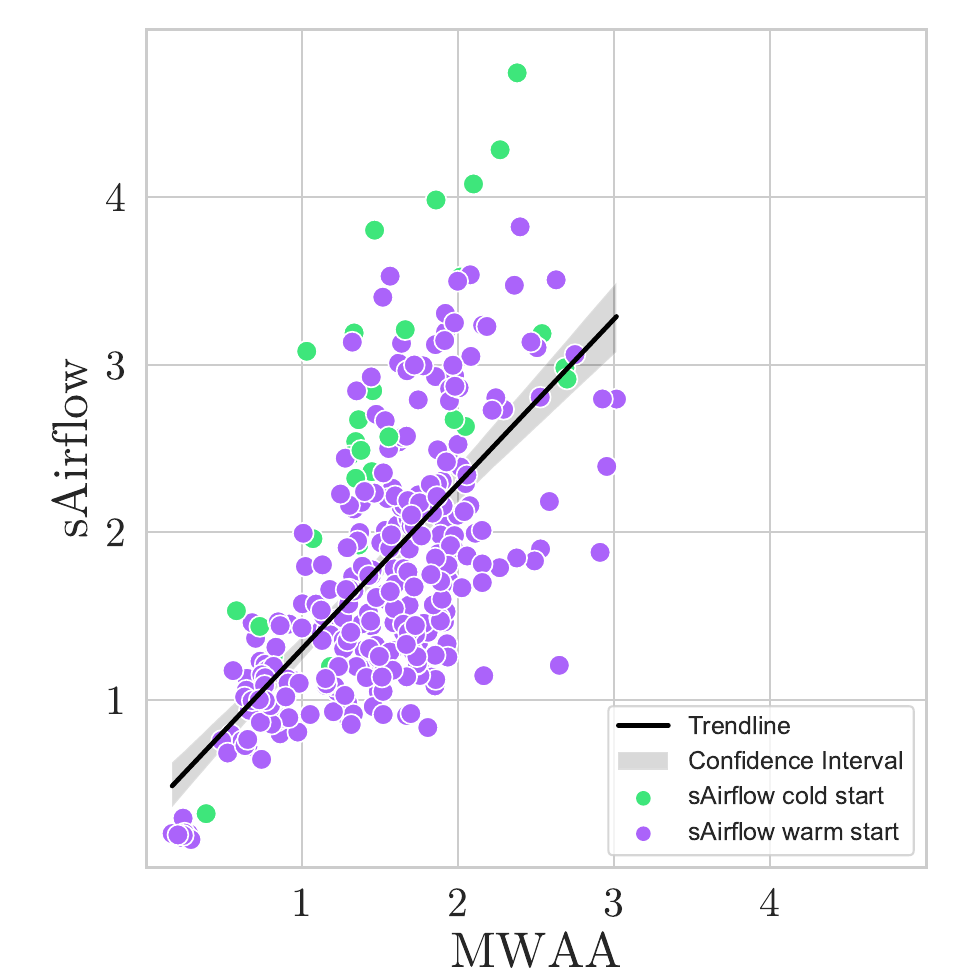}}%
  \subfloat[Normalized \label{fig:traces_dag_duration_overhead_by_parallel_longest}]{
    \includegraphics[width=0.327\textwidth]{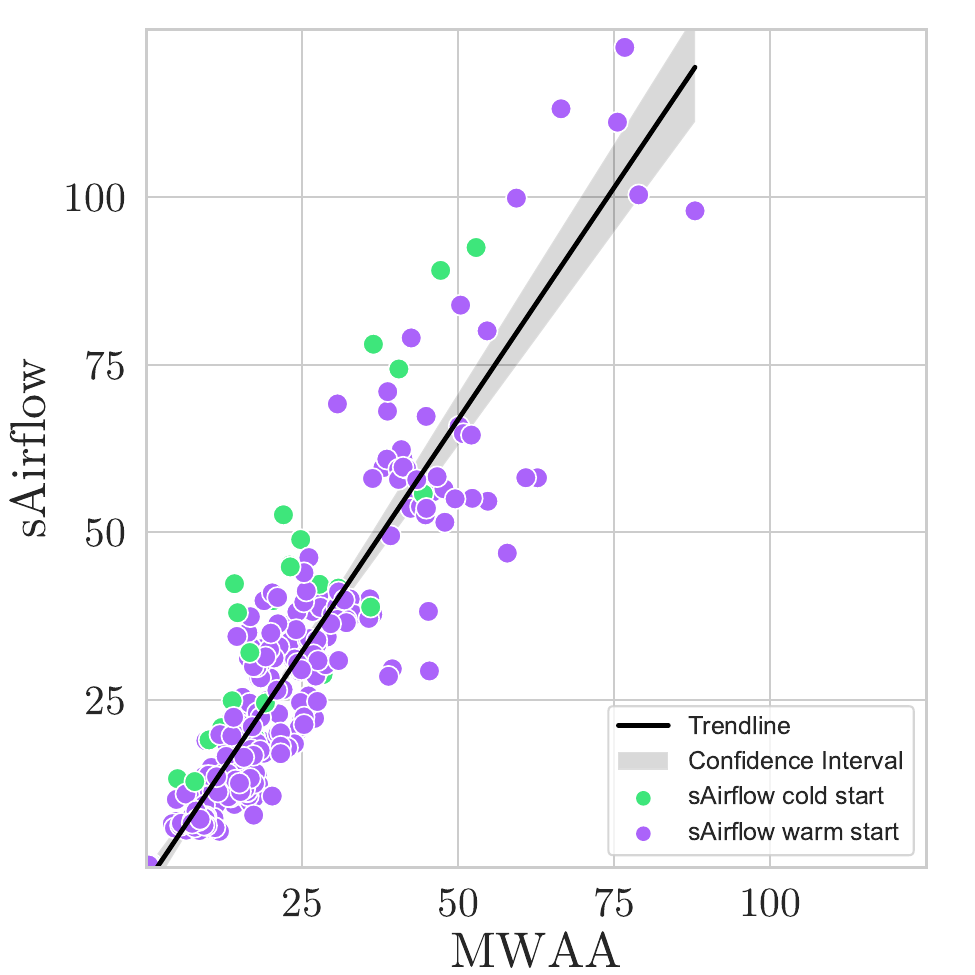}
  }
  \caption{Comparison of overhead for DAGs generated from Alibaba Cloud traces}
\end{figure}

\begin{figure}[tbp]
  \centering
  \includegraphics[width=1.0\textwidth]{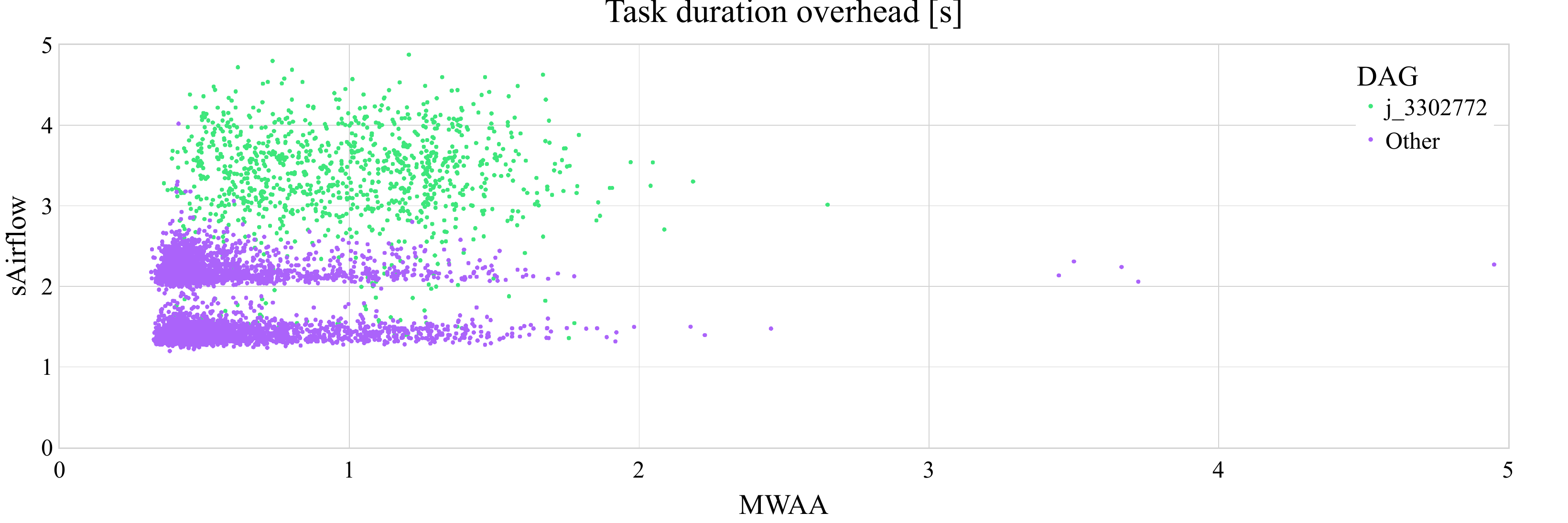}
  \caption{Comparison of task duration overheads for both systems. The overhead is calculated as the collected task execution time minus the task duration in the DAG. In an ideal system, the overhead for this metric is zero. }
  \label{fig:traces_duration_overhead}
\end{figure}

Both systems yield similar performance (Fig.~\ref{fig:traces_dag_run_raw_comparison}) concerning the DAG makespan. The difference between the DAG's critical path and the makespan (the overhead) is also similar (Fig.~\ref{fig:traces_dag_duration_overhead}). sAirflow's overhead is roughly 10\% higher (Fig.~\ref{fig:traces_dag_duration_overhead}) in comparison with MWAA, which we attribute to the longer task durations (Fig.~\ref{fig:traces_duration_overhead}). The DAG overhead metric averaged over all tasks in a DAG confirms this (Fig.~\ref{fig:traces_dag_duration_overhead_by_tasks_num}).

To better describe the performance, we normalize the DAG's overhead by a ratio between the DAG maximum parallelism and the longest path (number of nodes) in the DAG.
Assume the following notation, where $D$ is the DAG run of graph $d$: 
\begin{itemize}
    \item $C_{\max}(D)$: the DAG makespan: 
    \item $p_{d}$: the critical path duration, equal to \\
    $ \sum_{\textrm{i: task i } \in \textrm{ critical path of } d} p_i $
    \item $n_W$: the maximum parallelism - the maximum number of tasks that would run in parallel if the DAG is run on a system without any overhead and unlimited resources;
    \item $n_L$: the number of nodes on the longest path;
\end{itemize}
We normalize the performance by:
\begin{equation} \label{eq:overhead_normalized_custom}
\left( C_{\max} - p_{d} \right) * \left( \frac{n_L}{n_W} \right)
\end{equation}

\begin{figure}[t]
  \centering
  \begin{subfigure}[t]{0.49\linewidth}
      \includegraphics[width=1.0\textwidth]{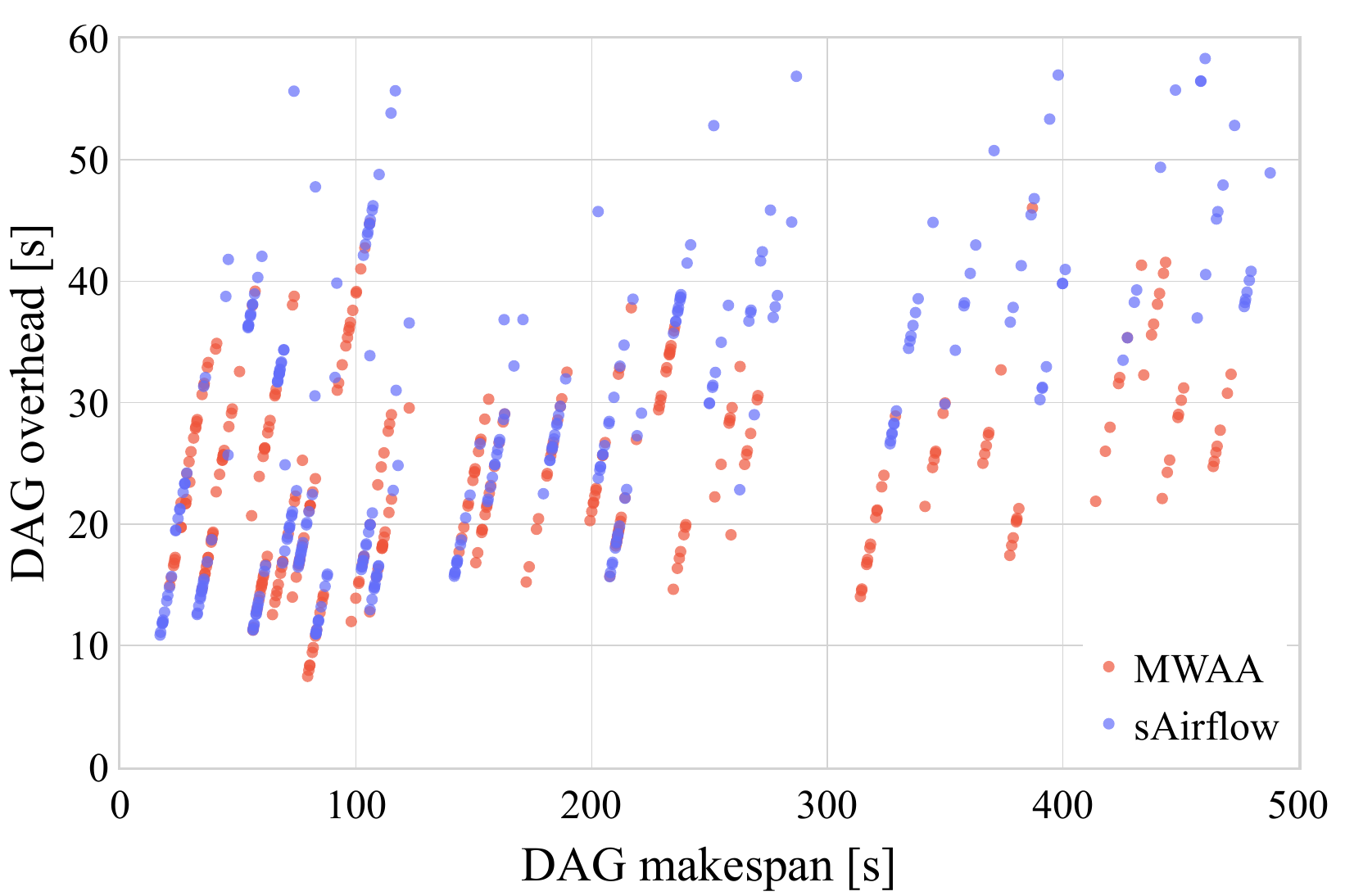}
      \caption{DAG overhead distribution.}
  \end{subfigure}
  \begin{subfigure}[t]{0.49\linewidth}
      \includegraphics[width=1.0\textwidth]{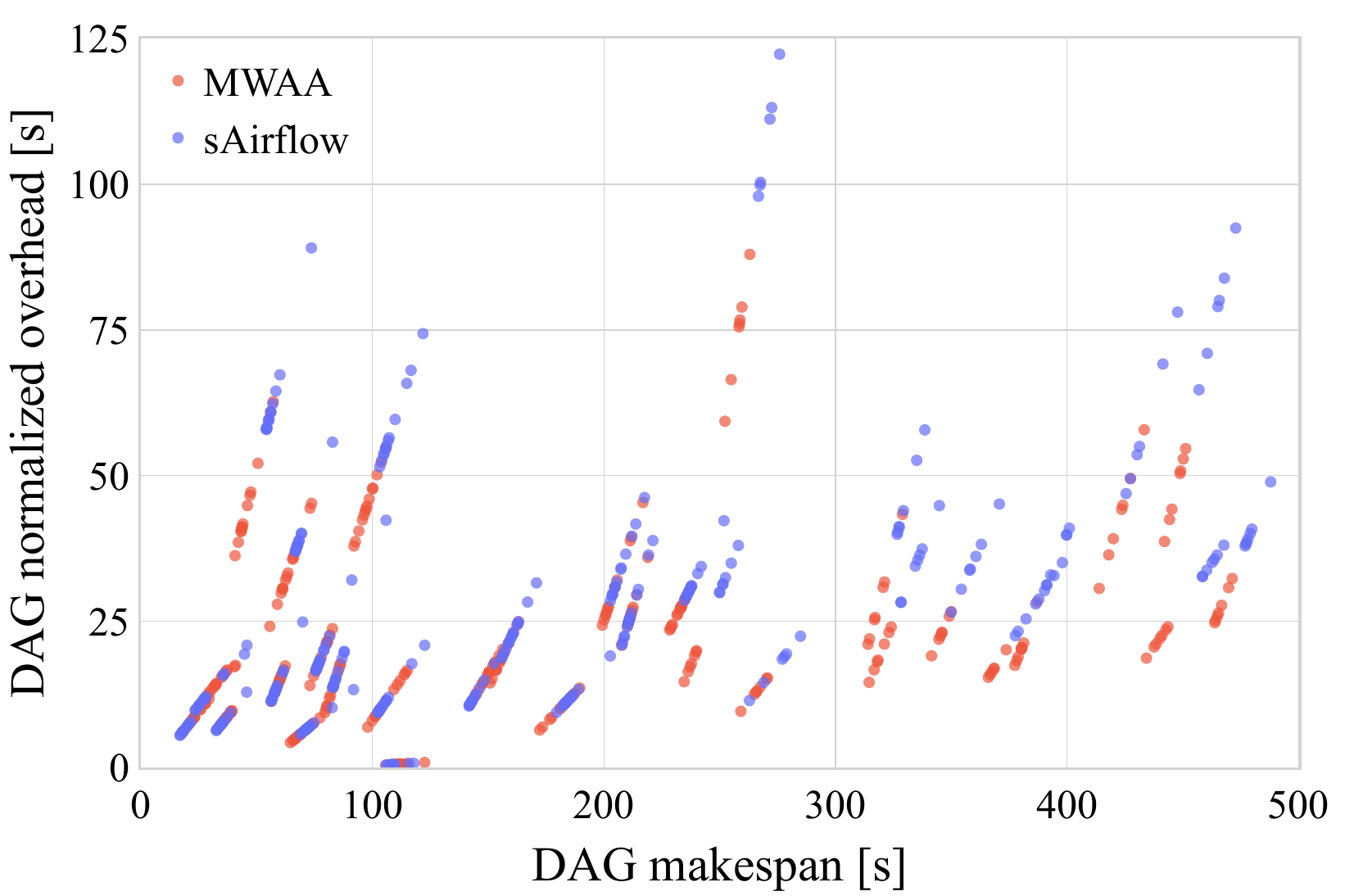}
      \caption{y-axis represents Equation \ref{eq:overhead_normalized_custom}.}
  \end{subfigure}
  \\
  \caption{DAG overhead metric on DAGs generated from Alibaba Cloud traces for each system, with different normalization methods applied.}
  \label{fig:dag_overhead_distribution}  
\end{figure}

\noindent The first component, $\left( C_{\max} - p_{d} \right)$ captures the system overhead on a DAG, whereas the second,$\left( \frac{n_L}{n_W} \right)$ represents the parallelizability of a DAG.
MWAA has less overhead on linear DAGs, and sAirflow performs better on more parallelizable DAGs. The metric aims to describe this
correlation backed by the distribution of the values (Fig.~\ref{fig:dag_overhead_distribution}). 

\newpage
\section{Container executor}
We now measure sAirflow using container workers (using AWS Batch with AWS Fargate). The task overhead depends on the number of tasks submitted in parallel and the size of the underlying Docker container image for the worker. \cite{9582324} comparises costs and performance of FaaS and CaaS on AWS; here, we confirm that similar patterns exist when these executors are exposed through sAirflow.

For all experiments, sAirflow requests 0.5 vCPU and 512MB of memory from AWS Fargate, the lowest configuration available. The typical latency for Batch with Fargate is 60--90~s of provisioning time and then 30~s of the start-up time (these numbers confirm measurements  
reported in \cite{airflow_native_aws_executors}). 
With the container executor, containers are not reused (no warm starts, in sharp contrast to the FaaS executor). Each time a container starts, its content is pulled from the registry (AWS ECR).

\begin{figure}[t]
\centering
\begin{minipage}[t]{1.0\linewidth}
    \centering
    \includegraphics[width=0.3\textwidth]{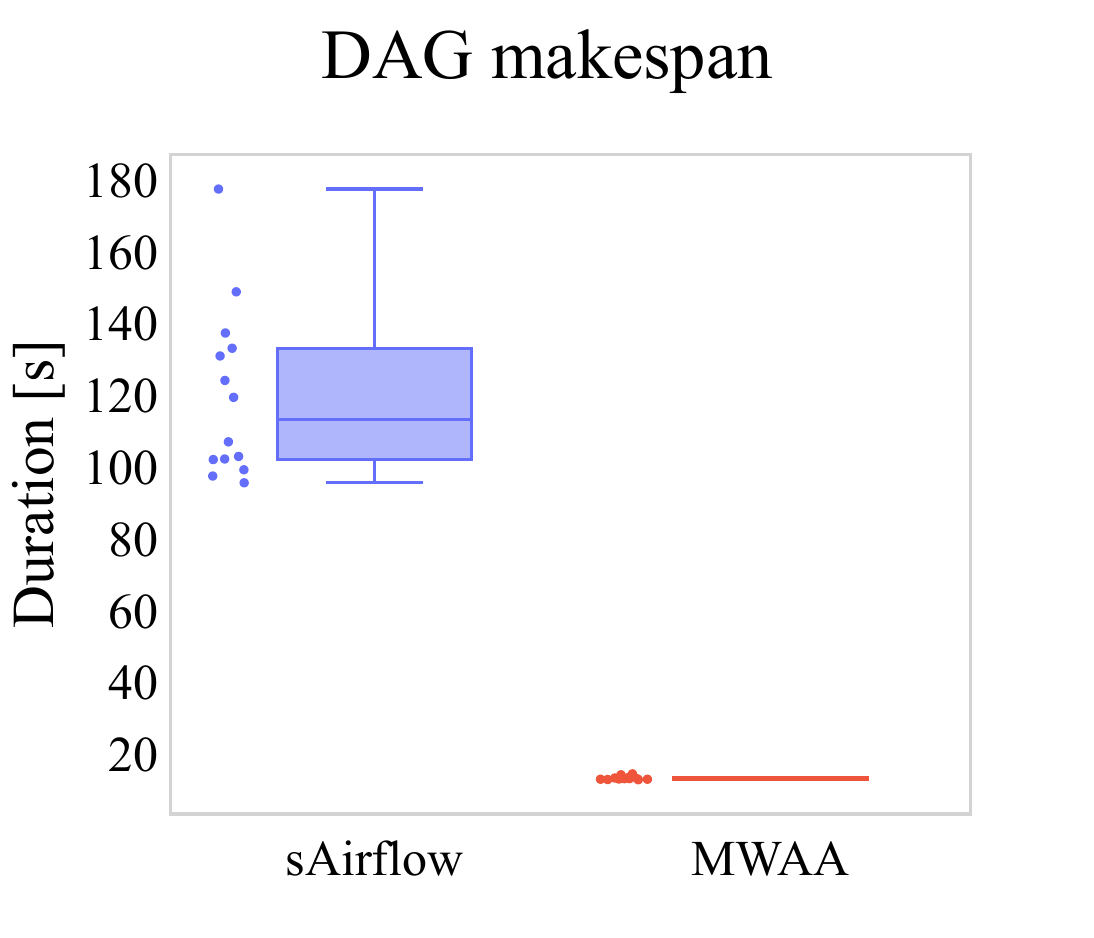}
    \includegraphics[width=0.3\textwidth]{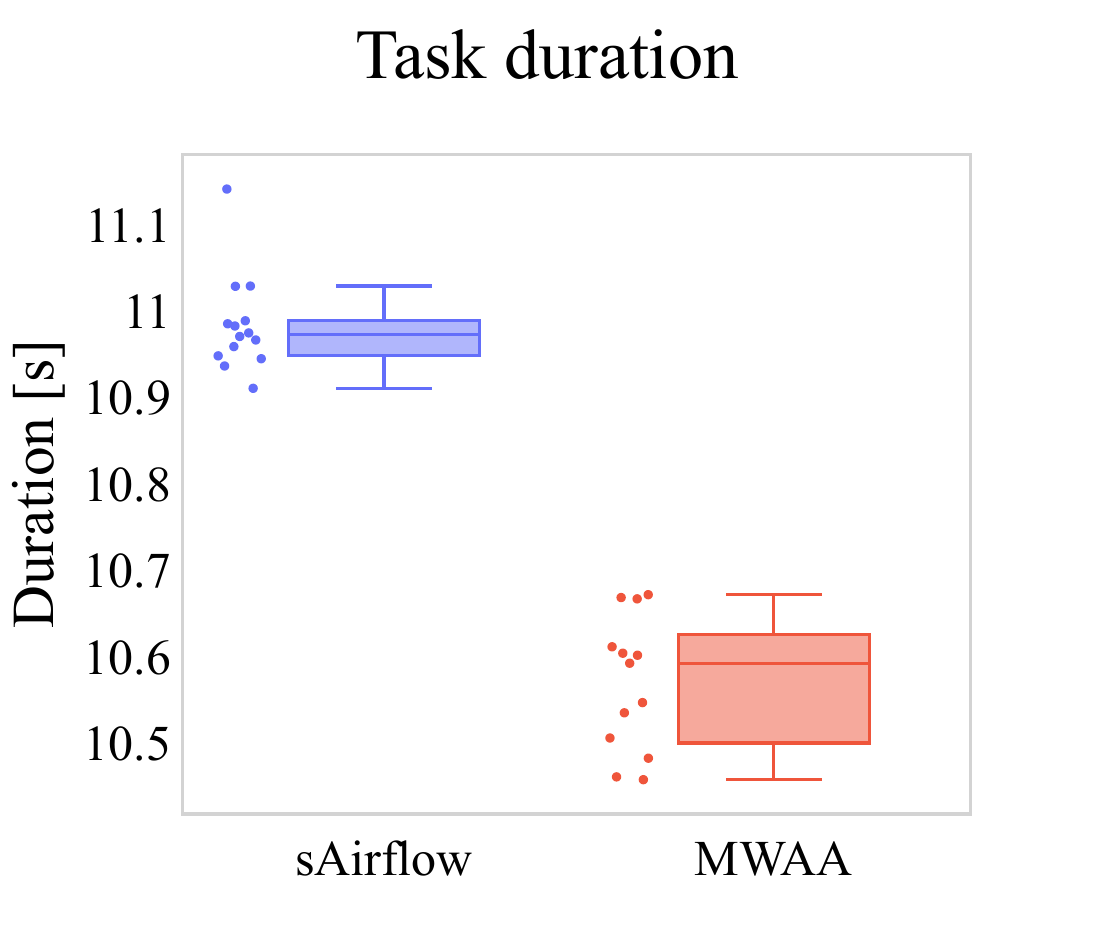}
    \includegraphics[width=0.3\textwidth]{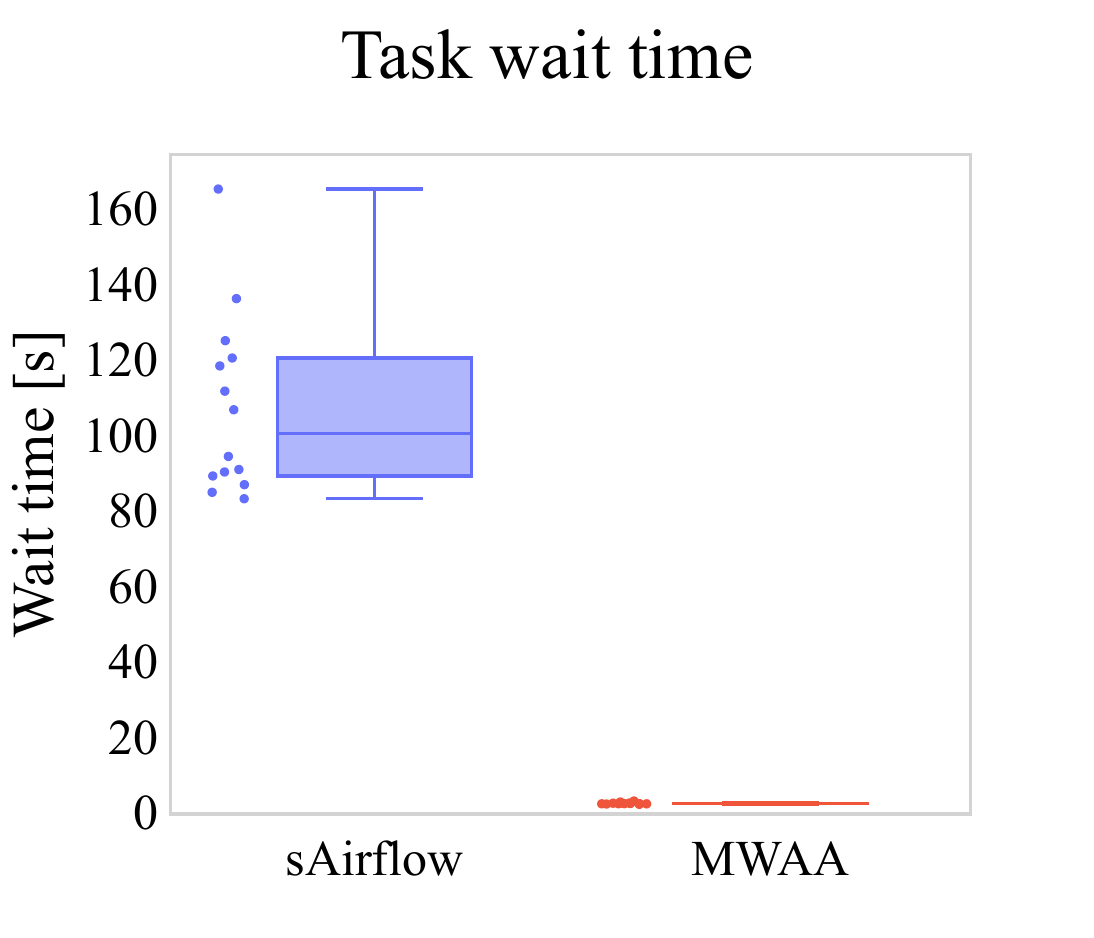}
    \caption{Chain DAG, $n=1$, $p=10$, $T=5$, sAirflow on CaaS. MWAA keeps at least one worker running, while sAirflow needs to create a new cold container for each task.}
    \label{fig:1task_linear_batch}
\end{minipage}
\end{figure}

\subsection{Chain DAGs: sAirflow has more overhead on a single task DAG for CaaS executor} 

This experiment shows the overhead that AWS Batch adds over a single task DAG (a chain DAG with $n=1$, $p=10$, $T=5$) on sAirflow with CaaS executor and compare the results with the previous setup, comparing sAirflow with FaaS executor to MWAA (Fig.~\ref{fig:1task_line}).

We start with an experiment on a single-task DAG (a chain DAG with $n=1$, $p=10$, $T=5$), Fig.~\ref{fig:1task_linear_batch}). As expected, the task wait time is higher in sAirflow, as MWAA has one active worker by default (which cannot be removed). 
In sAirflow, replacing AWS Lambda (the function executor) with AWS Batch (the container executor) results in the median wait time increasing from 2.5~s (Fig.~\ref{fig:1task_line_cold}) to 100.5~s. This follows from the AWS Fargate provisioning time and the container start-up time (e.g., loading the dependencies).
As we rely on Apache Airflow, all its dependencies, and other libraries required by sAirflow. This image needs to be propagated on each start up by AWS Fargate.
The expected latency on a single task is up to 2.5 minutes but in the end, this number might vary depending on the queuing in AWS Batch \cite{aws-batch-fargate}.
Yet, the task duration time is almost 1~s shorter (Fig.~\ref{fig:1task_linear_batch}, \ref{fig:1task_line}), which follows from the fact that the minimum configuration for AWS Fargate allocated more resources than using the AWS Lambda (the defaults for sAirflow specify 0.2vCPU and 400MB of memory for the function).

\begin{figure}[t]
  \centering
  \subfloat[$n=16$\label{fig:16task_parallel_batch}]{
    \includegraphics[width=0.19\textwidth]{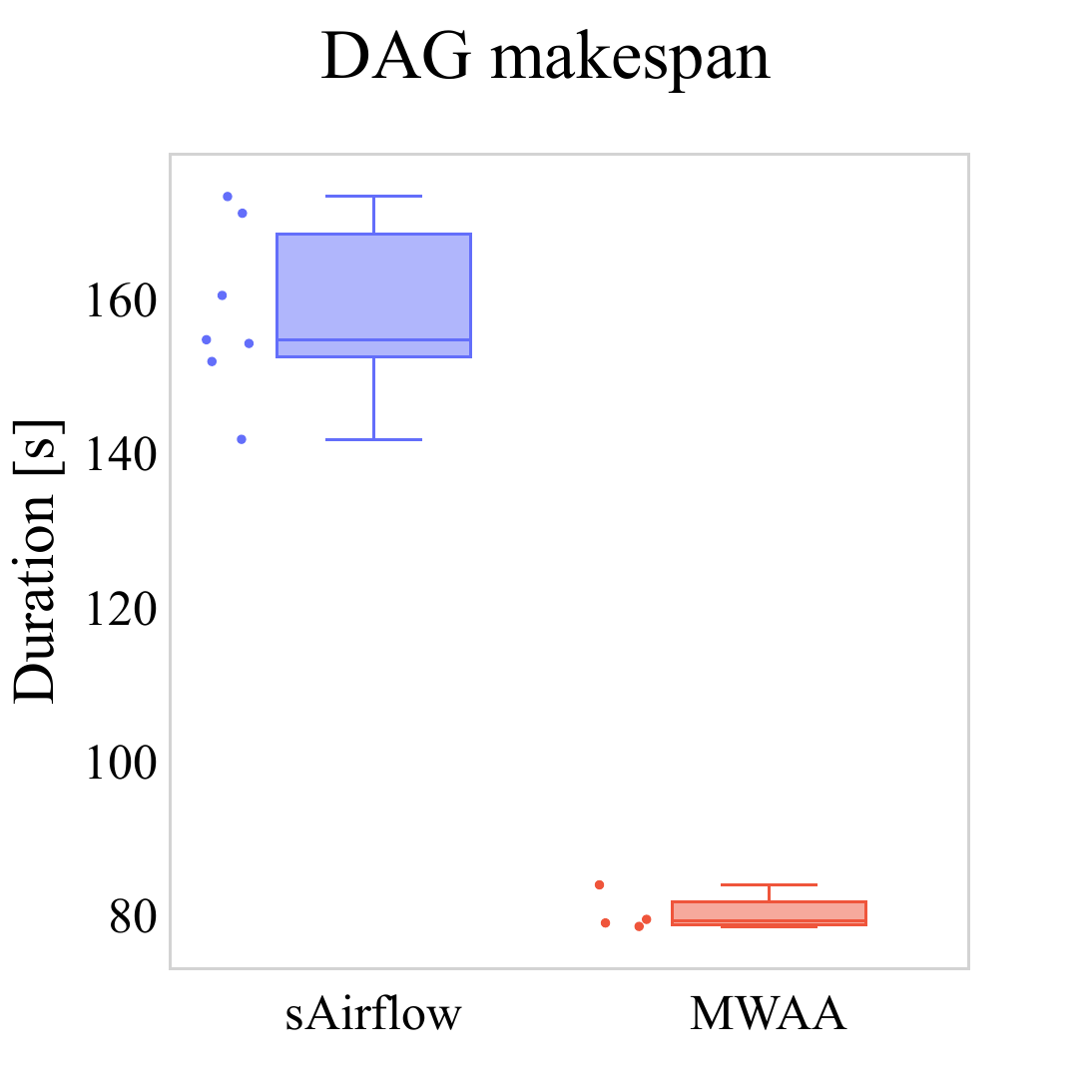}
    \includegraphics[width=0.19\textwidth]{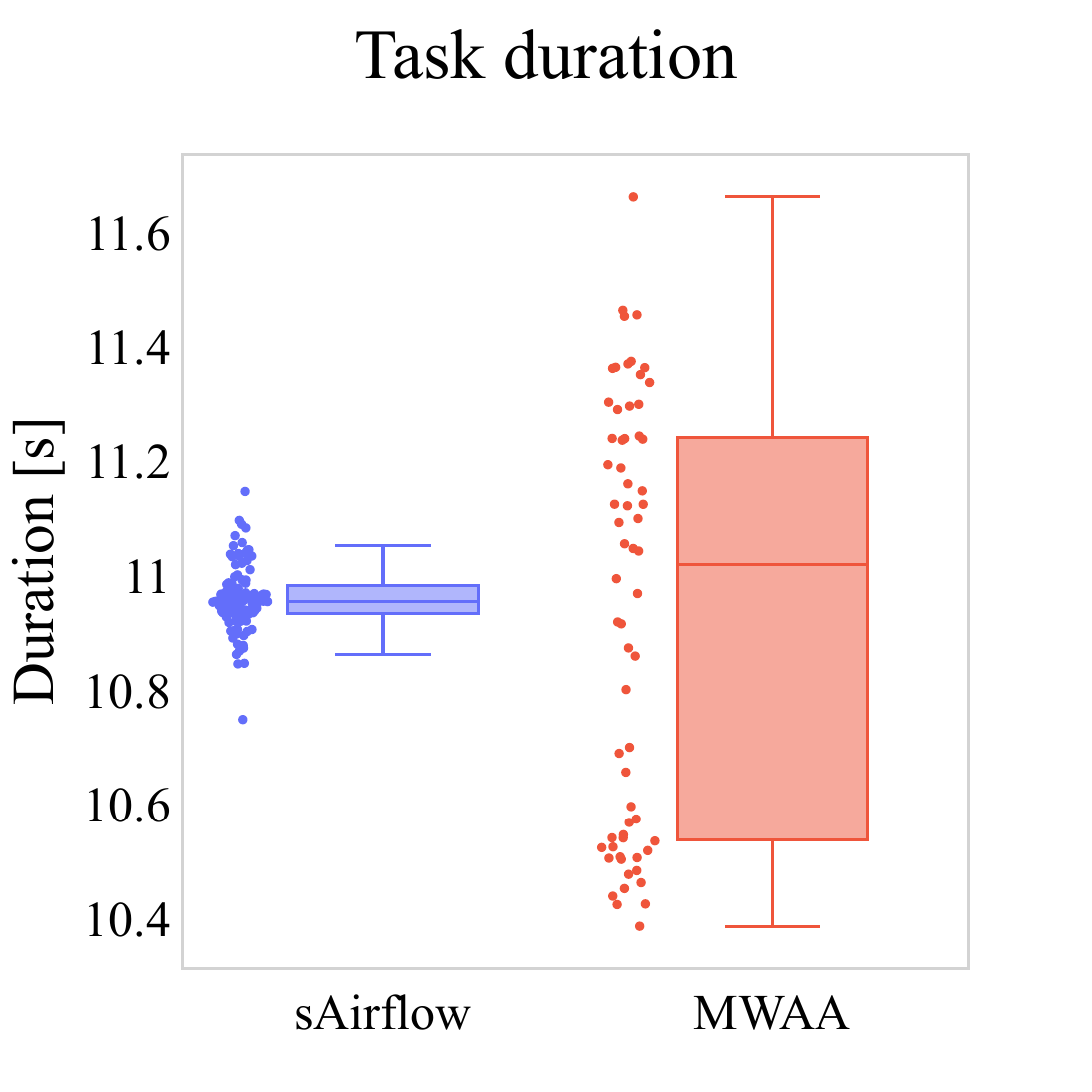}
    \includegraphics[width=0.19\textwidth]{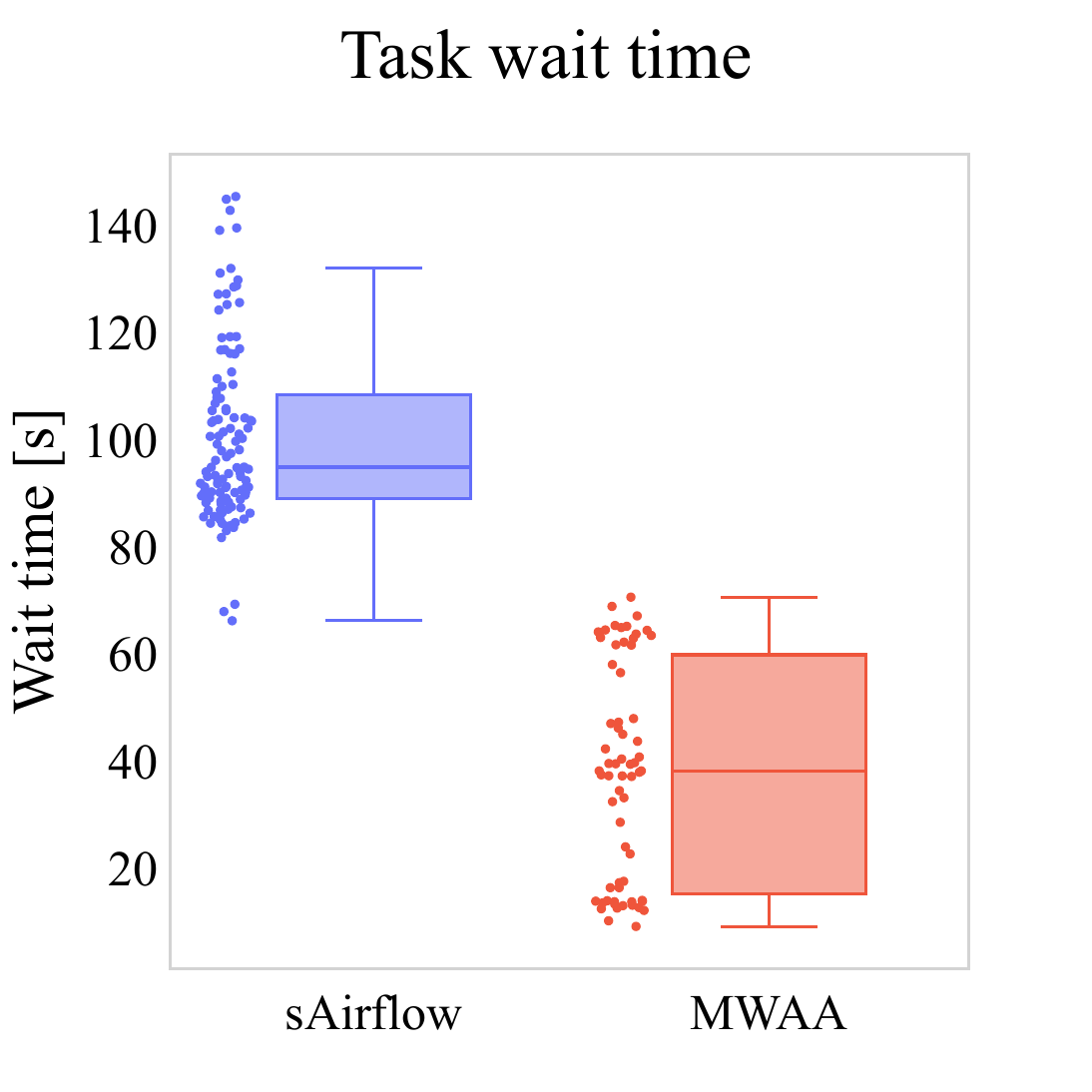}
    \includegraphics[width=0.19\textwidth]{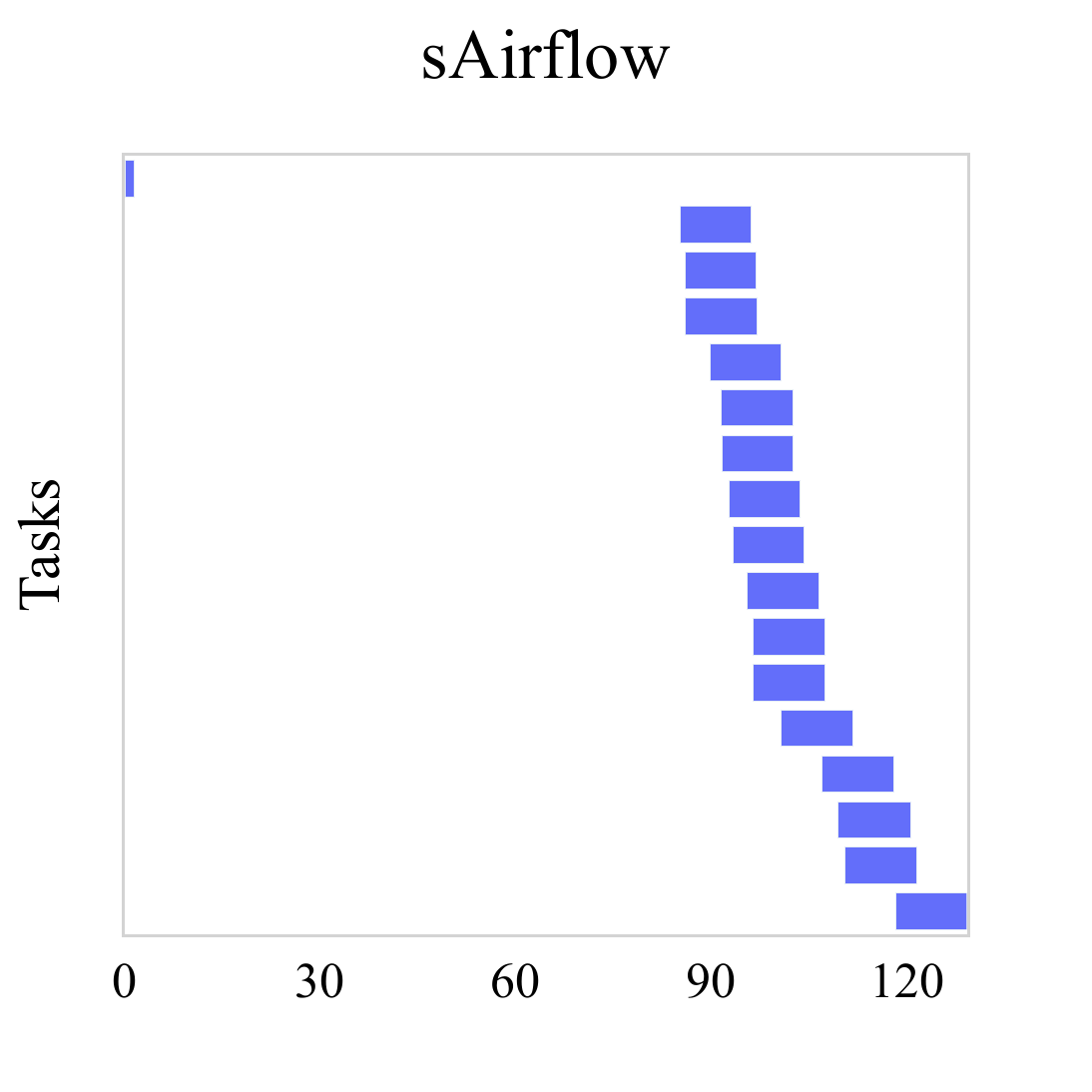}
    \includegraphics[width=0.19\textwidth]{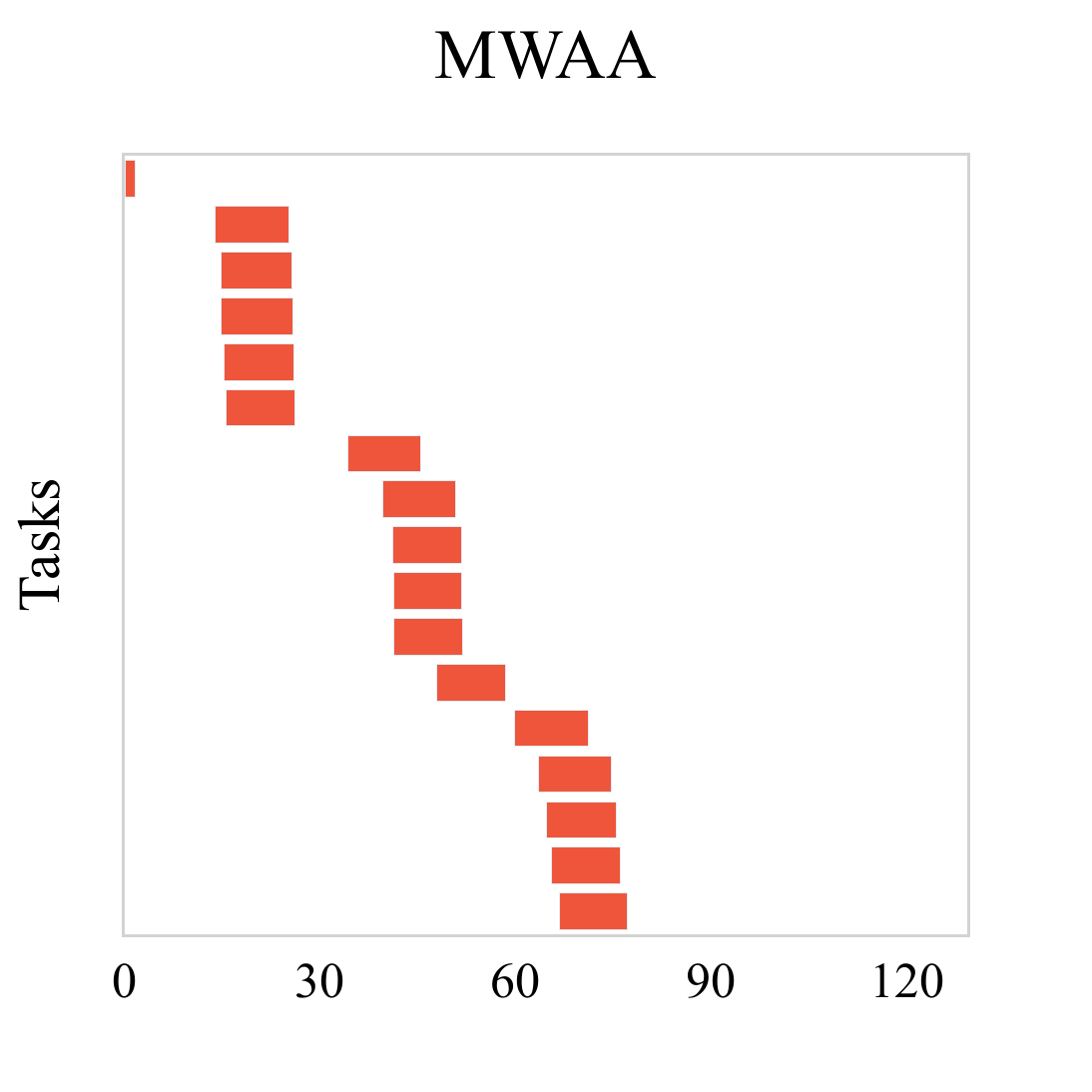}
  }
  \\
  \subfloat[$n=32$\label{fig:32task_parallel_batch}]{
    \includegraphics[width=0.19\textwidth]{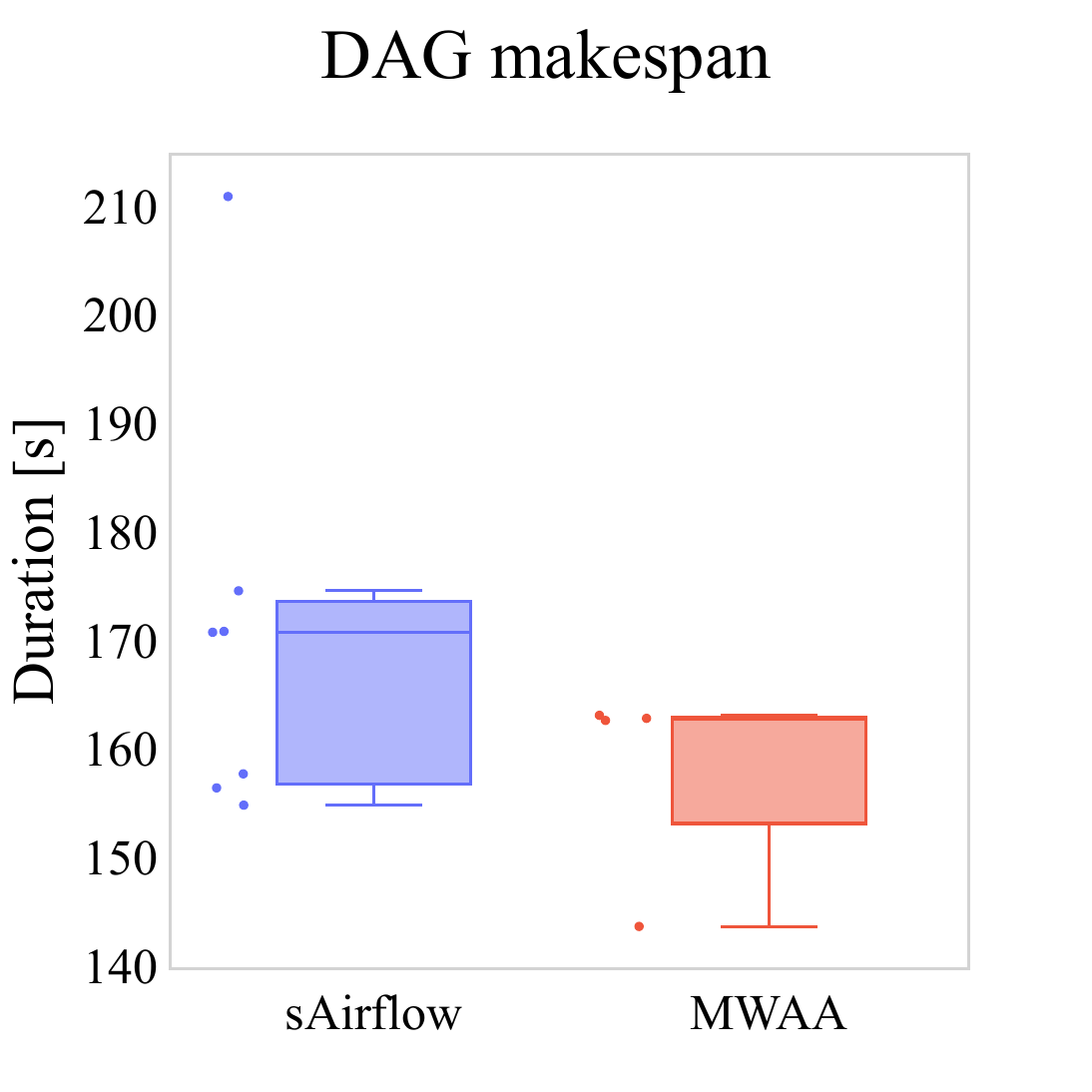}
    \includegraphics[width=0.19\textwidth]{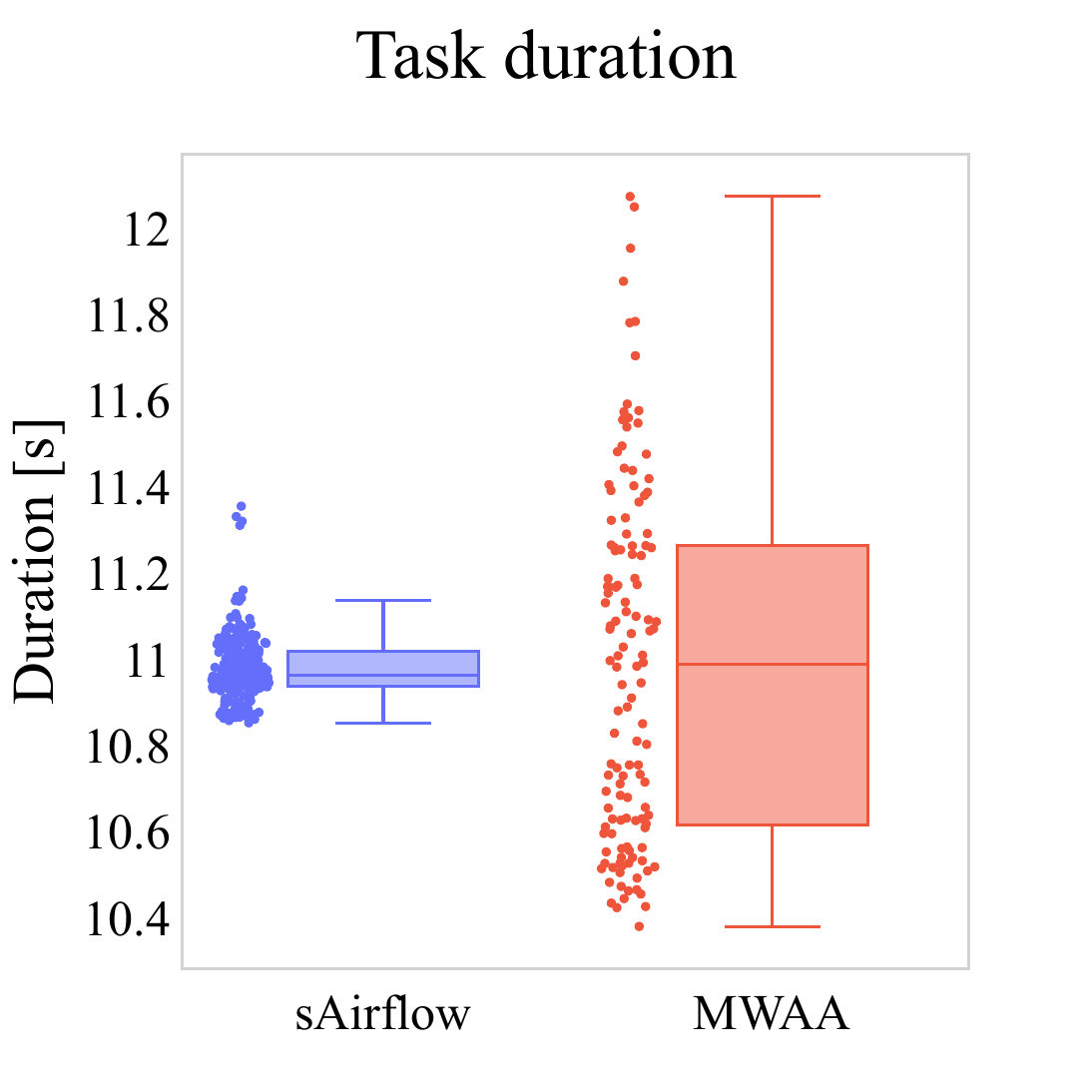}
    \includegraphics[width=0.19\textwidth]{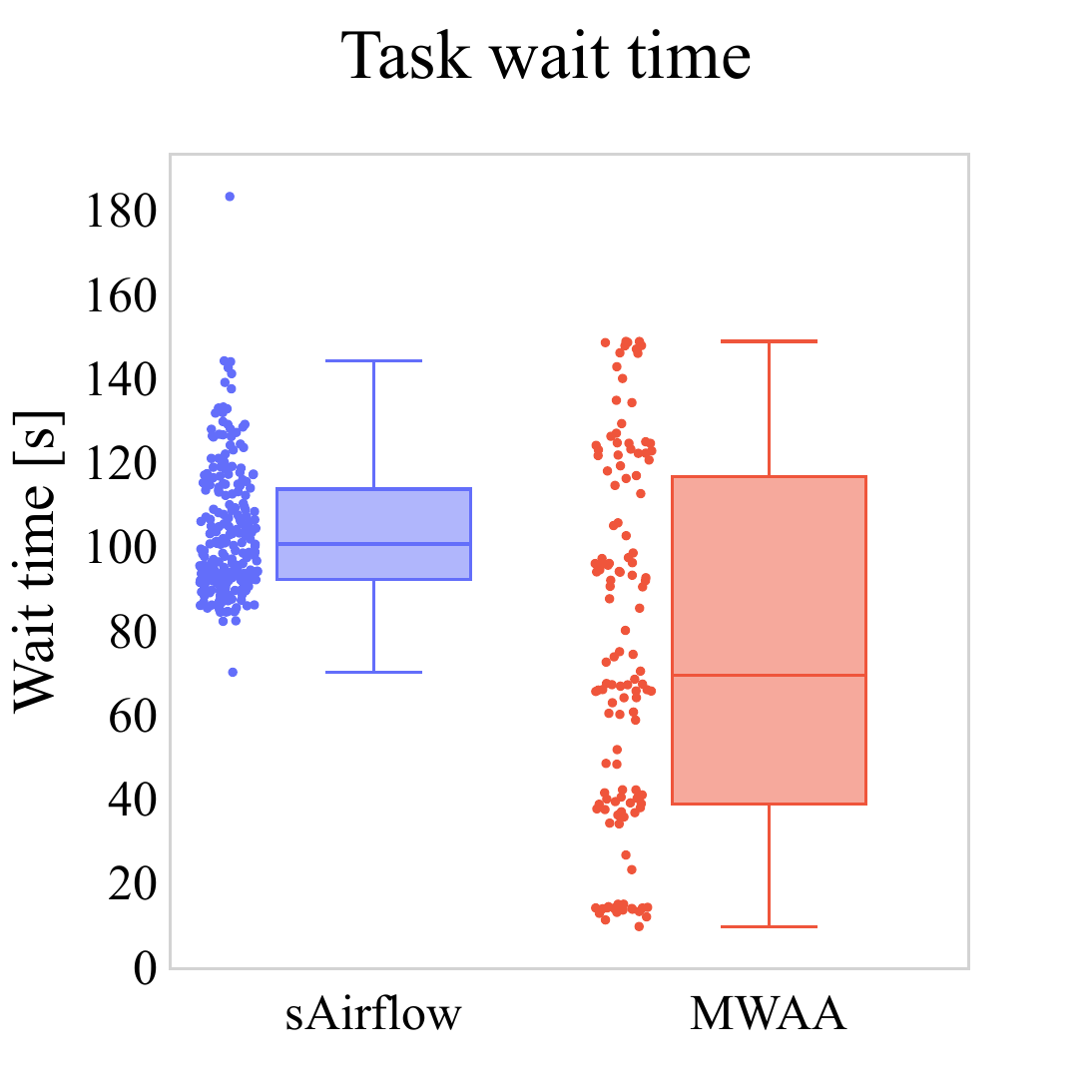}
    \includegraphics[width=0.19\textwidth]{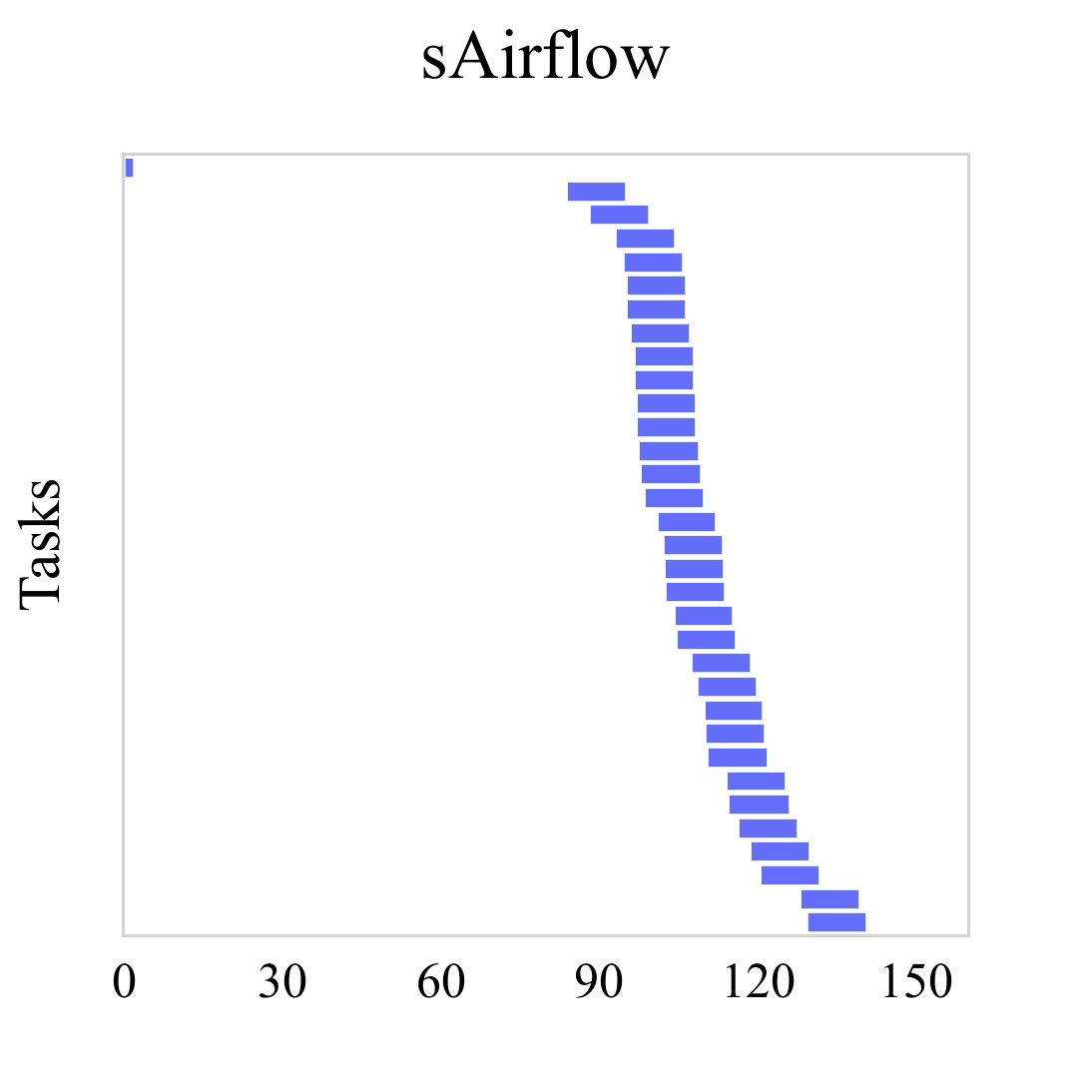}
    \includegraphics[width=0.19\textwidth]{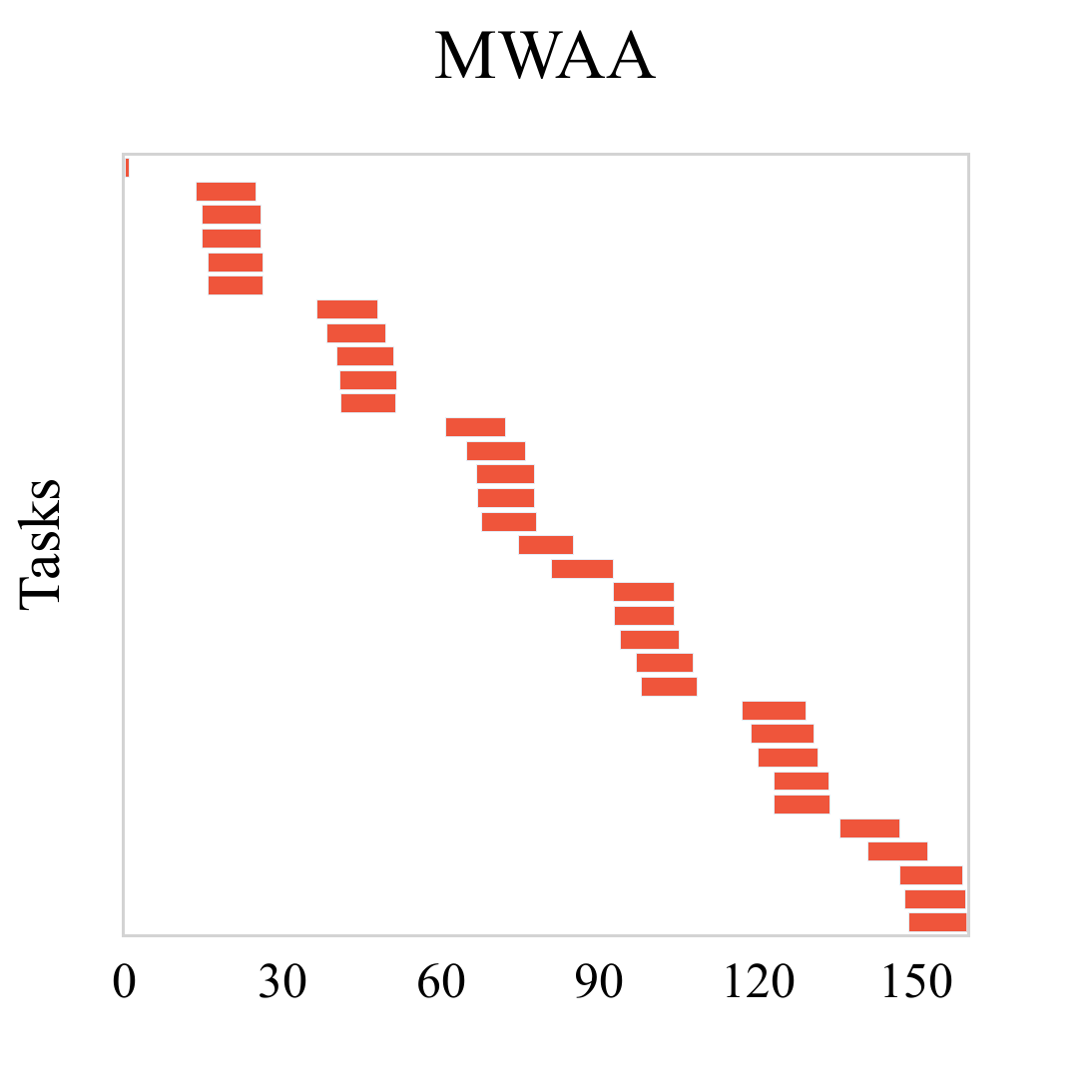}
  }
  \caption{Parallel DAG, $p=10$; sAirflow on CaaS (except for the immediately-completing DAG root executed on FaaS). MWAA results are from the cold start experiments.
  The Gantt charts on the right side shows a single DAG runs.
  MWAA does not manage to scale the cluster in time, so each job executes on the same worker node. sAirflow executes each task on a new container spawned by AWS Batch. sAirflow's start-up overhead (caused by AWS Batch queueing and loading the container image) heavily varies.}
\end{figure}

\subsection{Parallel DAGs: sAirflow with the container executor can match MWAA scaling} 

To measure how sAirflow with container workers scales horizontally, we run two experiments with the Parallel DAG with $p=10$, $T=10$, and $n \in \{16, 32\}$. The first, immediately-completing task of the workload is executed using the function executor, while the rest is executed on CaaS using the container executor. 
This configuration models a workload with a short coordinating task (running on FaaS), followed by long-running processing (while the experiment uses relatively short $p=10$ to reduce MWAA costs, the results would naturally extend for long-running tasks).
Consequently, only two tasks on the critical path, including one AWS Batch, mean the AWS Batch overhead is included only once.

We compare sAirflow with cold MWAA to measure sporadic, rather than continuous workloads. 
MWAA takes around 240–270~s to provision an additional worker. The difference between MWAA and sAirflow is that MWAA keeps the worker running in case there is more load (and there are notable issues in this approach due to the poor support for marking the removal of a worker \cite{aws_mwaa_autoscaling_downscaling_flaw}). 

Despite being slower for $n=16$ tasks (Fig.~\ref{fig:16task_parallel_batch}), sAirflow scales similarly when the number of tasks grows and outperforms the MWAA's slow autoscaling (Fig.~\ref{fig:32task_parallel_batch}). 
AWS Batch 
(Fig.~\ref{fig:16task_parallel_batch} and \ref{fig:32task_parallel_batch}) 
has worse scaling than AWS Lambda (Fig.~\ref{fig:mwaa_scale_out_64} and \ref{fig:mwaa_scale_out_125}). The parallel tasks do not start in the same burstable manner. As a consequence, there is no spike in the task duration metric (Fig.~\ref{fig:32task_parallel_batch}) as the load on the database during the tasks' startup is more evenly distributed in comparison with the previous experiments (Fig.~\ref{fig:parallel_cold}).

\newpage
\section{Monetary Cost Estimates}\label{appendix:monetary_cost}

\begin{table}[bp]
  \caption{Summary of the monetary cost comparison between MWAA and sAirflow in different scenarios. All cost is given in [\$], rounded up to two decimal places. Breakdown for sAirflow is in the following tables.} 
\label{table:sairflow_mwaa_scenario_cost}
\begin{tabularx}{\textwidth}{| p{.30\linewidth} | Y Y Y | Y Y Y Y |}
 \hline
 \multirow{ 2}{*}{Scenario} & MWAA & & & sAirflow & & & \\
 \cline{2-8}
 & Fixed & Workers & Total & Fixed & Variable & Executor & Total \\
\hline 
 \multirow{ 2}{*}{(1) Heavy} & \multirow{ 2}{*}{11.76} & \multirow{ 2}{*}{0.50} & \multirow{ 2}{*}{12.26} & \multirow{ 2}{*}{6.03} & 1.27 & FaaS & 7.30 \\
  & & & & & 0.89 & CaaS & 6.92 \\
  \hline
  (2) Distributed & 11.76 & 1.98 & 13.74 & 6.03 & 1.44 & FaaS & 7.47 \\ 
  \hline
  (3) Sporadic &  11.76 & 0 & 11.76 & 6.03 & 0.02 & FaaS & 6.05 \\
  \hline
  (4) Constant & 11.76 & 31.68 & 43.44 & 6.03 & 29.66 & CaaS & 35.69 \\
 \hline
\end{tabularx}
\end{table}

In this section, we estimate the monetary gains from serverless-ing Airflow.
When designing sAirflow, one of our goals was to reduce fixed costs. Yet, as Airflow relies on an always-available transactional database, sAirflow's cost must take into account that service --- and an associated CDC mechanism.
However, the resulting fixed cost is significantly lower.
The daily cost for running a small environment class for MWAA without any additional workers is \$11.76 \cite{aws_mwaa_pricing}, sAirflow amounts to \$6.03 in an equivalent configuration (Table \ref{table:serverless_airflow_cost_breakdown}). Most of the sAirflow's cost is attributed to running the database and the CDC subsystem. We compare the systems using four workloads with different characteristics:

\begin{enumerate}
    \item \textbf{Heavy Load}: A single DAG with 50 tasks in parallel, scheduled every 3 minutes, runs 20 times; each task takes 3 minutes. Thus, each DAG run takes 3 minutes, and the execution is finished in an hour.
    \item \textbf{Distributed Load}: A single DAG with 400 tasks, scheduled every 4 hours, runs 6 times. DAG execution time is always < 1 hour. Requires to scale to 35 tasks in parallel. Each task takes 1 minute to finish. In total, 6*400 tasks will run.
    \item \textbf{Sporadic Light Load}: A chain DAG with 20 tasks, scheduled every 24 hours, runs 1 time. Each task takes 30 seconds to finish.  
    \item \textbf{Constant Load}: A single DAG with 100 tasks in parallel, scheduled every 24 hours, runs only once. Each task takes 24 hours to finish.
\end{enumerate}

In every workload, we run both systems continuously for a 24-hour period.
We make the assumption that MWAA's autoscaling operates without any downtime(although this assumption is not accurate~\cite{aws_mwaa_autoscaling_downscaling_flaw}).
We assume that MWAA'a autoscaling works with zero downtime (which is not the case ). In all workloads, we run both systems for 24 hours. We exclude Free Tier from calculations for sAirflow. 

In general, sAirflow reduces the fixed cost by half. The final cost of sAirflow is lower by 17-48\% (Table \ref{table:sairflow_mwaa_scenario_cost}).
It's important to note that we do not include the Free Tier in our calculations for sAirflow.

Serverless offerings eliminate the necessity of optimizing deployment towards the worst-case scenario. By design, sAirflow is a more cost-effective option for sporadic and unpredictable workloads. The extent of cost savings varies depending on the specific nature of the workloads.

\renewcommand{\arraystretch}{1.3}

\begin{table}[tbp]
\caption{Cost breakdown of the major serverless components for running sAirflow in Scenario (1) with FaaS executor.}
\label{table:serverless_scenario1}
\begin{tabularx}{\textwidth}{|p{0.25\linewidth}  @{\hskip 2em} p{.5\linewidth} Y|}
 \hline
 Component & Notes & Cost [\$] \\ [0.5ex] 
 \hline
 Function Worker (Lambda) & 1000 invocations (one per task), 340MB memory, 3min each
 & 0.9963 \\
 Function Executor (Lambda) & 1000 invocations (one per task in the scheduled state), 256MB memory, 1s each & 0.0044 \\
 Scheduler (Lambda) & 1530 invocations, 512MB memory, 10s each; (input batch size is 10 events, there are $15 * 1000$ events for the tasks data and $20 * 15$ for the DAG schedules) & 0.1278 \\
  CDC event forwarded (Lambda) & 1530 invocations, 512MB memory, 1s each & 0.0131 \\
 Step functions & 1000 invocations, 4 state transitions each; \cite{aws-step-functions-pricing} & 0.1000 \\
 Dag files pull (S3) & 1000 GET requests in each task for the DAG file \cite{aws-s3-pricing} & 0.0004 \\
 Push task logs (S3) & 1000 PUT requests \cite{aws-s3-pricing} & 0.0050 \\
 Eventbridge & 1000 * 15 events ingested \cite{aws-eventbridge-pricing}  & 0.0150 \\
 SQS FIFO & 4320 calls (86400/20, seconds in the entire day per 20 seconds poll interval \cite{aws-sqs-short-long-polling}); Scheduler queue; \cite{aws-sqs-pricing} & 0.0022 \\
 SQS & 8640 calls (86400/10, seconds in the entire day per 10 seconds poll interval \cite{aws-sqs-short-long-polling}); Scheduler queue; \cite{aws-sqs-pricing} & 0.0035 \\ [1ex] 
\hline
Total &  & 1.2677 \\ [0.5ex] 
\hline
\end{tabularx}
\end{table}

\begin{table}[tbp]
  \caption{Cost breakdown of the major serverless components for running sAirflow in Scenario (2) with FaaS executor.}
\label{table:serverless_scenario2}

\begin{tabularx}{\textwidth}{|p{0.25\linewidth}  @{\hskip 2em} p{.5\linewidth} Y|}
 \hline
 Component & Notes & Cost [\$] \\ [0.5ex] 
 \hline
 Function Worker (Lambda) & 2400 invocations (one per task), 340MB memory, 1min each & 0.7974 \\
 Function Executor (Lambda) & 2400 invocations (one per task in the scheduled state), 256MB memory, 1s each & 0.0105 \\
 Scheduler (Lambda) & 3609 invocations, 512MB memory, 10s each; (input batch size is 10 events, there are 15 * 2400 events for the tasks data and 6 * 15 for the DAG schedules) & 0.3015 \\
  CDC event forwarded (Lambda) & 3609 invocations, 512MB memory, 1s each & 0.0308 \\
 Step functions & 2400 invocations, 4 state transitions each; \cite{aws-step-functions-pricing} & 0.24 \\
 Dag files pull (S3) & 2400 GET requests in each task for the DAG file \cite{aws-s3-pricing} & 0.001 \\
 Push task logs (S3) & 2400 PUT requests \cite{aws-s3-pricing} & 0.012 \\
 Eventbridge & 2400 * 15 events ingested \cite{aws-eventbridge-pricing}  & 0.036 \\
 SQS FIFO & 4320 calls (86400/20, seconds in the entire day per 20 seconds poll interval \cite{aws-sqs-short-long-polling}); Scheduler queue; \cite{aws-sqs-pricing} & 0.0022 \\
 SQS & 8640 calls (86400/10, seconds in the entire day per 10 seconds poll interval \cite{aws-sqs-short-long-polling}); Scheduler queue; \cite{aws-sqs-pricing} & 0.0035 \\ [1ex] 
\hline
Total &  & 1.4349 \\ [0.5ex] 
\hline
\end{tabularx}
\end{table}

\begin{table}[tbp]
  \caption{Cost breakdown of the major serverless components for running sAirflow in Scenario (3) with FaaS executor.}
\label{table:serverless_scenario3}

\begin{tabularx}{\textwidth}{|p{0.25\linewidth}  @{\hskip 2em} p{.5\linewidth} Y|}
 \hline
 Component & Notes & Cost [\$] \\ [0.5ex] 
 \hline
 Function Worker (Lambda) & 20 invocations (one per task), 340MB memory, 1min each & 0.0033 \\
 Function Executor (Lambda) & 20 invocations (one per task in the scheduled state), 256MB memory, 1s each & 0.0001 \\
 Scheduler (Lambda) & 32 invocations, 512MB memory, 10s each; (input batch size is 10 events, there are 1 * 20 events for the tasks data and 20 * 15 for the DAG schedules) & 0.0027 \\
  CDC event forwarded (Lambda) & 32 invocations, 512MB memory, 1s each & 0.0003 \\
 Step functions & 20 invocations, 4 state transitions each; \cite{aws-step-functions-pricing} & 0.002 \\
 Dag files pull (S3) & 20 GET requests in each task for the DAG file \cite{aws-s3-pricing} & 0 \\
 Push task logs (S3) & 20 PUT requests \cite{aws-s3-pricing} & 0.0001 \\
 Eventbridge & 20 * 15 events ingested \cite{aws-eventbridge-pricing}  & 0.0003 \\
 SQS FIFO & 4320 calls (86400/20, seconds in the entire day per 20 seconds poll interval \cite{aws-sqs-short-long-polling}); Scheduler queue; \cite{aws-sqs-pricing} & 0.0022 \\
 SQS & 8640 calls (86400/10, seconds in the entire day per 10 seconds poll interval \cite{aws-sqs-short-long-polling}); Scheduler queue; \cite{aws-sqs-pricing} & 0.0035 \\ [1ex] 
\hline
Total &  & 0.0145 \\ [0.5ex] 
\hline
\end{tabularx}
\end{table}

\begin{table}[tbp]
  \caption{Cost breakdown of the major serverless components for running sAirflow in Scenario (4) with CaaS executor.}
  \begin{tabularx}{\textwidth}{|p{0.25\linewidth}  @{\hskip 2em} p{.5\linewidth} Y|}
 \hline
 Component & Notes & Cost [\$] \\ [0.5ex] 
 \hline
 Container Worker (Batch) & 100 invocations (one per task), 0.25vCPU, 500MB memory, 24hours each 
 & 29.62 \\
 Container Executor (Lambda) & 100 invocations (one per task in the scheduled state), 256MB memory, 1s each & 0.0004 \\
 Scheduler (Lambda) & 152 invocations, 512MB memory, 10s each; (input batch size is 10 events, there are 15 * 100 events for the tasks data and 1 * 15 for the DAG schedules) & 0.0127 \\
  CDC event forwarded (Lambda) & 152 invocations, 512MB memory, 1s each & 0.0013 \\
 Step functions & 100 invocations, 4 state transitions each; \cite{aws-step-functions-pricing} & 0.01 \\
 Dag files pull (S3) & 100 GET requests in each task for the DAG file \cite{aws-s3-pricing} & 0 \\
 Push task logs (S3) & 100 PUT requests \cite{aws-s3-pricing} & 0.0005 \\
 Eventbridge & 100 * 15 events ingested \cite{aws-eventbridge-pricing}  & 0.0015 \\
 SQS FIFO & 4320 calls (86400/20, seconds in the entire day per 20 seconds poll interval \cite{aws-sqs-short-long-polling}); Scheduler queue; \cite{aws-sqs-pricing} & 0.0022 \\
 SQS & 8640 calls (86400/10, seconds in the entire day per 10 seconds poll interval \cite{aws-sqs-short-long-polling}); Scheduler queue; \cite{aws-sqs-pricing} & 0.0035 \\ [1ex] 
\hline
Total &  & 29.6521 \\ [0.5ex] 
\hline
\end{tabularx}
\label{table:serverless_scenario4}
\end{table}

\begin{table}[tbp]
  \caption{sAirflow's fixed price components breakdown. All cost is given in [\$], rounded up to two decimal places.}
  \label{table:serverless_airflow_cost_breakdown}
  \begin{tabularx}{\textwidth}{|p{.15\linewidth} @{\hskip 2em} p{.28\linewidth} Y Y Y Y|}
   \hline
   Component & Specification & Daily & Daily HA & Monthly & Monthly HA \\
   \hline
   RDS & instance: db.t3.small, SSD: 20GB  & 0.94 & 1.88 & 28.58 & 57.16 \\ 
   DMS & instance: t3.small, SSD: 10GB & 0.90 & 1.80 & 27.43 & 54.86 \\
   Kinesis & data streams & 0.72 & 0.72 & 21.90 & 21.90 \\
   NAT & instance: t2.micro, on-demand & 0.28 & 0.55 & 8.36 & 16.71 \\ 
   ECR & container images, 11*400MB & 0.02 & 0.02 & 0.50 & 0.50 \\ 
   SQL proxy &  & 0.72 & 0.72 & 21.90 & 21.90 \\ 
   AppRunner & 2GB of memory in a stopped state & 0.34 & 0.34 & 10.92 & 10.92 \\
  \hline 
  Total & & 3.92 & 6.03 & 119.59 & 183.95 \\  
  \hline 
  \end{tabularx}
  \end{table}
  
\end{document}